\documentclass[11pt]{article} 
\usepackage{amsmath,amssymb,mathrsfs,bbm,bm}
\usepackage[official]{eurosym}
\usepackage{comment}
\usepackage{natbib}
\usepackage[usenames, dvipsnames]{xcolor}
\usepackage{float}
\usepackage[colorlinks=true, citecolor=blue, urlcolor=blue, linkcolor=blue]{hyperref}
\usepackage{multirow}
\usepackage{setspace}
\usepackage{subcaption}
\usepackage{booktabs} 
\usepackage{fancyvrb}
\usepackage[framemethod=TikZ]{mdframed}
\usepackage{fancyhdr}
\usepackage[margin = 1in]{geometry}
\usepackage[space]{grffile}
\usepackage[nottoc]{tocbibind}

\numberwithin{figure}{section}
\numberwithin{table}{section}
\numberwithin{equation}{section}

\UseRawInputEncoding

\newcommand{\kSF}{0.60} 
\newcommand{\kDF}{0.45} 

\newcommand{\C}{c}
\newcommand{\D}{D}
\newcommand{\T}{T}
\newcommand{\bT}{\mathbf{T}}
\newcommand{\W}{\mathcal{W}}
\newcommand{\E}{\mathbb{E}}
\newcommand{\EW}{\mathbb{E}_\W}
\renewcommand{\P}{\mathbb{P}}
\newcommand{\PW}{\mathbb{P}_\W}

\newcommand{\I}{\mathbbm{1}}

\newcommand{\markedsection}[2]{\section[#2]{#2%
\sectionmark{#1}}
\sectionmark{#1}}

\newcounter{cajita}[section]
\renewcommand{\thecajita}{\arabic{section}.\arabic{cajita}}

\newcommand{\pathInputs}{./inputs}

\newcounter{snippet}[section]
\newcounter{figuras}[section]
\newcounter{tablas}[section]

\newcommand{\labelsnippet}[1]{%
  \newcounter{#1}%
  \setcounter{#1}{\value{snippet}}%
  \stepcounter{snippet}%
}%

\usepackage{listings}
\definecolor{codegreen}{rgb}{0,0.6,0}
\definecolor{codegray}{rgb}{0.5,0.5,0.5}
\definecolor{codepurple}{rgb}{0.58,0,0.82}
\definecolor{backcolour}{rgb}{0.95,0.95,0.95}
\definecolor{maincolour}{rgb}{0,0.1,0.1}

\lstdefinestyle{mystyle}{
	language=R,
    backgroundcolor=\color{backcolour},   
    commentstyle=\color{codegreen},
    keywordstyle=\color{magenta},
    keywords={},
    numberstyle=\tiny\color{codegray},
    stringstyle=\color{codepurple},
    basicstyle=\footnotesize\ttfamily\color{maincolour},
    breakatwhitespace=false,         
    breaklines=true,                 
    captionpos=b,                    
    keepspaces=true,                 
    numbers=none,                    
    numbersep=5pt,                  
    showspaces=false,                
    showstringspaces=false,
    showtabs=false,
    tabsize=2,
		xleftmargin=-15pt
}
\newcommand{\repliWeb}{https://rdpackages.github.io/replication}

\newcommand{\rsnip}[2] {
\VerbatimInput[frame=single,rulecolor=\color{JungleGreen},framesep=1.5mm, framerule=0.5mm,label=\fbox{#2}, fontsize=\small, xleftmargin=-0.6cm,xrightmargin=-0.6cm,baselinestretch=1.1,samepage=TRUE]{\pathInputs/#1}
}

\newcommand{\statasnip}[2] {
\VerbatimInput[frame=single,rulecolor=\color{Rhodamine},framesep=1.5mm, framerule=0.5mm, label=\fbox{#2}, fontsize=\small, xleftmargin=-0.6cm,xrightmargin=-0.6cm,baselinestretch=1.1,samepage=TRUE]{\pathInputs/#1-COMMAND-ONLY.txt}
}	
\renewcommand{\statasnip}[2] {}

\newcommand{\Rlink}[2]{
\href{\repliWeb/Vol-2-R-Snippet-#1.#2.R?attredirects=0}{R Snippet #1.#2}
}

\newcommand{\Slink}[2]{
\href{\repliWeb/Vol-2-Stata-Snippet-#1.#2.do?attredirects=0}{Stata Snippet #1.#2}
}

\begin{document}
\pagestyle{fancy}
\frenchspacing
\onehalfspacing

\begin{titlepage}
  \title{\bf A Practical Introduction to Regression Discontinuity Designs: Extensions}
  \author{
    Matias D. Cattaneo\thanks{Department of Operations Research and Financial Engineering, Princeton University.}
    \and Nicol\'as Idrobo\thanks{Department of Political Science, University of Pennsylvania.}     
    \and Roc\'{i}o Titiunik\thanks{Department of Politics, Princeton University.}
}

\date{\large{\vspace{0.5in} \today}}
\end{titlepage}
\maketitle

\thispagestyle{empty}

\begin{center}
Element prepared for\medskip\\
\textit{Cambridge Elements: Quantitative and Computational Methods for Social Science}\medskip\\
Cambridge University Press\bigskip\\
Published version:\medskip\\
\url{https://doi.org/10.1017/9781009441896}

\end{center}

\newpage\pagenumbering{roman}\setcounter{page}{1}

{\singlespacing\tableofcontents}

\newpage\pagenumbering{arabic}\setcounter{page}{1}
\setcounter{secnumdepth}{0}\section{Acknowledgments}\setcounter{secnumdepth}{3}

This monograph, together with its accompanying first part \citep*{Cattaneo-Idrobo-Titiunik_2020_Vol1}, collects and expands the instructional materials we prepared for more than $50$ short courses and workshops on Regression Discontinuity (RD) methodology that we taught between 2014 and 2023. These teaching materials were used at various institutions and programs, including the Asian Development Bank, the Philippine Institute for Development Studies, the International Food Policy Research Institute, the ICPSR's Summer Program in Quantitative Methods of Social Research, the Abdul Latif Jameel Poverty Action Lab, the Inter-American Development Bank, the Georgetown Center for Econometric Practice, the Universidad Cat\'{o}lica del Uruguay's Winter School in Methodology and Data Analysis, the Centre for Research in Economics and Management (NIPE), the Summer Institute of the Econometric Society (SIES), the International Initiative for Impact Evaluation (3ie), the National Bureau of Economic Research (NBER), the Bogot\'a Summer School in Economics, Amazon, Inc., the Summer School of the Italian Econometric Association (SIdE), the Northwestern Workshop on Research Design for Causal Inference, and the Mixtape Sessions. We also used these materials for teaching at the undergraduate and graduate level at Brigham Young University, Cornell University, Instituto Tecnol\'ogico Aut\'onomo de M\'exico, Pennsylvania State University, Pontificia Universidad Cat\'{o}lica de Chile, Princeton University, University of Michigan, University of Washington, and Universidad Torcuato Di Tella. We thank all these institutions and programs, as well as their many members and participants, for the interest, feedback, and support we received over the years.

The work collected in our two-volume monograph evolved and benefited from many insightful discussions with our current and former collaborators: Sebasti\'an Calonico, Rajita Chandak, Robert Erikson, Juan Carlos Escanciano, Max Farrell, Yingjie Feng, Brigham Frandsen, Sebasti\'an Galiani, Michael Jansson, Luke Keele, Marko Kla\v{s}nja, Xinwei Ma, Ricardo Masini, Kenichi Nagasawa, Brendan Nyhan, Filippo Palomba, Jasjeet Sekhon, Gonzalo Vazquez-Bare, Rae Yu, and Jos\'e Zubizarreta. Their intellectual contribution to our research program on RD designs has been invaluable, and certainly has made our monographs much better than they would have been otherwise. We also thank Alberto Abadie, Joshua Angrist, Ivan Canay, Richard Crump, Scott Cunningham, David Drukker, Jianqing Fan, Sebastian Galiani, Guido Imbens, Patrick Kline, Jason Lindo, Juliana Londo\~no-V\'elez, Justin McCrary, David McKenzie, Douglas Miller, Aniceto Orbeta, Zhuan Pei, and Andres Santos for the many stimulating discussions and criticisms we received from them over the years, which also shaped our work in important ways. The co-Editors Michael Alvarez and Nathaniel Beck offered insightful and constructive comments on several preliminary drafts of our manuscripts, including the suggestion of splitting the content into two stand-alone volumes. They were also infinitely patient with us when our plan to complete this volume was massively delayed due to the Covid-19 pandemic. Last but not least, we gratefully acknowledge the support of the National Science Foundation through grants \href{http://www.nsf.gov/awardsearch/showAward?AWD_ID=1357561}{SES-1357561}, \href{https://www.nsf.gov/awardsearch/showAward?AWD_ID=2019432}{SES-2019432}, and \href{https://www.nsf.gov/awardsearch/showAward?AWD_ID=2019432}{SES-2241575}.

The goal of our two-part monograph is purposely practical and hence we focus on the empirical analysis of RD designs. We do not seek to provide a comprehensive review of the methodological literature on RD designs, which we do in \cite{Cattaneo-Titiunik_2022_ARE}, nor discuss theoretical aspects in detail. As we did in the first volume, we mostly refrain from citing prior literature in the main text; instead, we provide a short list of references at the end of each section to guide readers who are interested in further methodological details and formal theoretical results. In this second part, we employ the replication data from \citet*{Cattaneo-Frandsen-Titiunik_2015_JCI}, \citet*{Lindo-Sanders-Oreopoulos_2010_AEJ}, \citet*{LondonoVelezRodriguezSanchez_2020_AEJ}, and \citet*{Keele-Titiunik_2015_PA} for empirical illustration of the different topics. We thank these authors for making their data and codes publicly available. We provide complete replication codes in \texttt{Python}, \texttt{R}, and \texttt{Stata} for all the empirical analyses discussed throughout the monograph. The general purpose, open-source software we use, as well as all replication codes and other supplementary materials, can be found at:
\begin{center}\url{https://rdpackages.github.io/}\end{center}

\clearpage
\section{Introduction}
\label{sec:intro}
\setcounter{figuras}{1}
\setcounter{snippet}{1}
\setcounter{tablas}{1}

The Regression Discontinuity (RD) design has emerged as one of the most credible research designs in the social, behavioral, biomedical, and statistical sciences for program evaluation and causal inference in the absence of an experimentally assigned treatment. In this manuscript, we continue the discussion in \citet*{Cattaneo-Idrobo-Titiunik_2020_Vol1}, covering practical topics in the analysis and interpretation of RD designs that were not included in our first monograph due to space constraints. While our discussion is meant to be self-contained, we recommend that readers who are unfamiliar with the basic features of the RD design consult our first monograph before reading this one, as several concepts and ideas discussed previously will be assumed known in this volume. In what follows, we refer to the first monograph as \textit{Foundations}, and to this monograph as \textit{Extensions}.

The RD design is defined by three fundamental ingredients: a score (also known as a running variable, forcing variable, or index), a cutoff, and a treatment rule that assigns units to treatment or control based on a hard-thresholding rule. All units receive a score, and the treatment is assigned to units whose value of the score exceeds the cutoff and not assigned to units whose value of the score is below the cutoff. This assignment rule implies that the probability of treatment assignment changes abruptly at the known cutoff. If units are not able to perfectly determine or manipulate the exact value of the score that they receive, the discontinuous change in the probability of treatment assignment can be used to study the effect of the treatment on outcomes of interest, at least locally, because units with scores barely below the cutoff can be used as comparisons or ``counterfactuals'' for units with scores barely above it.

To formalize, we assume that there are $n$ units, indexed by $i=1,2,\ldots,n$, and each unit receives a score $X_i$. Units with $X_i \geq \C$ are assigned to the treatment condition, and units with $X_i < \C$ are assigned to the untreated or control condition, where $\C$ denotes the cutoff. We summarize this in the treatment assignment rule $\T_i=\I(X_i \geq \C)$, where $\I(\cdot)$ is the indicator function. In \textit{Foundations}, we focused exclusively on the canonical Sharp RD design where the running variable is continuous and univariate, there is a single cutoff determining treatment assignment, compliance with treatment assignment is perfect, and the analysis is conducted using continuity-based methods (e.g., local polynomial approximations with robust bias correction inference). The goal of this manuscript is to discuss practical RD analysis when these assumptions are extended.

We adopt the potential outcomes framework---see \citet{Imbens-Rubin_2015_Book} for an introduction to potential outcomes and causality, and \citet{Abadie-Cattaneo_2018_ARE} for a review of program evaluation methodology. We assume that each unit has two potential outcomes, $Y_i(1)$ and $Y_i(0)$, which correspond, respectively, to the outcomes that would be observed under treatment and control. Treatment effects are therefore defined in terms of comparisons between features of (the distribution of) both potential outcomes, such as their mean or quantiles. If unit $i$ receives the treatment, we observe the unit's outcome under treatment, $Y_i(1)$, but $Y_i(0)$ remains unobserved, while if unit $i$ is untreated, we observe $Y_i(0)$ but not $Y_i(1)$. This is known as the fundamental problem of causal inference. The observed outcome $Y_i$ is therefore defined as
\begin{equation*}
Y_i = (1-\T_i) \cdot Y_i(0) + \T_i \cdot Y_i(1) =
\left\{ 
\begin{array}{ccl}
Y_i(0) && \text{if } X_i<\C \\ 
Y_i(1) && \text{if } X_i\geq \C \\ 
\end{array}
\right. .
\end{equation*}
\begin{figure}[ht]
	\centering
	\includegraphics[scale=\kSF]{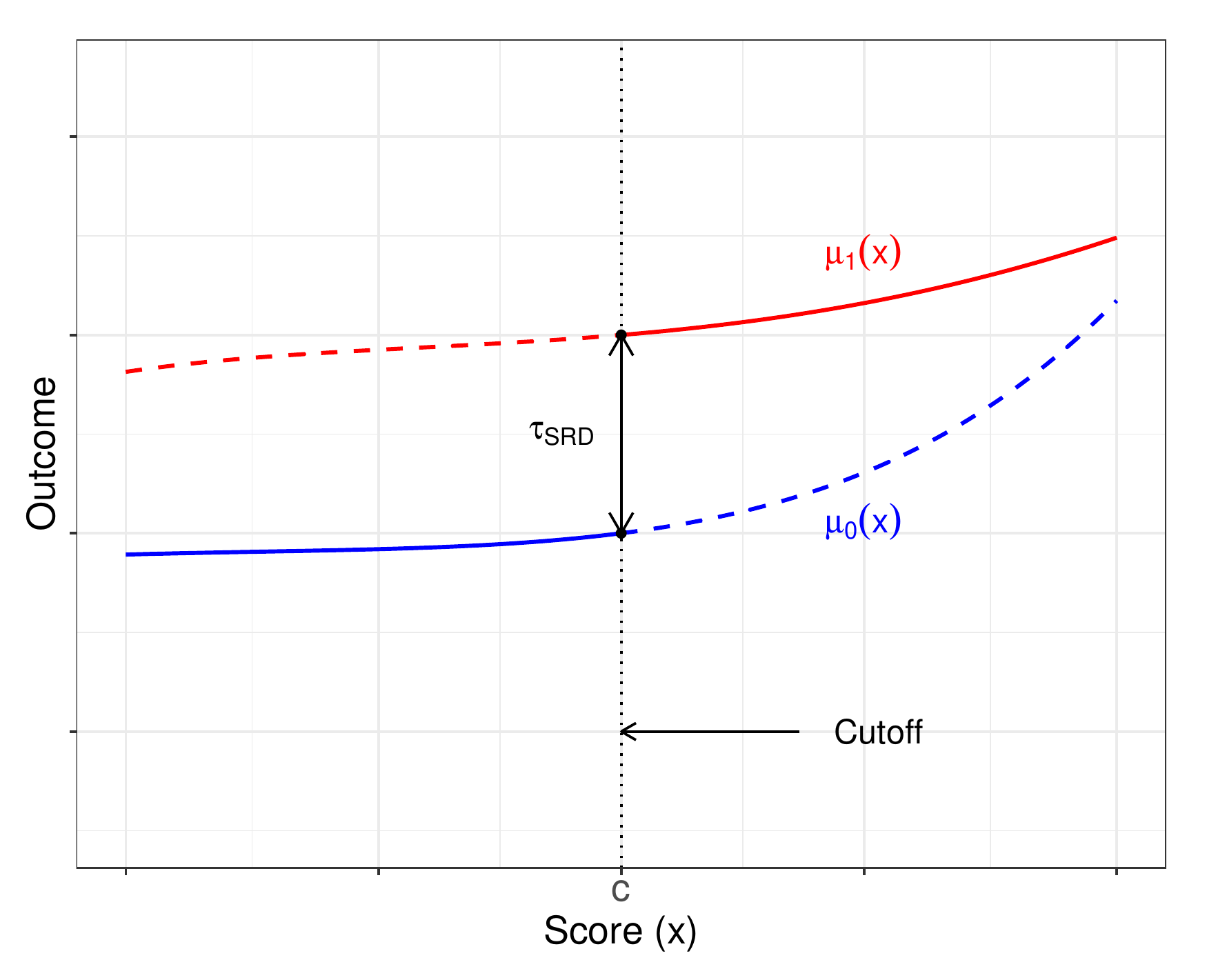}
	\caption{Canonical Sharp RD Effect}\label{fig:SRDeffect}
\end{figure}
The canonical Sharp RD design assumes that the potential outcomes $(Y_i(1),Y_i(0))$, $i=1,\dots,n$, are random variables by virtue of random sampling and focuses on the average treatment effect at the cutoff
\[\tau_{\mathtt{SRD}} \equiv \E[Y_i(1)-Y_i(0)|X_i=\C] = \mu_1(\C) - \mu_0(\C),\]
where $\mu_0(x)=\E[Y_i(0)|X_i=x]$ and $\mu_1(x)=\E[Y_i(1)|X_i=x]$. This parameter is called the Sharp RD treatment effect, and is depicted in Figure \ref{fig:SRDeffect}, where we also plot the regression functions $\mu_0(x)$ and $\mu_1(x)$ for values of the score $X_i=x$, with solid and dashed lines corresponding to their estimable and non-estimable portions, respectively. The continuity-based framework for RD analysis assumes that $\mu_0(x)$ and $\mu_1(x)$ are continuous at $x=\C$, which gives
\begin{equation}\label{HTV}
    \tau_{\mathtt{SRD}} = \lim_{x\downarrow{\C}} \E[Y_i|X_i=x] - \lim_{x\uparrow{\C}} \E[Y_i|X_i=x].
\end{equation}
Equation (\ref{HTV}) says that, if the average potential outcomes given the score are continuous functions of the score at $\C$, the difference between the limits of the treated and control average observed outcomes as the score approaches the cutoff is equal to the average treatment effect at the cutoff. This identification result is due to \citet*{Hahn-Todd-vanderKlaauw_2001_ECMA}, and has spurred a large body of methodological work on identification, estimation, inference, graphical presentation, and falsification for various RD design settings. In \textit{Foundations}, we focused exclusively on the canonical Sharp RD design, presenting a practical discussion of the methods developed by \citet*{Hahn-Todd-vanderKlaauw_2001_ECMA}, \citet{Lee_2008_JoE}, \citet{McCrary_2008_JoE}, \citet*{Calonico-Cattaneo-Titiunik_2014_ECMA,Calonico-Cattaneo-Titiunik_2015_JASA}, \citet*{Calonico-Cattaneo-Farrell_2018_JASA,Calonico-Cattaneo-Farrell_2020_ECTJ,Calonico-Cattaneo-Farrell_2022_Bernoulli}, \citet*{Calonico-Cattaneo-Farrell-Titiunik_2019_RESTAT}, and \citet*{Cattaneo-Jansson-Ma_2020_JASA}, among others. 

In this second monograph, we discuss several topics in RD methodology that build on and extend the analysis of RD designs introduced in \textit{Foundations}. We first present an alternative RD conceptual framework based on local randomization ideas, introduced by \citet*{Cattaneo-Frandsen-Titiunik_2015_JCI} and further developed by \citet*{Cattaneo-Titiunik-VazquezBare_2016_Stata,Cattaneo-Titiunik-VazquezBare_2017_JPAM}. This methodological approach can be useful in RD designs with discretely-valued scores, and can also be used more broadly as a complement to the continuity-based approach in other settings. Second, employing both continuity-based and local randomization approaches, we extend the canonical Sharp RD design in multiple directions: fuzzy RD designs, RD designs with discrete scores, and multi-dimensional RD designs. Most of the methods we discuss build on \citet*{Calonico-Cattaneo-Titiunik_2014_ECMA}, \citet*{Calonico-Cattaneo-Farrell-Titiunik_2019_RESTAT}, \citet{Keele-Titiunik_2015_PA}, \citet*{Cattaneo-Keele-Titiunik-VazquezBare_2016_JOP,Cattaneo-Keele-Titiunik-VazquezBare_2021_JASA}, \citet*{Cattaneo-Titiunik-VazquezBare_2020_Stata}, and references therein.

We start in Section \ref{sec:localrand} by introducing the local randomization framework for RD designs. In this framework, the score values are assumed to be as-if randomly assigned in a small window around the cutoff, so that placement above or below the cutoff and hence treatment assignment can be interpreted to be as-if experimental. This contrasts with the continuity-based approach, where extrapolation to the cutoff plays a predominant role. Once the local randomization assumption is invoked, the analysis can proceed by using tools from the analysis of experiments. This alternative approach, which we call the \textit{local randomization} approach to RD analysis, often requires stronger assumptions than the continuity-based approach discussed in \textit{Foundations}, and for this reason, it is not always applicable. Our discussion includes how to interpret the required assumptions, and how to perform estimation, inference, and falsification.

We continue in Section \ref{sec:FuzzyRD} with a discussion of the Fuzzy RD design where, in contrast to the Sharp RD design, compliance with the treatment assignment is imperfect: some units above the cutoff fail to take the treatment despite being assigned to take it, and/or some units below the cutoff take the treatment despite being assigned to the untreated condition. We define several parameters of interest that can be recovered under noncompliance, and discuss how to employ both continuity-based and local randomization approaches for analysis. We also discuss how to perform falsification analysis under noncompliance.

In Section \ref{sec:discrete}, we discuss RD designs where the running variable is discrete instead of continuous, and hence multiple units share the same value of the score. For example, the Grade Point Average (GPA) used by universities is often calculated up to one or two decimal places, and collecting data on all students in a college campus results in a dataset where hundreds or thousands of students have the same GPA value. In the RD design, the existence of such ``mass points'' in the score variable sometimes requires using alternative methods, as the standard continuity-based methods discussed in \textit{Foundations} are no longer generally applicable without modifications. We discuss when and why continuity-based methods may be inadequate to analyze RD designs with discrete scores, and also describe how the local randomization approach can be a useful alternative framework for analysis.

We devote the last section, Section \ref{sec:multiRD}, to generalize the assumption of a treatment assignment rule that depends on a single score and a single cutoff. We start by discussing \textit{Multi-Cutoff RD designs}, settings where units have a single score, but different subsets of units face different cutoff values. We then discuss RD designs with multiple running variables, which we refer to as \textit{Multi-Score RD designs}, where the treatment rule requires that two or more scores be above a cutoff in order to receive the treatment. Our discussion includes a particular case of the Multi-Score RD design where assignment to treatment changes discontinuously at the border that separates two geographic areas, typically known as the \textit{Geographic RD design}. Throughout this section, we explain how to generalize the methods discussed both in \textit{Foundations} and in the first sections of this monograph to both types of designs, which we call \textit{Multi-Dimensional RD designs}.

Each section illustrates the methods with a different empirical application. In Section \ref{sec:localrand}, we use the data employed by \citet{Cattaneo-Frandsen-Titiunik_2015_JCI} to study the incumbency advantage of political parties in elections for the U.S. Senate. In Sections \ref{sec:FuzzyRD} and \ref{sec:multiRD}, we use the data provided by \citet{LondonoVelezRodriguezSanchez_2020_AEJ} to study the effect of a government subsidy in Colombia on enrollment in higher-education institutions. In Section \ref{sec:discrete}, we use the data in \citet{Lindo-Sanders-Oreopoulos_2010_AEJ}, who analyze the effects of academic probation on subsequent academic achievement. Finally, in Section \ref{sec:multiRD}, we use the geographic data in \citet*{Keele-Titiunik_2015_PA}, who study the effect of campaign ads on voter turnout.

As in \textit{Foundations}, all the RD methods we discuss and illustrate are implemented using various general-purpose software packages (\url{https://rdpackages.github.io/}), which are free and available for \texttt{Python}, \texttt{R}, and \texttt{Stata}, three leading statistical software environments in the social sciences. Each numerical illustration we present includes an \texttt{R} command with its output, which we truncate when appropriate to conserve space. The analogous \texttt{Stata} and \texttt{Python} commands are not shown in the text but are available online. The codes that replicate all our analyses are available at \url{https://github.com/rdpackages-replication/CIT_2024_CUP}, which complement the replication codes for \textit{Foundations} available at \url{https://github.com/rdpackages-replication/CIT_2020_CUP}.

The local polynomial methods for continuity-based RD analysis are implemented in the package \texttt{rdrobust}, which is discussed in three companion software articles: \citet*{Calonico-Cattaneo-Titiunik_2014_Stata}, \citet*{Calonico-Cattaneo-Titiunik_2015_R}, and \citet*{Calonico-Cattaneo-Farrell-Titiunik_2017_Stata}; see also \citet*{Cattaneo-Titiunik-VazquezBare_2019_Stata} for power calculations and related methods. The \texttt{rdrobust} package has three functions specifically designed for continuity-based RD analysis: \texttt{rdbwselect} for data-driven bandwidth selection methods, \texttt{rdrobust} for local polynomial point estimation and inference, and \texttt{rdplot} for graphical RD analysis. In addition, the package \texttt{rddensity}, discussed by \citet*{Cattaneo-Jansson-Ma_2018_Stata}, provides manipulation tests of density discontinuity based on local polynomial density estimation methods. The accompanying package \texttt{rdlocrand}, which is discussed by \citet*{Cattaneo-Titiunik-VazquezBare_2016_Stata}, implements all the local randomization RD methods that we use throughout this volume. This package has two main functions: \texttt{rdwinselect} selects the local randomization window around the cutoff using pre-treatment covariates, and \texttt{rdrandinf} performs finite-sample and large sample inference in the selected window. Finally, to analyze multi-dimensional RD designs we employ the package \texttt{rdmulti}, which is discussed by \citet{Cattaneo-Titiunik-VazquezBare_2016_Stata}. This package has three main functions: \texttt{rdmc} for multi-cutoff estimation and inference, \texttt{rdmcplot} for multi-cutoff RD plots, and \texttt{rdms} for multi-score estimation and inference. 

We also provide further references for readers who wish to go beyond the contents we cover. For readers interested in other practical introductions to RD designs with additional references and empirical illustrations, we recommend \citet{Cattaneo-Titiunik-VazquezBare_2017_JPAM} in economics and public policy, \citet{Cattaneo-Titiunik-VazquezBare_2020_BookCh} in political science, and \citet{Cattaneo-Keele-Titiunik_2023_SIM} in biostatistics and medicine. For readers interested in the technical results underlying our practical discussion, we offer further references at the end of each section. We warn the reader, however, that these references are not meant to be exhaustive. A comprehensive review of the methodological RD literature can be found in \cite{Cattaneo-Titiunik_2022_ARE}.

\clearpage
\markedsection{Local Randomization RD Approach}{The Local Randomization Approach to RD Analysis}
\label{sec:localrand}
\setcounter{figuras}{1}
\setcounter{snippet}{1}
\setcounter{tablas}{1}

In \textit{Foundations}, we discussed in detail the continuity-based approach to RD analysis. That approach, which is based on assumptions of continuity (and further smoothness) of the regression functions $\mu_1(x)=\E[Y_i(1)|X_i=x]$ and $\mu_0(x)=\E[Y_i(0)|X_i=x]$, is by now the standard and most commonly used method to analyze RD designs. In this section, we discuss a different framework for RD analysis that is based on a formalization of the idea that the RD design can be interpreted as a randomized experiment near the cutoff $\C$. This alternative framework can be used as a complement and robustness check to the continuity-based analysis when the running variable is continuous (under appropriate assumptions), and is a natural framework for analysis when the running variable is discrete and has few mass points, a case we discuss in Section \ref{sec:discrete}.

When the RD design was first introduced by \citet{Thistlethwaite-Campbell_1960_JEP}, the justification for this then-novel research design was not based on approximation and extrapolation of smooth regression functions, but rather on the idea that the abrupt change in treatment status that occurs at the cutoff leads to a treatment assignment mechanism that, near the cutoff, resembles the assignment that we would see in a randomized experiment. Indeed, the authors described a hypothetical experiment where the treatment is randomly assigned near the cutoff as an ``experiment for which the regression-discontinuity analysis may be regarded as a substitute'' \citep[p. 310]{Thistlethwaite-Campbell_1960_JEP}. 

The idea that the treatment assignment is ``as good as'' randomly assigned in a neighborhood of the cutoff has been often invoked in the continuity-based framework to describe the required identification assumptions in an intuitive way, and it has also been used to develop formal results. However, within the continuity-based framework, the formal derivation of identification and estimation results always ultimately relies on continuity and differentiability of regression functions, and the idea of local randomization is used as a heuristic device only. In contrast, the \textit{local randomization approach} to RD analysis formalizes the idea that the RD design behaves like a randomized experiment near the cutoff by imposing explicit randomization-type assumptions that are stronger than the continuity-based conditions.

In a nutshell, the local randomization approach imposes conditions so that units above and below the cutoff whose score values lie in a small window around the cutoff are comparable to each other and thus can be studied ``as if'' they had been randomly assigned to treatment or control. The local randomization approach adopts this assumption explicitly, not as a heuristic interpretation, and builds a set of statistical tools directly based on this specific assumption. In most cases, the analysis proceeds conditionally on those units whose scores fall within a window near the cutoff. 

We discuss how adopting an explicit randomization assumption near the cutoff allows for the use of new methods of estimation and inference for RD analysis, highlighting the differences between this approach and the continuity-based approach. When the running variable is continuous, the local randomization approach typically requires stronger assumptions than the continuity-based approach; in these cases, it is natural to use the continuity-based approach for the main RD analysis, and to use the local randomization approach as a robustness check. But in settings where the running variable is discrete or other departures from the canonical RD framework occur, the local randomization approach no longer imposes the strongest assumptions and can be a natural and useful method for analysis. 

Recall that we are considering an RD design where the (continuous) score is $X_i$, the treatment assignment is $\T_i=\I(X_i \geq \C)$, and $Y_i$ is the observed outcome with underlying potential outcomes $Y_i(0)$ and $Y_i(1)$ under control and treatment, respectively. Throughout this section, we maintain the assumption that the RD design is sharp and thus compliance is perfect. We relax this assumption in Section \ref{sec:FuzzyRD}, where we discuss the Fuzzy RD design.

When the RD design is based on a local randomization assumption, instead of assuming that the unknown regression functions $\mu_1(x)=\E[Y_i(1)|X_i=x]$ and $\mu_0(x)=\E[Y_i(0)|X_i=x]$ are continuous at the cutoff, the researcher assumes that there is a small window around the cutoff, defined as $\W=[\C-w, \C+w]$ for a scalar $w>0$, such that for all units whose scores fall in that window their placement above or below the cutoff is assigned as it would have been assigned in a randomized experiment---an assumption that is sometimes called \textit{as if random assignment}. Formalizing this assumption requires careful consideration of the conditions that are guaranteed to hold in an actual experimental assignment.

\begin{figure}[ht]
	\centering
	\begin{subfigure}{0.48\textwidth}
		\centering
		\includegraphics[scale=\kDF]{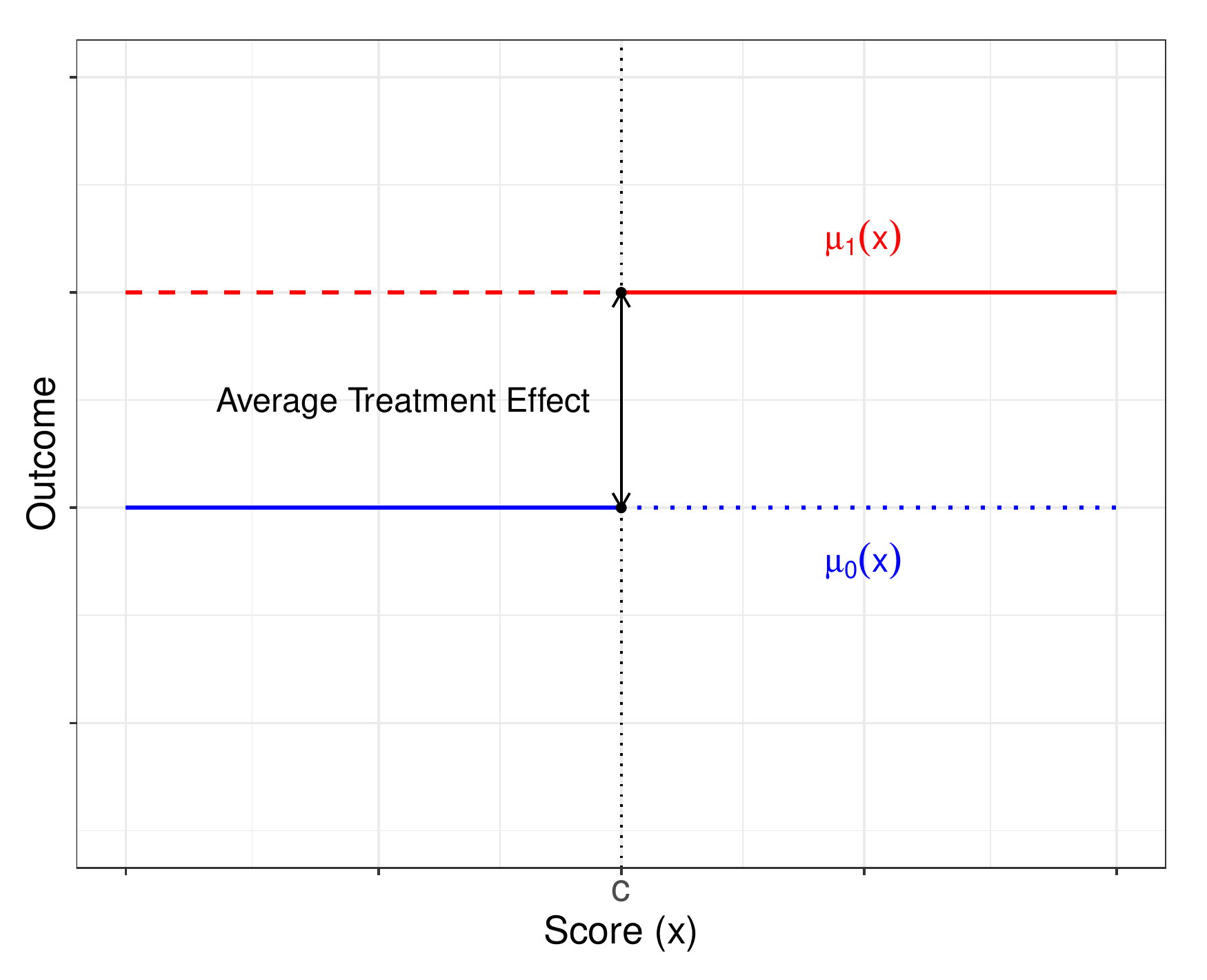}
		\caption{Randomized Experiment}\label{exp}		
	\end{subfigure}
	\begin{subfigure}{0.48\textwidth}
		\centering
		\includegraphics[scale=\kDF]{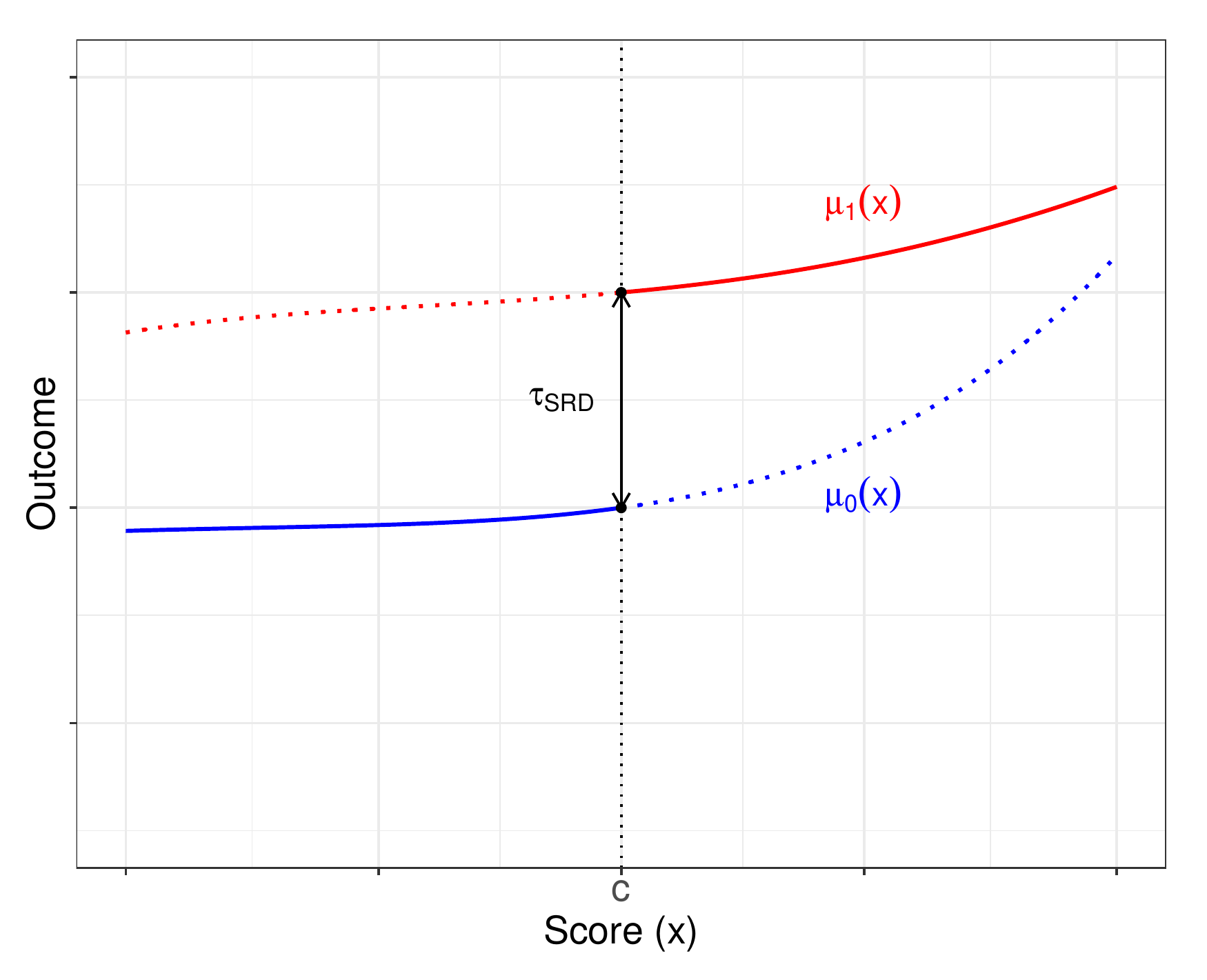}
		\caption{RD Design}\label{rd}
	\end{subfigure}
	\caption{Experiment vs. RD Design}\label{fig:RDvsExp}
\end{figure}

There are important differences between the RD design and an actual randomized experiment. To discuss such differences, we start by noting that any simple experiment can be recast as an RD design where the score is a randomly generated number, and the cutoff is chosen to ensure a certain probability of treatment. For example, consider an experiment in a student population that randomly assigns a scholarship with probability $1/2$. This experiment can be recast as an RD design where each student is assigned a random number with uniform distribution between $0$ and $100$, say, and the scholarship is given to students whose number is above $50$. We illustrate this scenario in Figure \ref{fig:RDvsExp}(\subref{exp}).

The crucial feature of a randomized experiment recast as an RD design is that the running variable, by virtue of being a randomly generated number, is unrelated to the potential outcomes. This is the reason why, in Figure \ref{fig:RDvsExp}(\subref{exp}), $\mu_1(x)=\E[Y_i(1)|X_i=x]$ and $\mu_0(x)=\E[Y_i(0)|X_i=x]$ are constant for all values of $x$. Since the regression functions are flat, the vertical distance between them can be recovered by the difference between the average observed outcomes among all units in the treatment and control groups, i.e. $\E[Y_{i} | X_i \geq 50] - \E[Y_{i} | X_i < 50] = \E[Y_i(1) | X_i \geq 50] - \E[Y_i(0) | X_i < 50] = \E[Y_i(1)] - \E[Y_i(0)]$.

In contrast, in the standard continuity-based RD design there is no requirement that the potential outcomes be unrelated to the running variable over its support. Figure \ref{fig:RDvsExp}(\subref{rd}) illustrates a continuity-based RD design where the average treatment effect at the cutoff, $\tau_\mathtt{SRD}$, is the same as in the experimental setting in Figure \ref{fig:RDvsExp}(\subref{exp}) but where the average potential outcomes are non-constant functions of the score. This non-constant relationship between running variable and potential outcomes is characteristic of most RD designs: the RD score (e.g. poverty index, vote share, or exam grade) is often strongly related to units' ability, resources, or performance, so that units with higher scores are often systematically different from units whose scores are lower. For example, in an RD design where the score is a party's vote share in a given election and the outcome of interest is the party's vote share in the following election, the relationship between the score and the outcome will likely exhibit a positive slope, as districts that strongly support a party in one election are likely to continue to support the same party in the near future.

The crucial difference between the scenarios in Figures \ref{fig:RDvsExp}(\subref{exp}) and \ref{fig:RDvsExp}(\subref{rd}) is our knowledge of the functional form of the regression functions. In a continuity-based approach, the RD treatment effect in \ref{fig:RDvsExp}(\subref{rd}) can be estimated by calculating the limit of the conditional average of the observed outcomes given the score as the score approaches the cutoff for the treatment and control groups separately, $\lim_{x\downarrow{\C}} \E[Y_i|X_i=x] - \lim_{x\uparrow{\C}} \E[Y_i|X_i=x]$. As we discussed extensively in \textit{Foundations}, the estimation of these limits requires that the researcher approximate the regression functions, and this approximation will typically contain an error that may directly affect estimation and inference. This is in stark contrast to the experiment depicted in Figure \ref{fig:RDvsExp}(\subref{exp}), where estimation does not require functional form assumptions: by construction, the regression functions are constant in the entire region where the score is randomly assigned. This shows that RD designs are not canonical randomized experiments but rather natural experiments \citep{Titiunik_2021_HandbookCh}, and thus belong to the toolkit of observational studies methods.

A point often overlooked is that the known functional form of the regression functions in a true experiment does not follow from the random assignment of the score per se, but rather from the lack of relationship between the score and the potential outcomes that is assumed to be a consequence of the randomization. If the value of the score were randomly assigned but had a direct effect on the average outcomes, the regression functions in Figure \ref{fig:RDvsExp}(\subref{exp}) would not necessarily be flat. Such direct effects are common and occur in any study where the score affects the outcome directly, separately from the treatment. For example, if 70 is the passing grade in a 100-point exam, students who receive a score of 68 or 69 might feel discouraged because they failed to pass by a narrow margin, while students who scored 70 or 71 would not experience this adverse psychological effect. Imagine that we send a congratulatory certificate to all students who score 70 and above, and we are interested in the effect of the certificate on future academic performance. If this discouragement affected future academic achievement, we might observe a difference in outcomes between students who scored 68-69 and students who scored 70-71, even if the certificate itself had no effect. Importantly, the spurious effect would occur even if the true grades of students who originally scored between 68 and 71 were randomly shuffled and students were notified of their ``randomly selected'' grade. This kind of direct effect is the reason why many medical trials are ``double blind'' and do not reveal to patients whether they are treated or control until the end of the experiment.

A local randomization approach to RD analysis must thus be based not only on the assumption that placement above or below the cutoff is randomly assigned within a window of the cutoff, but also on the assumption that the value of the score within this window is unrelated to the potential outcomes---a condition that is not guaranteed by the random assignment of the score $X_i$ (nor by the random assignment of the treatment $T_i$). To formalize, let $\W = [\C - w , \C + w]$ for some window length $w>0$, and $\mathbf{X}_\W$ be the vector of scores for all $i$ such that $X_i\in\W$, with analogous notation for the vectors of potential outcomes under control and treatment status, $\mathbf{Y}_\W(0)$ and $\mathbf{Y}_\W(1)$, respectively. 

The basic local randomization framework can be summarized by the two following conditions:
\begin{itemize}	
	\item[(LR1)] The joint distribution of the scores is unconfounded and the joint distribution of the treatment assignments is known within $\W$.
	\item[(LR2)] The potential outcomes are not affected by the score within $\W$.
\end{itemize}

The first condition implies that, inside the window, the treatment assignment mechanism is known and not a function of the potential outcomes, as would happen in a randomized experiment. Importantly, in the local randomization framework, all probability and moment calculations as well as all parameter definitions are often done conditionally on those units whose scores fall within the window $\W$. Define $\P_{\W}[\cdot]$ to be the probability computed conditionally for units with $X_i\in\W$. With these conventions, LR1 requires that $\P_\W[\mathbf{X}_\W \le \mathbf{x} | \mathbf{Y}_\W(0),\mathbf{Y}_\W(1)]=\P_\W[\mathbf{X}_\W \le \mathbf{x}]$, which is not a function of the potential outcomes inside the window. The second condition, LR2, is an exclusion restriction ensuring that the potential outcomes are not a function of the score for those units with score inside $\W$, as would be expected in a true double-blind randomized experiment. To formalize, let $Y_i(0,x)$ and $Y_i(1,x)$ denote the potential outcomes with now explicit dependence on the score variable only through their second argument. Then, if the potential outcomes are non-random, LR2 means that $Y_i(0,x')=Y_i(0,x)$ and $Y_i(1, x') = Y_i(1,x)$, for all $x,x'\in\W$ and all units such that $X_i\in\W$. If the potential outcomes are random, LR2 means $\P_\W[Y_i(0,x')=Y_i(0,x)]=1$ and $\P_\W[Y_i(1,x')=Y_i(1,x)]=1$ for all $x,x'\in\W$.

\begin{figure}[ht]
	\centering
	\includegraphics[scale=\kSF]{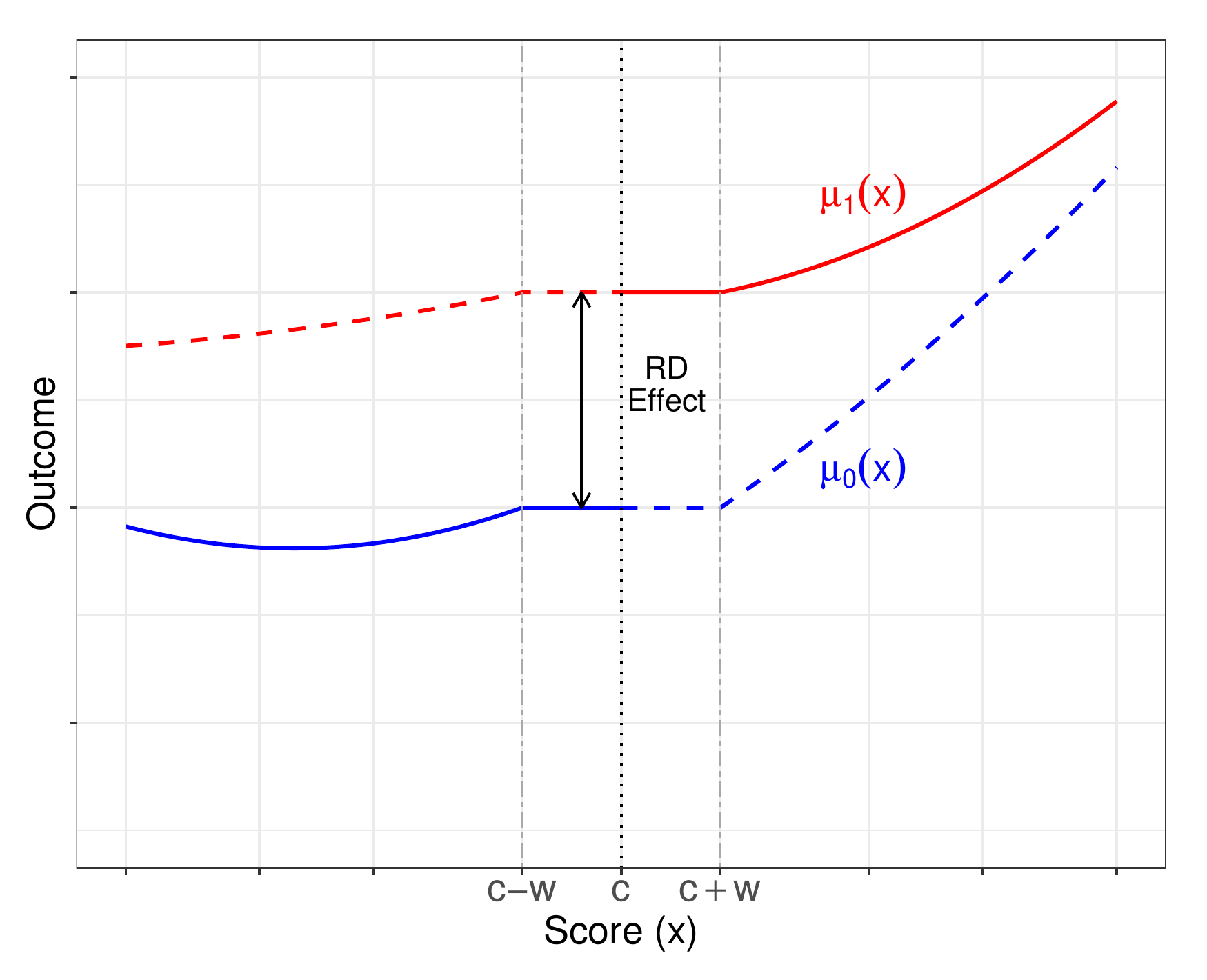}
	\caption{Local Randomization RD}\label{fig:RD-locran}
\end{figure}

Under LR1 and LR2, for all units with $X_i\in\W=[\C-w,\C+w]$, placement above or below the cutoff is unrelated to the potential outcomes, and the potential outcomes are unrelated to the running variable; therefore, the regression functions are flat inside $\W$. This is illustrated in Figure \ref{fig:RD-locran}, where for the case of random potential outcomes $\mu_1(x)=\E[Y_i(1)|X_i=x]$ and $\mu_0(x)=\E[Y_i(0)|X_i=x]$ are constant for all values of $x$ in $\W$, but can have non-zero slopes outside of $\W$.

The contrast between Figures \ref{fig:RDvsExp}(\subref{exp}), \ref{fig:RDvsExp}(\subref{rd}), and \ref{fig:RD-locran} illustrates the differences between an actual randomized experiment, a continuity-based RD design, and a local randomization RD design. In the actual experiment, the potential outcomes are unrelated to the score for all possible score values, and the functional forms of $\E[Y_i(1)|X_i=x]$ and $\E[Y_i(0)|X_i=x]$ are constant---and therefore known. In the continuity-based RD design, the potential outcomes can be related to the score everywhere; the functions $\E[Y_i(1)|X_i=x]$ and $\E[Y_i(0)|X_i=x]$ are unknown but assumed to be smooth, and estimation and inference are based on approximating them near the cutoff. Finally, in the local randomization RD design, the potential outcomes can be related to the running variable far from the cutoff, but there is a window around the cutoff where this relationship ceases; the functions $\E[Y_i(1)|X_i=x]$ and $\E[Y_i(0)|X_i=x]$ are unknown over the entire support of the running variable, but inside the window $\W$ they are assumed to be constant functions of $x$.

In many applications, assuming that the score has no effect on the potential outcomes near the cutoff may be regarded as unrealistic or too restrictive. However, such an assumption can be taken as an approximation, at least for the very few units with scores closest to the cutoff. As we discuss below, a key advantage of the local randomization approach is that it enables finite sample inference methods, which remain valid and can be used even when only a handful of observations very close to the cutoff are included in the analysis. Furthermore, the restriction that the score cannot directly affect the (average) potential outcomes near the cutoff could be relaxed under additional assumptions \citep{Cattaneo-Titiunik-VazquezBare_2017_JPAM}.

\subsection{The Effect of Winning Elections on Future Vote Shares}

We illustrate the local randomization methods with the study originally conducted by \citet*{Cattaneo-Frandsen-Titiunik_2015_JCI}, which uses a Sharp RD design in the United States to study the effect of the electoral advantages of incumbent political parties in U.S. Senate elections between $1914$ and $2010$. In winner-takes-all elections, there is a discontinuous relationship between the incumbency status of a political party and the vote share that the party obtains in an election: if there are only two parties competing for a seat, the party that gets just above $50\%$ of the vote wins the election and becomes the incumbent, while the opponent loses. Thus, party incumbency advantages can be studied with an RD design.

In the U.S., there are two U.S. Senate seats in each of the $50$ states, for a total of $100$ seats. Each seat is up for election every six years, but the seats are staggered so that one-third of seats are up for election every two years, and the two seats in the same state are never up for election simultaneously. We estimate the RD effect of the Democratic party winning a Senate seat on its vote share in the following election for that seat. In this RD design, the unit of analysis is the U.S. state, and the score is the Democratic party's margin of victory at election $t$---defined as the difference between the vote share obtained by the Democratic party minus the vote share obtained by its strongest opponent. The outcome of interest is the vote share of the Democratic party in the following election for that same seat; we denote this election $t+2$ because the election immediately following election $t$, which we denote $t+1$, is for the other Senate seat in the same state.

The Democratic margin of victory can be positive or negative, and the cutoff that determines a Democratic party victory is located at zero. The treatment indicator is equal to one when the Democratic margin of victory at $t$ is zero or above.  Thus, the treatment group is the set of states that elect a U.S. Senator from the Democratic party at $t$, and the control group is the set of states that elect a U.S. Senator from another party (the Republican party in most cases). The index $t$ covers every even year between $1914$ and $2010$, inclusive. In all our codes, we rename the score, outcome, and treatment variables to \texttt{X}, \texttt{Y}, and \texttt{T}, respectively. 

The dataset also contains several predetermined covariates: the Democratic vote share obtained (i) in the previous presidential election in that state, (ii) in the Senate election immediately prior to election $t$ (which we denote $t-1$ and is for the other seat in the state), and (iii) in the prior Senate election for the same seat (which we denote $t-2$); and indicators for Democratic Party victory at the $t-1$ and $t-2$ Senate elections, for midterm election year, and for no incumbent candidate running at $t$.

\begin{figure}[ht]
	\centering
	\includegraphics[scale=0.40]{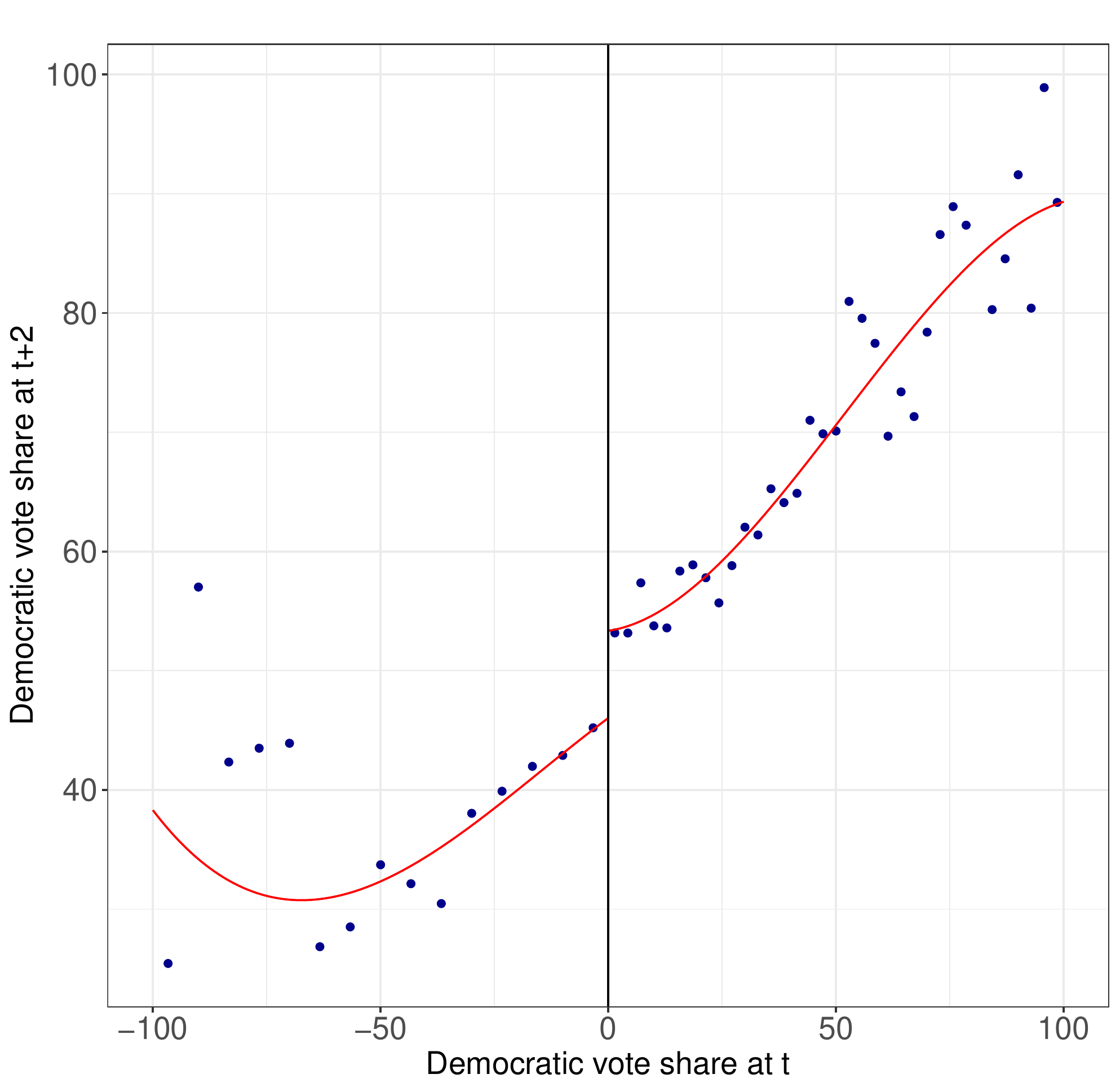}
	\caption{RD Plot ($p=3$)---U.S. Senate Data}\label{fig:MeyerssonRDplot}
\end{figure}

Figure \ref{fig:MeyerssonRDplot} presents an RD plot of the outcome \texttt{Y} against the score \texttt{X} that illustrates the continuity-based average treatment effect at the cutoff. The solid line is a third-order global polynomial fit and the dots represent local means---see Section 3 in \textit{Foundations} for details. The observations above the cutoff correspond to elections where the Democratic party won at $t$, while observations below the cutoff are elections where the Democratic party lost. At the cutoff, the average Democratic vote share is lower for states where the Democratic party loses than for states where the Democratic party wins. Employing the continuity-based analysis discussed in Section 4 in \textit{Foundations}, we use \texttt{rdrobust} to fit a local linear polynomial on each side of the cutoff within a mean-squared-error (MSE) optimal bandwidth and find that this effect is large and positive: states where the Democratic party barely wins the U.S. Senate election at $t$ receive on average $7.4$ additional percentage points in their vote share in the following election for the same seat at $t+2$---compared to states where the Democratic party barely loses at $t$. (We show an abbreviated output for future comparisons with the local randomization results.)

\labelsnippet{rdrobustA}
\rsnip{Vol-2-R_senate_rdrobust_triangular_mserd_p1_rhofree_regterm1.txt}{\Rlink{\thesection}{\therdrobustA}}
\statasnip{Vol-2-STATA_senate_rdrobust_triangular_mserd_p1_rhofree_regterm1}{\Slink{\thesection}{\therdrobustA}}

\subsection{Local Randomization Estimation and Inference}

The practical implementation of the local randomization approach to RD analysis requires knowledge or estimation of two ingredients: (i) the window $\W$ where the local randomization assumption is invoked; and (ii) the randomization mechanism that is needed to approximate the assignment of units within $\W$ to treatment or control. In applications, $\W$ is often unknown and must be selected by the researcher. (\citet{Hyytinen-etal-_2018_QE} discuss an interesting empirical example where $\W$ is known.) Once $\W$ has been chosen, the choice of the randomization mechanism can be guided by the structure of the data generating process.

Given a choice of $\W$ and assignment mechanism, under a local randomization RD approach, we can analyze the data as we would analyze a randomized experiment. If the number of observations inside $\W$ is large, researchers can use the full menu of standard methods for the analysis of experiments, which are often based on large sample approximations for point estimators and test statistics. These methods may or may not involve the assumption of random sampling, and may or may not require LR2 per se (though removing LR2 will change the interpretation of the RD parameter in general). In contrast, if the number of observations inside $\W$ is small, as is the case in many RD applications, estimation and inference based on large sample approximations may be invalid; in this case, under appropriate assumptions, researchers can still employ randomization-based inference methods that are exact in finite samples and do not require large sample approximations for their validity. We review both types of approaches below, assuming for simplicity that $\W$ is known. We discuss a data-driven method to choose $\W$ in Section \ref{subsec:chooseW}.

Throughout, we assume that there are $N_{\W}$ total units with $X_i \in \mathcal{W}$, of which $N_{\W,+}$ are assigned to the treatment condition and $N_{\W,-}=N_{\W}-N_{\W,+}$ are assigned to the control condition.

\subsubsection{Finite Sample Methods: Fisherian Inference}

In many RD applications, a local randomization assumption will only be plausible in a very small window around the cutoff, and by implication, this small window will likely contain very few observations. In this case, it is natural to employ a Fisherian inference approach, which is valid in any finite sample and thus leads to correct inferences even when the sample size in $\W$ is small.

The Fisherian approach sees the potential outcomes as fixed or non-stochastic. (This stands in contrast to the approach in the continuity-based RD framework, where the potential outcomes are random variables as a consequence of random sampling.) The hypothesis of interest is the so-called \textit{sharp null hypothesis} that the treatment has no effect for any unit:
\[ \mathsf{H}_0^\mathtt{F}: Y_i(0)=Y_i(1)\; \text{for all $i$ such that $X_i \in \mathcal{W}$}.\] 

The combination of non-stochastic potential outcomes and the sharp null hypothesis leads to inferences that are (type-I error) correct for any sample size because, under $\mathsf{H}_0^\mathtt{F}$, the observed outcome of each unit is equal to the unit's two potential outcomes, $Y_i = Y_i(1)= Y_i(0)$. When the assignment mechanism is known, the full knowledge of all potential outcomes under the null hypothesis allows us to derive the null distribution of any test statistic from the randomization distribution of the treatment assignment alone. Since the latter distribution is known exactly in finite samples, the Fisherian framework allows researchers to make inferences without relying on large sample approximations.

The implementation of the local randomization approach requires specifying a particular window $\W$ and the particular way in which the treatment assignment is ``randomized'' within $\W$. Naturally, the distribution of the treatment assignment within $\W$ is unknown; in practice, it has to be approximated by assuming a particular assignment mechanism within $\W$. Implementation also requires choosing a particular test statistic.

 We define the assignment mechanism within $\W$ as $\PW[\bT_\W = \mathbf{t}]$, where $\bT_\W$ denotes the vector of treatment assignment for all units with $X_i\in\W$, and $\mathbf{t}\in\mathcal{T}_{\W}$, with $\mathcal{T}_{\W}$ collecting all possible treatment assignment vectors. A common choice of assignment mechanism is to assume that $N_{\W,+}$ units are assigned to treatment and $N_{\W}- N_{\W,+}$ units are assigned to control within $\W$, where each treatment assignment vector has probability $\P_{\W}[\textbf{T}_{\W}=\mathbf{t}]= \binom{N_{\W}}{N_{\W,+}}^{-1}$ of being chosen for $\mathbf{t}\in\mathcal{T}_{\W}$, where $\mathcal{T}_{\W}$ now collects all vectors of length $N_{\W}$ with $N_{\W,+}$ ones and $N_{\W,-}$ zeros. This is commonly known as a complete or fixed-margins randomization mechanism. Under this mechanism, the numbers of units assigned to treatment and control are always fixed to $N_{\W,+}$ and $N_{\W}-$, respectively. For example, if there are 5 units in $\W$ of which 3 are treated and 2 are control, under complete randomization the number of elements in $\mathcal{T}_{\W}$ is $\binom{5}{3}=10$ and $\P_{\W}[\textbf{T}_{\W}=\mathbf{t}]= 1/10$.

 We collect in $\mathbf{Y}_{\W}$ the $N_{\W}$ observed outcomes for units with $X_i \in \W$. We also need to choose a test statistic, which we denote $S=S(\mathbf{\T}_{\W},\mathbf{Y}_{\W})$, a function of $\mathbf{\T}_{\W}$ and $\mathbf{Y}_{\W}$. Of all the possible values of the treatment vector $\mathbf{\T}_{\W}$ that can occur, only one will have occurred; we call this value the observed treatment assignment, $\mathbf{t}_{\W}^\mathtt{obs}$, and we denote $S^\mathtt{obs}$ the observed value of the test-statistic associated with it, i.e. $S^\mathtt{obs} = S(\mathbf{t}^\mathtt{obs}_{\W},\mathbf{Y}_{\W})$. Then, the finite-sample exact p-value associated with a test of the sharp null hypothesis $\mathsf{H}_0^\mathtt{F}$ is the probability that the test-statistic is larger than or equal to the observed value:
\[p^\mathtt{F}
= \P_\W(S(\mathbf{\T}_{\W},\mathbf{Y}_{\W}) \geq S^\mathtt{obs})
= \sum_{\mathbf{t}_{\W} \in \mathcal{T}_{\W}} \I(S(\mathbf{t}_{\W},\mathbf{Y}_{\W}) \geq S^\mathtt{obs} )
\cdot \P_\W(\mathbf{\T}_{\W}=\mathbf{t}_{\W}).
\]
Under $\mathsf{H}_0^\mathtt{F}$, all potential outcomes are known and can be imputed, $\mathbf{Y}_{\W}=\mathbf{Y}_{\W}(1)=\mathbf{Y}_{\W}(0)$, so that $S(\mathbf{\T}_{\W},\mathbf{Y}_{\W})$ can be computed for all treatment assignments. Thus, under $\mathsf{H}_0^\mathtt{F}$, the only randomness in $S(\mathbf{T}_{\W},\mathbf{Y}_{\W})$ comes from the random assignment of the treatment, which is assumed to be known.

In practice, it often occurs that the total number of different treatment vectors $\mathbf{t}_{\W}$ that can occur inside the window $\W$ is too large, and enumerating them exhaustively is unfeasible. For example, assuming a fixed-margins randomization inside $\W$ with $15$ observations on each side of the cutoff, there are $\binom{N_{\W}}{N_{\W,+}} = \binom{30}{15} = 155,117,520$ possible treatment assignments. When exhaustive enumeration is unfeasible, we can approximate $p^\mathtt{F}$ using simulations by randomly sampling different vectors of treatment assignment.

Fisherian confidence intervals can be obtained by specifying sharp null hypotheses about treatment effects and then inverting these tests. This requires specifying a treatment effect model and testing hypotheses about the specified parameters. A simple choice is a constant treatment effect model, $Y_i(1) = Y_i(0) + \tau$, which leads to the null hypothesis $\mathsf{H}^\mathtt{F}_{\tau_0}:\tau = \tau_0$. (Note that $\mathsf{H}^\mathtt{F}_0$ is a special case of $\mathsf{H}^\mathtt{F}_{\tau_0}$ when $\tau_0 = 0$.) Under this model, a $1-\alpha$ confidence interval for $\tau$ can be obtained by collecting the set of all the values $\tau_0$ that fail to be rejected when we test $\mathsf{H}^\mathtt{F}_{\tau_0}:\tau = \tau_0$ with an $\alpha$-level test.

To test $\mathsf{H}^\mathtt{F}_{\tau_0}$, we build test statistics based on an adjustment to the potential outcomes that renders them constant under this null hypothesis. Under $\mathsf{H}^\mathtt{F}_{\tau_0}$, the observed outcome is $Y_i = \T_i \cdot \tau_0 + Y_i(0)$ and the adjusted outcome $\ddot{Y}_i \equiv Y_i - T_i \tau_0 = Y_i(0)$ is constant. A randomization-based test of $\mathsf{H}^\mathtt{F}_{\tau_0}$ proceeds by first calculating the adjusted outcomes $\ddot{Y}_i$ for all the units in the window, and then computing the test statistic using the adjusted outcomes instead of the raw outcomes, i.e. computing $S(\mathbf{\T}_{\W},\mathbf{\ddot{Y}}_{\W})$. Once the adjusted outcomes are used to calculate the test statistic for all possible treatment assignments, a test of $\mathsf{H}^\mathtt{F}_{\tau_0}:\tau = \tau_0$ can be implemented as a test of the sharp null hypothesis $\mathsf{H}^\mathtt{F}_0$, using $S(\mathbf{T}_{\W},\mathbf{\ddot{Y}}_{\W})$ instead of $S(\mathbf{T}_{\W},\mathbf{Y}_{\W})$. We use $p^\mathtt{F}_{\tau_0}$ to refer to the p-value associated with a randomization-based test of $\mathsf{H}^\mathtt{F}_{\tau_0}$.

In practice, assuming that $\tau$ takes values in $[\tau_\mathtt{min}, \tau_\mathtt{max}]$, computing these confidence intervals requires building a grid $G^{\tau_0} = \left\{\tau_0^1, \tau_0^2, \ldots, \tau_0^G \right\}$, with $\tau_0^1 \geq \tau_\mathtt{min}$ and $\tau_0^G \leq \tau_\mathtt{max}$, and collecting all $\tau_{0} \in G^{\tau_0}$ that fail to be rejected with an $\alpha$-level test of $\mathsf{H}^\mathtt{F}_{\tau_0}$. Thus, the Fisherian $(1-\alpha) \times 100 \%$ confidence intervals is
\[ \mathtt{CI}_\mathtt{LRF} =  \left\{ \tau_0 \in G^{\tau_0}: p^\mathtt{F}_{\tau_0} > \alpha \right\} \text{.}\]

Although it is common to implement the approach with a difference-in-means test statistic and a fixed-margins mechanism, the general principle of Fisherian inference works for any appropriate choice of test statistic and randomization mechanism. Other possible test statistics include the Kolmogorov-Smirnov (KS) statistic and the Wilcoxon rank sum statistic. Other randomization mechanisms include the Bernoulli assignment, where each unit is assigned independently to treatment with the same probability---for implementation, it is common to choose either $1/2$ or the proportion of treated units in $\W$. Complete randomization and Bernoulli randomization often lead to similar conclusions, and they are the most commonly used.

Finally, while the main goal of Fisherian methods is inference and not point estimation, it is possible to define parameters of interest and point estimate them. However, any point estimator based on the Fisherian framework requires assuming a sharp treatment effect model that allows full imputation of all potential outcomes under the null hypothesis, as we did to build confidence intervals by test inversion. In particular, because the average treatment effect is not sharp, Fisherian methods do not provide a general way to estimate this parameter. This is sometimes seen as a limitation, since most common parameters do not allow for null hypotheses that are sharp.

\subsubsection{Large Sample Methods: Neyman and Super-population Estimation and Inference}\label{subsec:locrand-estimation}

In some RD applications, even the smallest windows have many observations. In these cases, although Fisherian methods continue to be valid and can certainly be used, researchers may choose to use more standard methods that rely on large sample approximations. Compared to Fisherian methods, the main advantage of large sample methods is that they provide consistent point estimators of parameters of interest, in addition to leading to statistical inferences based on asymptotic distributional approximations.

All large sample methods assume that the sample size is ``large'' (the formal requirement is that the sample size tends to infinity). The application of these methods to the local randomization RD context thus requires that the number of observations within the window $\W$, $N_{\W}$, be large enough. The are two kinds of frameworks for large sample methods, Neyman and super-population, depending on whether the potential outcomes are seen as fixed or random. 

In the Neyman framework, the potential outcomes $(Y_i(0),Y_i(1))$, $i=1,2,\dots,n$, are non-stochastic, so all parameters are, in this sense, conditional on the potential outcomes. Neyman envisions an urn model of assignment, where there is one urn per treatment condition and each urn has the potential outcomes corresponding to that treatment condition for each unit. In the binary treatment case, and proceeding conditionally for those units with $X_i\in\W$, the treatment urn contains the $N_\W$ ``balls'' $Y_1(1), Y_2(1), \ldots, Y_{N_\W}(1)$, and the control urn contains $Y_1(0), Y_2(0), \ldots, Y_{N_\W}(0)$. Estimates of average potential outcomes, $\mu_{\W,+}\equiv\frac{1}{N_\W} \sum_{i:X_i \in \W} Y_i(1)$ and $\mu_{\W,-}\equiv\frac{1}{N_\W} \sum_{i:X_i \in \W} Y_i(0)$, are created by drawing balls from the urns. For example, in a fixed margins randomization, $N_{\W,+}$ balls are taken from the treated urn, and $N_{\W,-}=N_{\W}-N_{\W,+}$ are taken from the control urn, in such a way that once a ball is taken from one urn, it disappears from the other. Because the sampling is without replacement, the draws are not independent. The Neyman approach relies on large sample approximations, imagining that the urn model is used many times to produce different assignments of units to treatment and control.

In the super-population framework, the units are assumed to be drawn from a larger population using independently and identically distributed (i.i.d.) sampling. This sampling scheme results in the potential outcomes $(Y_i(0),Y_i(1))$, $i=1,2,\dots,N_\W$, being random variables rather than fixed quantities. Thus, there are two sources of randomness: the sampling from the super-population, and the assignment of the sampled units to treatment or control. In contrast, in the Neyman framework (and also in Fisher's) the only source of randomness is the treatment assignment. Table \ref{tab:frameworks} compares the three approaches.

\begin{table}
	\centering
    \resizebox{\textwidth}{!}{\begin{tabular}{lccccc}
	\toprule
	& Sampling & Potential Outcomes & Sample size & Null hypothesis & Inferences\\
	\midrule
	Fisher  & None & Non-random & Fixed & Sharp & Exact\\
	Neyman  & Urn model & Non-random & Large& Non-sharp & Approximate\\
	Super-population & i.i.d. & Random & Large& Non-sharp & Approximate\\
	\bottomrule
	\end{tabular}}
	\caption{Comparison of Statistical Frameworks: Fisherian, Neyman, and Super-population}\label{tab:frameworks}
\end{table}

Regardless of whether a Fisher, Neyman or super-population approach is adopted, we can now define parameters of interest. Let $\EW[\cdot ]$ denote the expectation computed with respect to the probability $\PW$, that is, the expectation computed conditionally for those units with $X_i\in\W$. The local randomization Sharp RD treatment effect is the average treatment effect inside $\W$:
\[\theta_{\mathtt{SRD}} \equiv \frac{1}{N_{\W}}\sum_{i: X_i \in \W} \EW[Y_i(1)-Y_i(0)].\]
The definition of $\theta_{\mathtt{SRD}}$ is designed to cover both random and non-random potential outcomes under different sampling schemes. In a Neyman framework, it reduces to 
$\theta_{\mathtt{SRD}} = \frac{1}{N_{\W}}\sum_{i: X_i \in \W} [Y_i(1)-Y_i(0) ]$ because the potential outcomes are fixed and the (conditional) expectation integrates to one. In the super-population framework under i.i.d. sampling, we have $\theta_{\mathtt{SRD}} = \E[Y_i(1)-Y_i(0) | X_i \in \W]$.

The parameter $\theta_{\mathtt{SRD}}$ is different from the continuity-based RD parameter $\tau_{\mathtt{SRD}}$ defined in the introduction and discussed in \textit{Foundations}. While $\theta_{\mathtt{SRD}}$ is an average effect inside an interval (the window $\W$), $\tau_{\mathtt{SRD}}$ is an average at a single point (the cutoff $\C$) where the number of observations is zero whenever the score is continuously distributed. This means that the decision to adopt a continuity-based approach versus a local randomization approach directly affects the definition of the parameter of interest. Naturally, the smaller the window $\W$ is, the more conceptually similar $\theta_{\mathtt{SRD}}$ and $\tau_{\mathtt{SRD}}$ become.

Under the local randomization assumptions invoked within $\W$, we have
\[\theta_{\mathtt{SRD}}
= \frac{1}{N_\W} \sum_{i:X_i\in\W} \E_\W\Big[ \frac{T_i Y_i}{\P_\W[T_i=1]} \Big] 
- \frac{1}{N_\W} \sum_{i:X_i\in\W} \E_\W\Big[ \frac{(1-T_i) Y_i}{1-\P_\W[T_i=1]} \Big],
\]
regardless of whether the potential outcomes are fixed or random. This identification result expresses the counterfactual RD effect $\theta_{\mathtt{SRD}}$ as a function of observed random variables, and suggests the weighted difference-in-means estimator
\[\widehat{\theta}_\mathtt{SRD} = \bar{Y}_{\W,+} - \bar{Y}_{\W,-}, \qquad \bar{Y}_{\W,+} = \frac{1}{N_{\W,+}} \sum_{i:X_i\in\W} \omega_i T_i Y_i, \qquad 
\bar{Y}_\mathcal{W,-} = \frac{1}{N_{\W,-}} \sum_{i:X_i\in\W} \omega_i (1-T_i) Y_i,\]
where $\omega_i$ denotes an appropriately defined weighting scheme for unit $i$.

For example, when the assignment mechanism is Bernoulli, we have $\P_\W[T_i=1]=p\in(0,1)$ for all units with $X_i \in \W$. In this case, defining the weights as
\[\omega_i = \frac{N_{\W,+}}{N_\W\cdot p} \cdot T_i + \frac{N_{\W,-}}{N_\W\cdot (1-p)} \cdot (1-T_i),\]
we have $\E_\W[ \widehat{\theta}_\mathtt{SRD}] = \E_\W[\bar{Y}_{\W,+}] - \E_\W[\bar{Y}_{\W,-}] = \theta_\mathtt{SRD}$, that is, $\widehat{\theta}_\mathtt{SRD}$ is unbiased for $\theta_\mathtt{SRD}$. This result follows from $\E_\W[ T_i Y_i ]=\E_\W[ T_i] \E_\W[Y_i(1)]$ and $\E_\W[ (1-T_i) Y_i ]=\E_\W[ T_i] \E_\W[Y_i(0)]$ from fixed potential outcomes in the Neyman framework or from independence between the treatment assignment and the potential outcomes in the super-population framework.

The standard difference-in-means estimator is a particular case of $\widehat{\theta}_\mathtt{SRD} $ with $\omega_i=1$ for all units. When the assignment mechanism follows a fixed-margins randomization, this choice of weighting scheme makes $\widehat{\theta}_\mathtt{SRD}$ unbiased for $\theta_\mathtt{SRD}$, that is, $\EW[\widehat{\theta}_\mathtt{SRD}]=\theta_\mathtt{SRD}$. This follows from $\EW[ T_i ] = \PW[T_i=1]= \frac{N_{\W,+}}{N_\W}$ for all $i$ with $X_i \in \W$, and under the specific conditions on the potential outcomes imposed in each framework. By implication, whenever the assignment mechanism does not follow a fixed-margins randomization, the unweighted difference-in-means estimator is not unbiased for $\theta_{\mathtt{SRD}}$, although it is consistent under standard large sample arguments. Thus, whenever the randomization mechanism is assumed to be different from a fixed-margins randomization, the use of the unweighted difference-in-means estimator must be justified based on large sample approximations.

For inference, both the Neyman and the super-population approaches rely on a Gaussian approximation justified by appropriate central limit theorems. A possibly conservative estimator of the variance of $\widehat{\theta}_{\mathtt{SRD}}$ can be constructed using standard least squares results. A $100(1-\alpha)\%$ confidence interval can be constructed in the usual way by relying on a Gaussian large sample approximation to the statistic of interest. For example, an approximate two-sided $95\%$ confidence interval is
\[\mathtt{CI}_\mathtt{LS} = \left[\hat\theta_{\mathtt{SRD}} \pm 1.96 \cdot \sqrt{\widehat{V}} \right],\]
where $\widehat{V}$ denotes an appropriate choice of variance estimator, which can depend on the specific framework considered. A conservative choice is obtained if the so-called HC2 or HC3 heteroskedastic-robust variance estimators are used. Hypothesis testing is based on Gaussian approximations as well. The Neyman or super-population null hypothesis is
\[ \mathsf{H}_0: \frac{1}{N_{\W}}\sum_{i: X_i \in \W} \EW[Y_i(1) ] = \frac{1}{N_{\W}}\sum_{i: X_i \in \W} \EW[Y_i(0)].\]
In contrast to Fisher's sharp null hypothesis $\mathsf{H}_0^\mathtt{F}$, this null hypothesis does not allow us to calculate the full profile of potential outcomes for every possible realization of the treatment assignment vector. Thus, unlike the Fisherian approach, the large sample approach to hypothesis testing relies on an approximation and is therefore not exact but, when valid, it allows us to rely on well-known methods for estimation and inference based on least squares and related approaches.

\subsubsection{Local Randomization Estimation and Inference in Practice}

We start the local randomization analysis of the U.S. Senate application using the function \texttt{rdrandinf}, which is part of the \texttt{rdlocrand} library. The main arguments of \texttt{rdrandinf} include the outcome variable \texttt{Y}, the running variable \texttt{X}, and the upper and lower limits of the window where inferences will be performed (\texttt{wr} and \texttt{wl}). We first choose the ad-hoc window $[-2.5, 2.5]$, postponing the discussion of automatic data-driven window selection until the next section. To make inferences in $\mathcal{W}=[-2.5, 2.5]$, we set $\texttt{wl}=-2.5$ and $\texttt{wr}=2.5$. Since the implementation of Fisherian methods is based on simulations, in order to ensure the replicability of the results at a later time, we set the random seed using the \texttt{seed} argument.

\labelsnippet{rdrandinfA}
\rsnip{Vol-2-R_senate_rdrandinf_adhoc_p0.txt}{\Rlink{\thesection}{\therdrandinfA}}
\statasnip{Vol-2-STATA_senate_rdrandinf_adhoc_p0}{\Slink{\thesection}{\therdrandinfA}}

By default, \texttt{rdrandinf} uses the following specifications: a polynomial of order zero (outcomes are not transformed), a uniform kernel (the test statistic is computed using the unweighted observations), 1,000 simulations for Fisherian inference, null hypothesis set to $\tau_0=0$ (i.e. a test of $\mathsf{H}_0^\mathtt{F}$ and $\mathsf{H}_0$), a fixed margins randomization mechanism, and a difference-in-means test statistic.

Although there is a total of 595 control observations and 702 treated observations, the number of observations in the window $[-2.5,2.5]$ is much smaller, with only 63 elections below the cutoff and 57 elections above it. 

The last panel reports the results. The first column reports the type of test statistic employed for testing the Fisherian sharp null hypothesis, and the column labeled \texttt{T} reports its value. In this case, the difference-in-means is $9.167$; given the information in the \texttt{Mean of outcome} row, we see that this is the difference between a Democratic vote share of $53.235$ percentage points in elections where the Democratic party barely wins and $44.068$ percentage points in elections where the Democratic party barely loses. The \texttt{Finite sample} column reports the p-value ($p^\mathtt{F}$) associated with a randomization-based test of the Fisherian sharp null hypothesis $\mathsf{H}_0^\mathtt{F}$ (or the alternative sharp null hypothesis $\mathsf{H}_{\tau_0}^\mathtt{F}$ based on a constant treatment effect model if the user sets $\tau_0\neq 0$ via the option \texttt{nulltau}). This p-value is $0.000$, which means we reject the sharp null hypothesis at $5\%$, $1\%$, $0.1\%$, and even lower levels.

Finally, the \texttt{Large sample} columns in the bottom panel report inferences based on the large sample approximate behavior of the (distribution of the) statistic---based on Neyman or superpopulation approaches discussed above. The p-value reported here is thus the one associated with a test of the null hypothesis $\mathsf{H}_0$ that the average treatment effect is zero. The last column reports the power of the test to reject a true average treatment effect equal to \texttt{d}, where by default \texttt{d} is set to one half of the standard deviation of the outcome variable for the control group, which in this case is $10.627$ percentage points. As for the p-value, the calculation of the power versus the alternative hypothesis \texttt{d} is based on the Gaussian approximation. The large sample p-value is $0.000$, indicating that this null hypothesis is also easily rejected at conventional levels. The estimated average effect of 9.167 is large, approximately similar to one standard deviation of the control outcome.

We note the different interpretations of the difference-in-means test statistic in the Fisherian versus large sample framework. In Fisherian inference, the difference-in-means is simply one of the various test statistics that can be chosen to test the sharp null hypothesis, and should not be interpreted as an estimated effect; this is because the focus is on hypothesis testing, not on point estimation. In contrast, in the large sample framework (Neyman or super-population), the focus is on the sample average treatment effect; since the difference-in-means is a consistent estimator of this parameter under the assumptions we have made, it can be appropriately interpreted as an estimated effect under those assumptions.

To illustrate how robust Fisherian inferences can be to the choice of randomization mechanism and test statistic, we modify our call to \texttt{randinf} to use a binomial randomization mechanism, where every unit in the ad-hoc window $[-2.5, 2.5]$ has a $1/2$ probability of being assigned to treatment. For this, we first create an auxiliary variable that contains the treatment assignment probability of every unit in the window; this variable is then passed as an argument to \texttt{rdrandinf}.

\labelsnippet{rdrandinfB}
\rsnip{Vol-2-R_senate_rdrandinf_adhoc_p0_bernoulli.txt}{\Rlink{\thesection}{\therdrandinfB}}
\statasnip{Vol-2-STATA_senate_rdrandinf_adhoc_p0_bernoulli}{\Slink{\thesection}{\therdrandinfB}}

We omit the output to conserve space, as the Fisherian inference results do not change. The Fisherian p-value is again $0.000$, the same p-value obtained above under the assumption of a fixed margins randomization. The conclusion of rejection of $\mathsf{H}_0^\mathtt{F}$ is therefore unchanged. This robustness of the Fisherian p-value to the choice of fixed margins versus Bernoulli randomization is typical in applications. The large sample results are of course exactly the same as before, since the choice of randomization mechanism does not affect the large sample inferences.

We could also change the test statistics used to test the Fisherian sharp null hypothesis. For example, to use the Kolmogorov-Smirnov test statistic instead of the difference-in-means, we can use the option \texttt{statistic = "ksmirnov"} (not shown). 

To obtain confidence intervals, we must specify a grid $G^{\tau_0}$ of treatment effect values to invert tests of the sharp null hypothesis. The function \texttt{rdrandinf} tests the null hypotheses $\mathsf{H}_{\tau_0}^{F}:Y_i(1) - Y_i(0) = \tau_0$ for all values of $\tau_0$ in the grid and collects in the confidence interval all the hypotheses that fail to be rejected in a randomization-based test of the desired level (default is $0.05$). To calculate these confidence intervals, we create the grid, and then call \texttt{rdrandinf} with the \texttt{ci} option. For this example, we choose a grid of values for $\tau_0$ between $-20$ and $20$, with $0.10$ increments. Thus, we test $\mathsf{H}_{\tau_0}$ for all $\tau_0 \in G^{\tau_0} = \left\{-20, -19.90, -19.80,\ldots, 19.80, 19.90, 20 \right\}$. 

\labelsnippet{rdrandinfD}
\rsnip{Vol-2-R_senate_rdrandinf_adhoc_p0_ci.txt}{\Rlink{\thesection}{\therdrandinfD}}
\statasnip{Vol-2-STATA_senate_rdrandinf_adhoc_p0_ci}{\Slink{\thesection}{\therdrandinfD}}

The Fisherian $95\%$ confidence interval is $[5.7,12.6]$. This confidence interval assumes a constant treatment effect model. The interpretation is therefore that, given the assumed randomization mechanism and the constant treatment effect model $Y_i(1) = Y_i(0) + \tau$, all values of $\tau$ between $5.7$ and $12.6$ fail to be rejected with a $5\%$-level randomization-based Fisherian test. 

\subsubsection{How to Choose the Window}\label{subsec:chooseW}

In practice, the window $\W$ is almost always unknown and must be chosen; this is an important step in the implementation of the local randomization RD approach. For simplicity, the windows we consider are symmetric around the cutoff, i.e. $\W=[c-w,c+w]$ for $w\geq 0$. One option is to choose $\W$ in an \textit{ad hoc} way. For example, a scholar may believe that elections decided by $0.5$ percentage points or less are essentially decided as if by the flip of a coin, and choose $\W=[\C-0.5,\C+0.5]$. The disadvantage of an ad-hoc method is that it lacks transparency and objectivity.

\begin{figure}[ht]
	\centering
	\includegraphics[scale=\kSF]{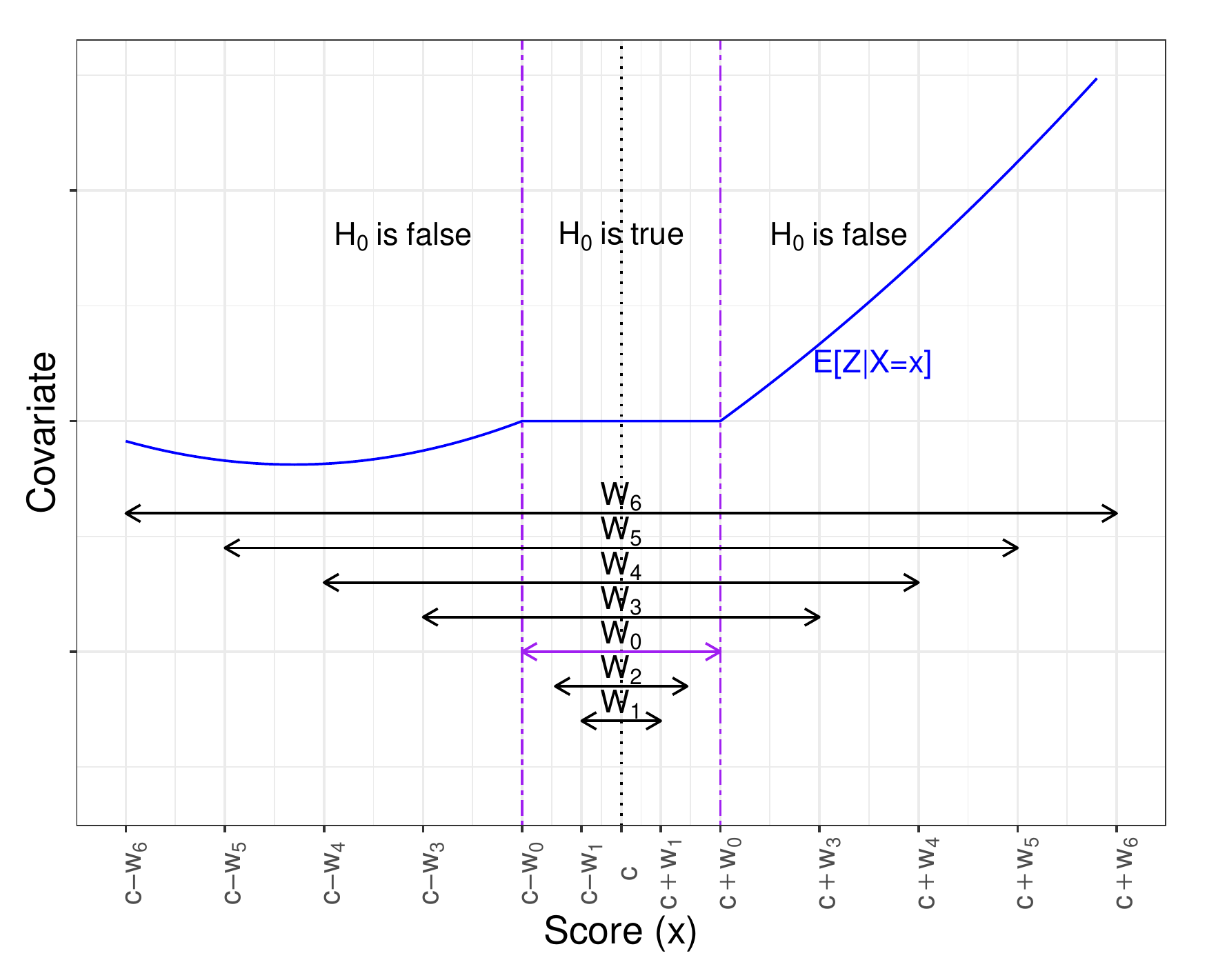}
	\caption{Window Selector Based on Covariate Balance}\label{fig:RD-locwin}
\end{figure}

A preferred alternative is to use a principled data-driven procedure. A leading example is based on predetermined covariates---variables that capture important characteristics of the units and whose values are determined before the treatment is assigned and received. This approach requires assuming that there exists at least one predetermined covariate, $Z_i$, that is associated with the running variable only outside the window $\W=W_0$ where the local randomization assumptions hold. Specifically, the requirement is that $Z_i$ be associated with the score in windows larger than $W_0$, possibly due to correlation between the score and another characteristic that also affects $Z_i$, but independent of the score in $W_0$ and all smaller windows. Moreover, because $Z_i$ is a predetermined covariate, the effect of the treatment on $Z_i$ is zero by construction. Figure \ref{fig:RD-locwin} shows a hypothetical illustration based on the conditional expectation of $Z_i$ given the score. (We focus on the conditional expectation of a random covariate for illustration purposes only, but the idea applies more generally.)

This motivates a data-driven method to choose $\W$. We define a generic null hypothesis $\mathcal{H}_0$ stating that the treatment is unrelated to $Z_i$ (or that $Z_i$ is ``balanced" between the groups). This hypothesis could be the Fisherian hypothesis $\mathsf{H}_0^\mathtt{F}$ or the large sample hypothesis $\mathsf{H}_0$. The procedure starts with the smallest possible window---$W_1$ in Figure \ref{fig:RD-locwin}---and tests $\mathcal{H}_0$. Since there is no treatment effect inside $W_1$, $\mathcal{H}_0$ will fail to be rejected. A larger window $W_2$ is selected, and the null hypothesis is tested again inside $W_2$. The procedure keeps increasing the length of the window and re-testing $\mathcal{H}_0$ in each larger window until a window is reached where $\mathcal{H}_0$ is rejected at the chosen significance level $\alpha^\star \in (0,1)$. In the figure, assuming the test has perfect power, $\mathcal{H}_0$ will not be rejected in $W_0$, nor will it be rejected in $W_2$ or $W_1$. The chosen window is the largest window such that $\mathcal{H}_0$ fails to be rejected inside that window and in all windows contained in it.\footnote{It is possible to imagine a covariate that is related with the potential outcomes outside of $\W$ in such a way that the relationship above $c+w$ is identical to the relationship below $c-w$, and the hypothesis of balance fails to be rejected. This does not invalidate the method. For practical purposes, a covariate that behaves in this way is not a confounder, and would correctly lead to a choice of window equal to the entire support of the score. We thank a reviewer for this observation.} 

The practical implementation of the procedure requires several choices:
\begin{itemize}

\item\textit{Null hypothesis}. Since the procedure will typically involve some windows with very few observations and point estimation is not the goal, we recommend using the Fisherian methods for the sharp null hypothesis, $\mathsf{H}_0^\mathtt{F}: Z_i(1)=Z_i(0)$ for all $i$.

	\item \textit{Relevant covariates}. The covariates employed should be related to both the outcome and the treatment assignment. If multiple covariates are chosen, the procedure can be applied using either the p-value of an omnibus test statistic, or by testing $\mathcal{H}_0$ for each covariate separately and using the minimum p-value across all covariates.
	
	\item \textit{Test statistic}. Typical choices of the statistic used to test $\mathcal{H}_0$ include the difference-in-means, the Kolmogorov-Smirnov statistic, and the Wilcoxon rank-sum statistic.
	
	\item \textit{Randomization mechanism}. If Fisherian inference is used, typical choices of randomization mechanisms are complete randomization and Bernoulli assignment. Large sample methods do not require the full specification of the assignment mechanism, but do require general assumptions about the type of assignment such as no stratification.
	
	\item \textit{Minimum number of observations in the smallest window}. If the smallest window where $\mathcal{H}_0$ is tested is too small, it will contain too few observations and the power to reject the null hypothesis when it is false will be too low. The smallest window should contain a minimum number of observations to ensure acceptable power; we recommend at least roughly ten observations on either side of the cutoff.
	
	\item \textit{Level $\alpha^\star$}. Because the main concern is failing to reject a false $\mathcal{H}_0$, the threshold significance level that determines when $\mathcal{H}_0$ is rejected should be higher than the usual 0.05. When we test $\mathcal{H}_0$ at a higher level, we tolerate a higher probability of Type I error and a lower probability of concluding that the covariate is unrelated to the treatment assignment when in fact it is. We recommend setting $\alpha^\star \geq 0.15$ if possible, and ideally no smaller than $0.10$.
\end{itemize}

We use this procedure to select a window in the U.S. Senate application using the predetermined covariates described above. We use the function \texttt{rdwinselect}, which is part of the \texttt{rdlocrand} library. The main arguments are the score variable \texttt{X}, the matrix of predetermined covariates, and the sequence of nested windows. By default, \texttt{rdwinselect} starts with the smallest symmetric window that has at least $10$ observations on either side of the cutoff. To control the sequence of windows where we test $\mathcal{H}_0$, we set the option \texttt{wobs=2}, which uses a  sequence of symmetric windows where the number of observations in each step increases by at least two observations on either side of the cutoff. 

\labelsnippet{rdwinselectA}
\rsnip{Vol-2-R_senate_rdwinselect_automatic_p0_wobs2_secondtry.txt}{\Rlink{\thesection}{\therdwinselectA}}
\statasnip{Vol-2-STATA_senate_rdwinselect_automatic_p0_wobs2_secondtry}{\Slink{\thesection}{\therdwinselectA}}

The default method of inference is Fisherian, though this can be changed with the \texttt{approximate} option. Our results use the difference-in-means as the test statistic, but this can be changed with the \texttt{statistic} option; the available options are the Kolmogorov-Smirnov statistic (\texttt{ksmirnov}), the Wilcoxon-Mann-Whitney studentized statistic (\texttt{ranksum}), and Hotelling's T-squared statistic (\texttt{hotelling}).

For every window, the \texttt{p-value} column reports the minimum of all the p-values associated with the tests of the null hypothesis performed for each covariate ($p_\mathtt{min}$). The column $\texttt{Var. name}$ reports the covariate associated with the minimum p-value---that is, the covariate $Z_k$ such that $p_k= p_\mathtt{min}$. The \texttt{Bin. test} column uses a Binomial test to calculate the probability of observing $N_{\W,+}$ successes out of $N_{\W}$ trials; we postpone the discussion of this test until the upcoming section on falsification.

\begin{figure}[ht]
	\centering
	\includegraphics[scale=0.6]{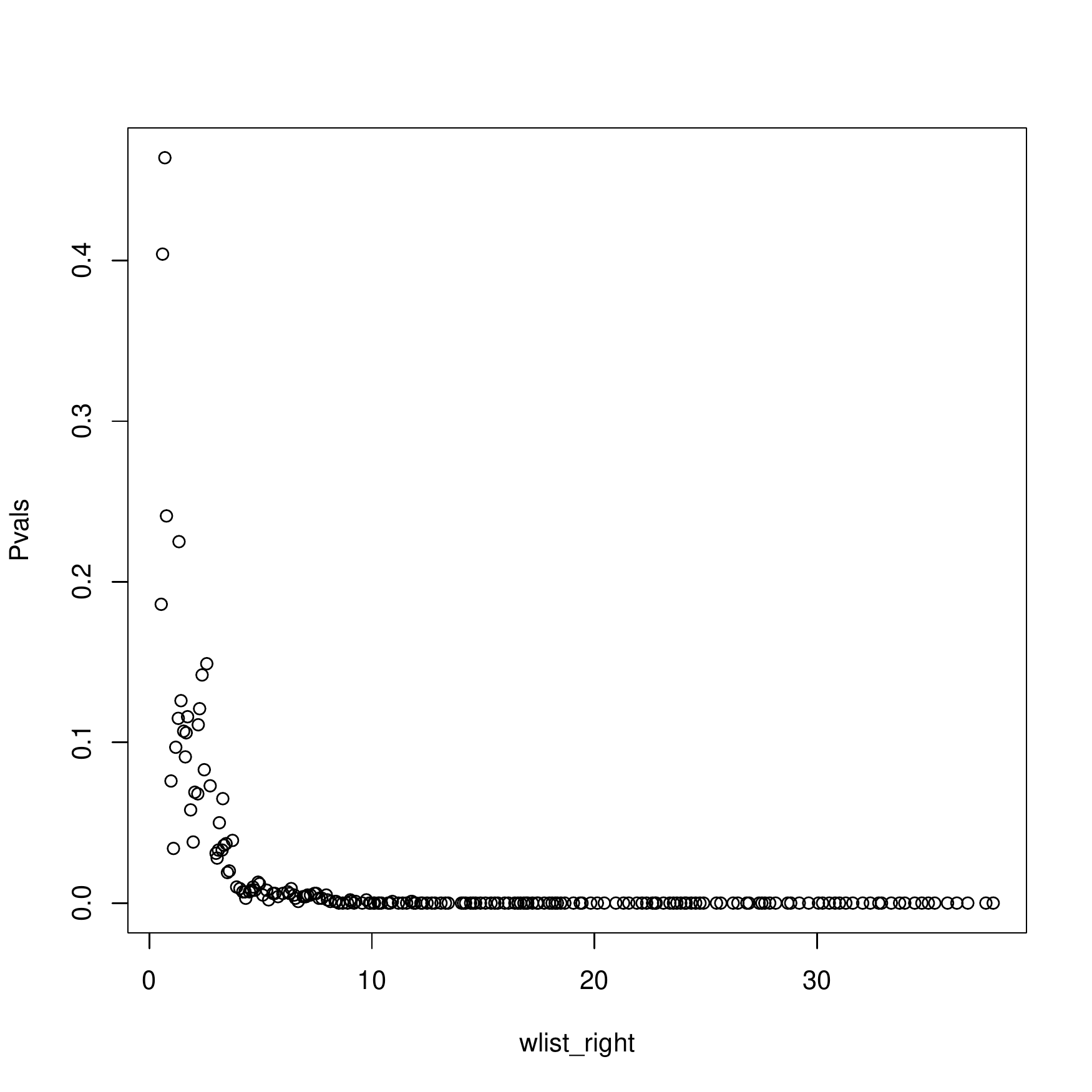}
	\caption{P-value for Different Windows---U.S. Senate data}\label{fig:pvals}
\end{figure}

The output indicates that the p-values are above $0.15$ in all windows between the minimum window $[-0.5287,0.5287]$ and the window $[-0.7652,0.7652]$. In the window immediately after $[-0.7652,0.7652]$, the p-value drops to 0.076, considerably below the suggested $0.15$ threshold. The chosen data-driven window is, therefore, $W_0 =[-0.7652,0.7652]$. After this window, the p-values start decreasing, albeit initially this decrease is not monotonic. By default, \texttt{rdwinselect} only shows the first $20$ windows, but this number can be increased with the option \texttt{nwindows}. We can also set the option \texttt{plot=TRUE} to create a plot of the minimum p-values associated with the length of each window considered; we show the plot in Figure \ref{fig:pvals} for the first 200 windows.

Figure \ref{fig:pvals} shows that the minimum p-value decreases sharply with window length, staying below $0.10$ for all windows approximately larger than $[-2.5,2.5]$. Although the p-values increase above $0.10$ for some windows between $[-1,1]$ and $[-2.5,2.5]$, they decrease sharply once windows larger than $[-3,3]$ are considered. The pattern in this plot is common in most applications: a strong negative relationship between p-values and window length, with high p-values for the smallest windows that decrease rapidly (albeit not necessarily monotonically) and stay at zero once the window length is large enough. Although in this example the absolute value of the running variable ranges from $0$ to $100$, the p-values become approximately zero for windows larger than $[-3,3]$. This shows that there are sharp differences between states where the Democratic party wins and states where the Democratic party loses even for elections decided by moderate margins. For this reason, the window selector chose $[-0.7652, 0.7652]$, suggesting that the local randomization assumptions, if they hold at all, hold in a very small window near the cutoff.

To assess the sensitivity of the window selector, we can call \texttt{rdwinselect} with \texttt{wstep=0.1}. This option starts at the minimum window and increases the length by $0.1$. The suggested window in this case is $[-0.8287, 0.8287]$, very similar to the $[-0.7652, 0.7652]$ window chosen above with \texttt{wobs=2}. We omit the output to conserve space.

We can now use \texttt{rdrandinf} to perform a local randomization analysis in the chosen window, using the options \texttt{wl} and \texttt{wr} to input, respectively, the lower and upper limits of the window. We also use the option $\mathtt{d}=7.414$ to calculate the power of a large sample test to reject the null hypothesis of a zero average treatment effect when the true average difference is $7.414$, which is the continuity-based point estimate shown in \texttt{R} Snippet 2.1.

\labelsnippet{rdrandinfE}
\rsnip{Vol-2-R_senate_rdrandinf_specific1.txt}{\Rlink{\thesection}{\therdrandinfE}}
\statasnip{Vol-2-STATA_senate_rdrandinf_specific1}{\Slink{\thesection}{\therdrandinfE}}

The difference-in-means in $W_0 = [-0.7652, 0.7652]$ is $10.203$, larger than the continuity-based local linear point estimate of $7.414$ but leading to the same conclusion of a positive advantage. Both the large sample and Fisherian approaches reject the null hypothesis of, respectively, no average treatment effect and no treatment effect for any unit. As shown in the last column, the large sample power to detect a difference of around $7$ percentage points is high, at $87.2\%$. In accordance with these results, the Fisherian $95\%$ confidence interval under a constant treatment effect model is $[ 5,15.3]$, showing positive effects of Democratic victory on future vote share. To calculate these confidence intervals, we use the option \texttt{ci} to pass a grid of treatment effect values, each of which is used as a null hypothesis in a Fisherian test to collect all hypotheses that fail to be rejected. 

The procedure of first choosing the window using \texttt{rdwinselect} and then estimating outcome effects using \texttt{rdrandinf} has the advantage of separating the window selection step from the effect estimation step. Because \texttt{rdwinselect} will never show outcome results, following the analysis in this order will reduce the possibility of choosing the window where the outcome results are in the ``expected'' direction. In other words, choosing the window without looking at the outcome results minimizes pre-testing and specification-searching issues.

\subsection{Validation and Falsification Analysis}

In \textit{Foundations}, we discussed the importance of conducting falsification tests to provide evidence in support of the RD assumptions. Falsification and validation analyses are as important in the local randomization framework as they are in the continuity-based framework; the difference lies in their implementation. Instead of providing empirical evidence in favor of continuity assumptions as in the continuity-based approach, the main goal in a local randomization approach is to provide evidence consistent with the local randomization assumptions. 

We now discuss four types of empirical falsification tests for a local randomization RD design, all of which were discussed in \textit{Foundations} in the context of the continuity-based approach: (i) tests of a null treatment effect on pre-treatment covariates or placebo outcomes, (ii) tests to assess the density of the score around the cutoff, (iii) treatment effect estimation at artificial cutoff values, and (iv) sensitivity to neighborhood choices. 

\subsubsection{Predetermined Covariates and Placebo Outcomes}

This crucial falsification test focuses on two types of variables: predetermined covariates---variables that are set and measured before the treatment is assigned, and placebo outcomes---variables that are determined after the treatment is assigned but are known to be unaffected by the treatment for scientific reasons. The idea is that, in a valid RD design, there should be no systematic differences between treated and control groups at or near the cutoff in terms of both placebo outcomes and predetermined covariates, because these variables could not have been affected by the treatment. For implementation, the researcher conducts a test of the hypothesis that the treatment effect is zero for each predetermined covariate and placebo outcome. If the treatment does have an effect on these variables, the plausibility of the RD assumptions is called into question. 

An important principle behind this type of falsification analysis is that all predetermined covariates and placebo outcomes should be analyzed in the same way as the outcome of interest. In the local randomization approach, this means that the null hypothesis of no treatment effect should be tested within the window where the assumption of local randomization is assumed to hold, using the same inference procedures and the same treatment assignment mechanism and test statistic used for the analysis of the outcome. Since the local randomization assumptions are assumed to hold in $\W=W_0$, all covariates and placebo outcomes should be analyzed within this window. This illustrates a fundamental difference between the approaches: in the continuity-based approach, estimation and inference requires approximating unknown regression functions, which requires estimating different bandwidths for each covariate or placebo variable analyzed; in contrast, in the local randomization approach, all analyses occur within the same window.

In order to test if the predetermined covariates are balanced within our chosen window $W_0= [-0.7652,0.7652]$, we employ Fisherian methods with a difference-in-means statistic, the same statistic we used for the outcome. Under the local randomization assumptions, we expect the difference-in-means between treated and control groups for each covariate to be indistinguishable from zero within $W_0$. Naturally, we already know that the covariates used to choose the window are balanced in $W_0$. In this sense, the window selector procedure is itself a validation procedure. We note, however, that it is possible (and indeed common) for researchers to choose the window based on a given set of covariates, and then assess balance on a different set.

In order to test this formally, we use the \texttt{rdrandinf} function, using each covariate as the outcome of interest. Table \ref{tab:ttest_covariates} contains the results. We cannot conclude that the control and treatment means are different for any covariate since all p-values are greater than or equal to $0.24$. For example, when we study \texttt{presdemvoteshlag1}, we see that the difference-in-means statistic is relatively small ($46.415-44.463 = 1.952$), and the finite sample p-value is large ($0.461$), showing that this covariate is balanced inside the chosen window. The number of observations is fixed in all cases (the small changes are due to missing values for particular covariates) because the window is set to the same $W_0$ for all covariates. 

\begin{table}[ht]
	\centering
	\resizebox{\textwidth}{!}{\begin{tabular}{lccccc}
			\toprule
			\multicolumn{1}{c}{\multirow{2}{*}{Variable}} & Mean of & Mean of & Diff-in-Means & Fisherian & Number of \\
			& Controls & Treated & Statistic     & p-value    & Observations\\
			\midrule
			Democratic presidential vote share at t-1 & $44.46$ & $46.42$ & $1.95$ & $0.46$ & $41$ \\
Democratic vote share at t-1 & $48.41$ & $52.48$ & $4.07$ & $0.37$ & $40$ \\
Democratic vote share at t-2 & $49.14$ & $50.53$ & $1.39$ & $0.58$ & $40$ \\
=1 if Democratic won at t-1 & $0.47$ & $0.64$ & $0.17$ & $0.47$ & $40$ \\
=1 if Democratic won at t-2 & $0.47$ & $0.52$ & $0.05$ & $0.95$ & $40$ \\
=1 if midterm election at t & $0.56$ & $0.56$ & $0.00$ & $1.00$ & $41$ \\
=1 if presidential election at t & $0.50$ & $0.40$ & $-0.10$ & $0.77$ & $41$ \\
=1 if open seat & $0.38$ & $0.16$ & $-0.22$ & $0.24$ & $41$ \\ \bottomrule

	\end{tabular}}
	\caption{Local Randomization Analysis for Covariates---U.S. Senate data}\label{tab:ttest_covariates}
\end{table}

 \subsubsection{Density of Running Variable}

Another important falsification test analyzes whether the number of observations just above the cutoff is roughly similar to the number of observations just below the cutoff---the so-called density test. The idea is that, if units lack the ability to control precisely the value of the score they receive, they should be just as likely to receive a score value just above the cutoff as they are to receive a score value just below it. In a local randomization approach, it is implemented by testing the null hypothesis that, within the window $\W$ where the treatment is assumed to be randomly assigned, the number of treated and control observations is consistent with whatever assignment mechanism is assumed inside $\W$. 

For example, assuming a simple ``coin flip'' or Bernoulli trial with probability of success $q$, we would expect the control sample size, $N_{\W,-}$, and treatment sample size, $N_{\W,+}$, within $\W$ to be compatible with the numbers generated by these $N_{\W,-} + N_{\W,+} = N_{\W} $ Bernoulli trials. In this case, the number of treated units in $\W$ follows a binomial distribution, and the null hypothesis of the test is that the probability of success in the $N_{\W}$ Bernoulli experiments is $q$. As discussed, the true probability of treatment is unknown. In practice, researchers can choose $q=1/2$ (a choice that can be justified from a large sample perspective when the score is continuous). 

The binomial test is implemented in all common statistical software and is also part of the \texttt{rdlocrand} package via the \texttt{rdwinselect} command. Using the Senate data, we can implement this falsification test in our selected window $W_0=[ -0.7652, 0.7652]$, where there are $16$ control observations and $25$ treated observations. The p-value of a binomial test that uses a success probability equal to $1/2$ is $0.211$, so we find no evidence against the null hypothesis: the difference in the number of treated and control observations is generally consistent with what would be expected if states were assigned to a Democratic win or loss by the flip of an unbiased coin inside the window.

\subsubsection{Placebo Cutoffs}

This falsification test chooses one or more artificial cutoff values at which the probability of treatment assignment does not change, and analyzes the outcome of interest at these cutoffs using the same methods used to conduct the analysis at the actual cutoff. The expectation is that no effect should be found at any of the artificial cutoffs. To avoid contamination from the actual treatment effect, only treated observations are included for artificial cutoffs above the actual cutoff, and only control observations are included for cutoffs below the actual cutoff. 

In the local-randomization approach, one possible implementation is to choose several artificial cutoff values, and then conduct an analysis of the outcome using a symmetric window of the same length as the original window $W_0$ around each of the cutoffs. Another possibility is to choose a symmetric window around the artificial cutoff with at least the same number of observations as the window chosen for the main analysis. Since our chosen window in the U.S. Senate application is $W_0=[-0.7652,0.7652]$, we consider windows of length $\pm 0.7652$ around each artificial cutoff. For example, for the cutoff $c=1$, we analyze the outcome in the window $[0.235, 1.765]$ (only using treated observations). We perform a similar analysis for the artificial cutoff $c=-1$ with a window given by the cutoff $-1 \pm 0.7652$, this time only using control observations. 

We summarize the results in Table \ref{tab:rdrandinf_placebocutoffs}, where we see that the point estimate of roughly $10$ percentage points that we saw around the real cutoff is dramatically reduced to $2.30$ and $-0.33$ at the artificial cutoffs, with very large Fisherian p-values. Reassuringly, in contrast to the true cutoff, there is no evidence of a treatment effect at the artificial cutoffs.

\begin{table}[ht]
	\centering
	\resizebox{.8\textwidth}{!}{\begin{tabular}{lccccc}
			\toprule
			\multicolumn{1}{c}{\multirow{2}{*}{Cutoff}} & Mean of  & Mean of & Diff-in-Means & Fisherian  & Number of \\
			& Controls & Treated & Statistic     & p-value    & Observations\\
			\midrule
			Cutoff at -1 & $42.17$ & $41.84$ & $-0.33$ & $0.93$ & $42$ \\
Cutoff at 1 & $50.65$ & $52.94$ & $2.30$ & $0.35$ & $37$ \\ \bottomrule

	\end{tabular}}
	\caption{Local Randomization Analysis for Placebo Cutoffs---U.S. Senate data}\label{tab:rdrandinf_placebocutoffs}
\end{table}

\subsubsection{Sensitivity to Window Choice}

Just like in a continuity-based approach researchers are interested in the sensitivity of the results to the bandwidth choice, in a local randomization approach we are often interested in sensitivity to the window choice. To assess this sensitivity, researchers can consider different windows and repeat the randomization-based analysis for the outcome of interest as conducted in the original window---that is, using the same test-statistic, same randomization mechanism, etc. 

This analysis should be implemented carefully, however. If $W_0$ was chosen based on covariate balance as we recommend, results in windows larger than $W_0$ will not be reliable because in such windows the treated and control groups will be imbalanced in important covariates. Thus, the sensitivity analysis should only consider windows smaller than $W_0$; unfortunately, in many applications, this analysis will be limited by the small number of observations that are likely to occur in these windows. In the Senate example, our chosen window $W_0=[ -0.7652, 0.7652]$ has only $25$ and $16$ observations on either side of the cutoff, so our ability to explore smaller windows is limited. Nonetheless, we consider the smaller window $\W=[-0.6934, 0.6934]$. The results (omitted) show that the conclusion of a positive party advantage remains unchanged when this smaller window is considered (the effect is 9.124, with large-sample and Fisherian p-values well below 0.01.

\labelsnippet{windowsensitivity}
\rsnip{Vol-2-R_senate_windowsensitivity.txt}{\Rlink{\thesection}{\thewindowsensitivity}}
\statasnip{Vol-2-STATA_senate_windowsensitivity}{\Slink{\thesection}{\thewindowsensitivity}}

\subsection{When To Use The Local Randomization Approach}
The RD treatment assignment rule $\I(X_i\geq c)$ does not imply that the treatment is randomly assigned within some window. Like the continuity assumptions, the local randomization assumptions must be made \textit{in addition} to the RD assignment mechanism, and are inherently untestable. When the score is continuous, the local randomization assumptions are strictly stronger than the continuity assumptions, in the sense that if there is a window around $\C$ in which the regression functions are constant functions of the score, these regression functions will also be continuous functions of the score at $\C$. However, the converse is not true: there may be applications where the regression functions satisfy the continuity assumptions even though there is no window around the cutoff that satisfies the local randomization assumptions. Why, then, would researchers want to impose stronger assumptions to make their inferences?

Although the continuity-based approach relies on the weaker condition of continuity, it unavoidably requires extrapolation because there are no observations with score exactly equal to the cutoff. The extrapolation consists of using observations in a neighborhood of the cutoff to approximate the unknown regression functions, and then calculating the value of the regression functions exactly at the cutoff using the approximated functional form. Although the smoothness assumptions required for this approximation to be valid do not impose parametric restrictions, the approximation does introduce an error that is only negligible if the sample size is large enough. This makes the continuity-based approach more appealing when there are enough observations near the cutoff to approximate the regression functions with reasonable accuracy---but possibly inadequate when the number of observations is small. In applications with few observations, the local randomization approach has the advantage of requiring minimal extrapolation and avoiding the use of smoothing methods. In addition, local randomization can also be useful in cases when the number of observations is not prohibitively small, but researchers are interested in assessing the robustness of their continuity-based analysis. Our Senate example belongs to this category: although it has enough observations to conduct a continuity-based analysis, our analysis shows that the continuity-based conclusions remain generally the same even after discarding most observations and imposing local randomization conditions.

Finally, a situation in which a local randomization approach may be preferable to a continuity-based approach is when the running variable is discrete---i.e., when multiple units share the same value of the score. When the score is discrete, the continuity-based approach is not directly applicable, and local randomization is often a natural and useful alternative. We consider this issue in Section \ref{sec:discrete}, where we discuss the analysis of RD designs with discrete scores.

\subsection{Further Reading}

Textbook reviews of Fisherian and Neyman estimation and inference methods in the context of the analysis of experiments are given by \citet{Rosenbaum_2010_Book} and \citet{Imbens-Rubin_2015_Book}; the latter also discusses super-population approaches and their connections to finite population inference methods. \citet{Ernst_2004_SS} discusses the connection and distinctions between randomization and permutation inference methods. \citet{Cattaneo-Frandsen-Titiunik_2015_JCI} propose Fisherian randomization-based inference to analyze RD designs based on a local randomization assumption, and the window selection procedure based on covariate balance tests. \citet{Cattaneo-Titiunik-VazquezBare_2017_JPAM} use transformations of the potential outcomes to relax the local randomization assumption and allow for a weaker exclusion restriction; they also compare RD analysis in continuity-based and randomization-based approaches. See also \citet*{Cattaneo-Titiunik-VazquezBare_2016_Stata}. The interpretation of the RD design as a local experiment and its connection to the continuity-based framework is also discussed by \citet{Sekhon-Titiunik_2016_ObsStud,Sekhon-Titiunik_2017_AIE}. Other refinements are surveyed in \citet{Cattaneo-Titiunik_2022_ARE}. For an RD application where the treatment is truly randomized in a window around the cutoff, see \cite{Hyytinen-etal-_2018_QE}.

\clearpage
\section{The Fuzzy RD Design}
\label{sec:FuzzyRD}
\setcounter{figuras}{1}
\setcounter{snippet}{1}
\setcounter{tablas}{1}

We now discuss how to modify the analysis and interpretation of the RD design when some units fail to comply with the treatment condition that is assigned to them. In all RD designs, the assignment of treatment follows the rule $\T_i=\I(X_i \geq \C)$, which assigns all units whose score is below the cutoff $\C$ to the control condition, and all units whose score is above $\C$ to the treatment condition. In the Sharp RD design, all units assigned to the treatment condition do in fact take the treatment, and no units assigned to the control condition take the treatment. In this case, the rule $\T_i=\I(X_i \geq \C)$ indicates not only the treatment assigned to the units, but also the treatment received by the units. 

However, it is common in practice to encounter RD designs where either some of the units with $X_i \geq \C$ fail to receive the treatment or some of the units with $X_i < \C$ receive the treatment anyway---or both. The phenomenon of units receiving a treatment condition different from the condition that is originally assigned to them is generally known as \textit{imperfect compliance} or \textit{non-compliance}. The RD design with imperfect compliance is usually referred to as the Fuzzy RD design, to distinguish it from the Sharp RD design where compliance is perfect. Imperfect compliance is common in randomized experiments, and is no less common in RD designs.

The Fuzzy RD treatment assignment rule is still $\T_i=\I(X_i \geq \C)$ but compliance with this assignment is imperfect. As a consequence, although the probability of receiving treatment still jumps abruptly at the cutoff, it no longer changes from $0$ to $1$ as in the Sharp RD case. (Naturally, the probability of being assigned to treatment still jumps from $0$ to $1$ at $\C$.) We use the binary variable $\D_i$ to denote whether the treatment was actually received by unit $i$. Our notation now distinguishes between the treatment assigned, $\T_i$, and the treatment received, $\D_i$. We can thus say that the key characteristic of the Fuzzy RD design is that there are some units for which $\T_i\neq \D_i$.

\begin{figure}[ht]
	\centering
	\begin{subfigure}{0.48\textwidth}
		\centering
		\includegraphics[scale=\kDF]{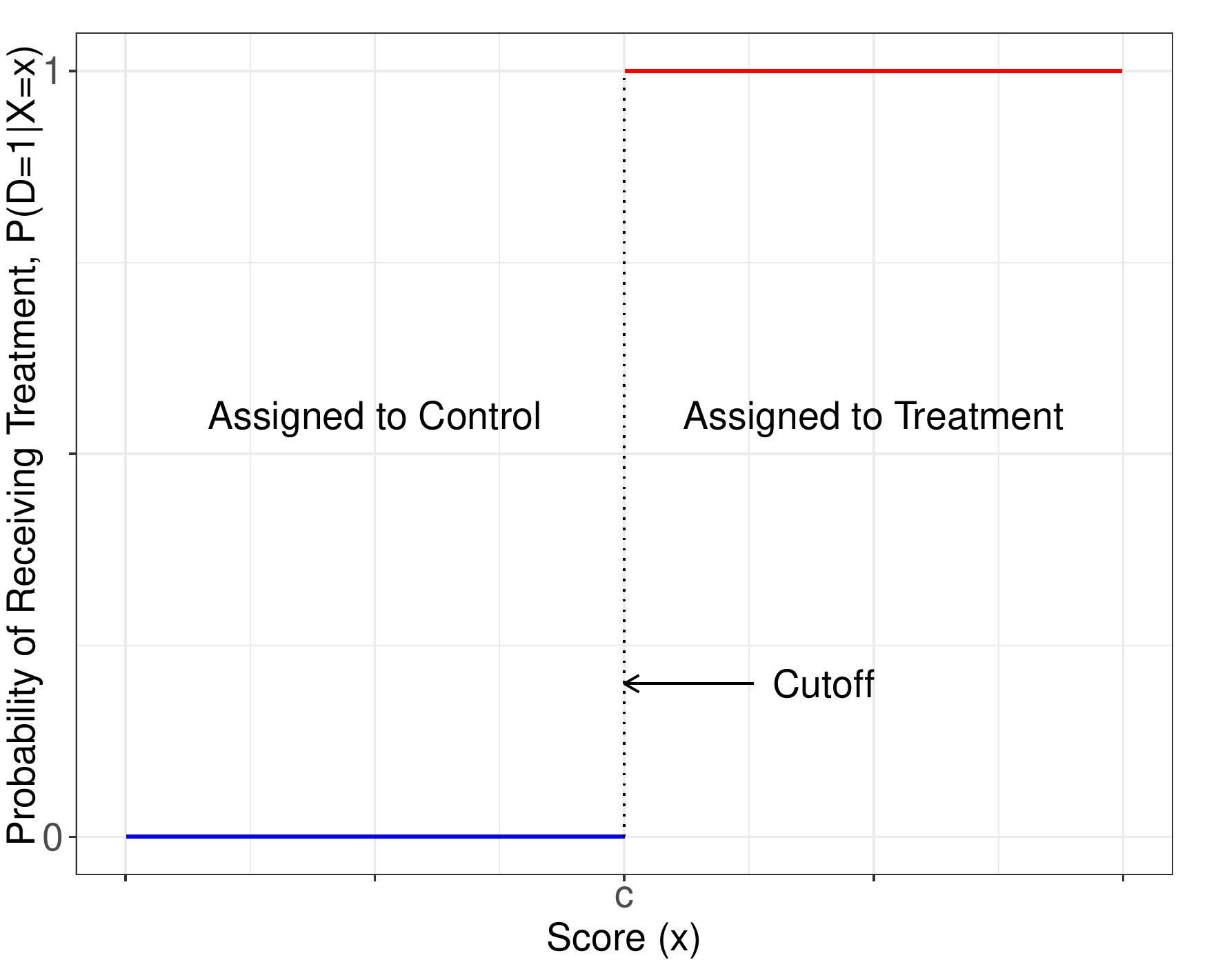}
		\caption{Sharp RD}\label{sharp}		
	\end{subfigure}
	\begin{subfigure}{0.48\textwidth}
		\centering
		\includegraphics[scale=\kDF]{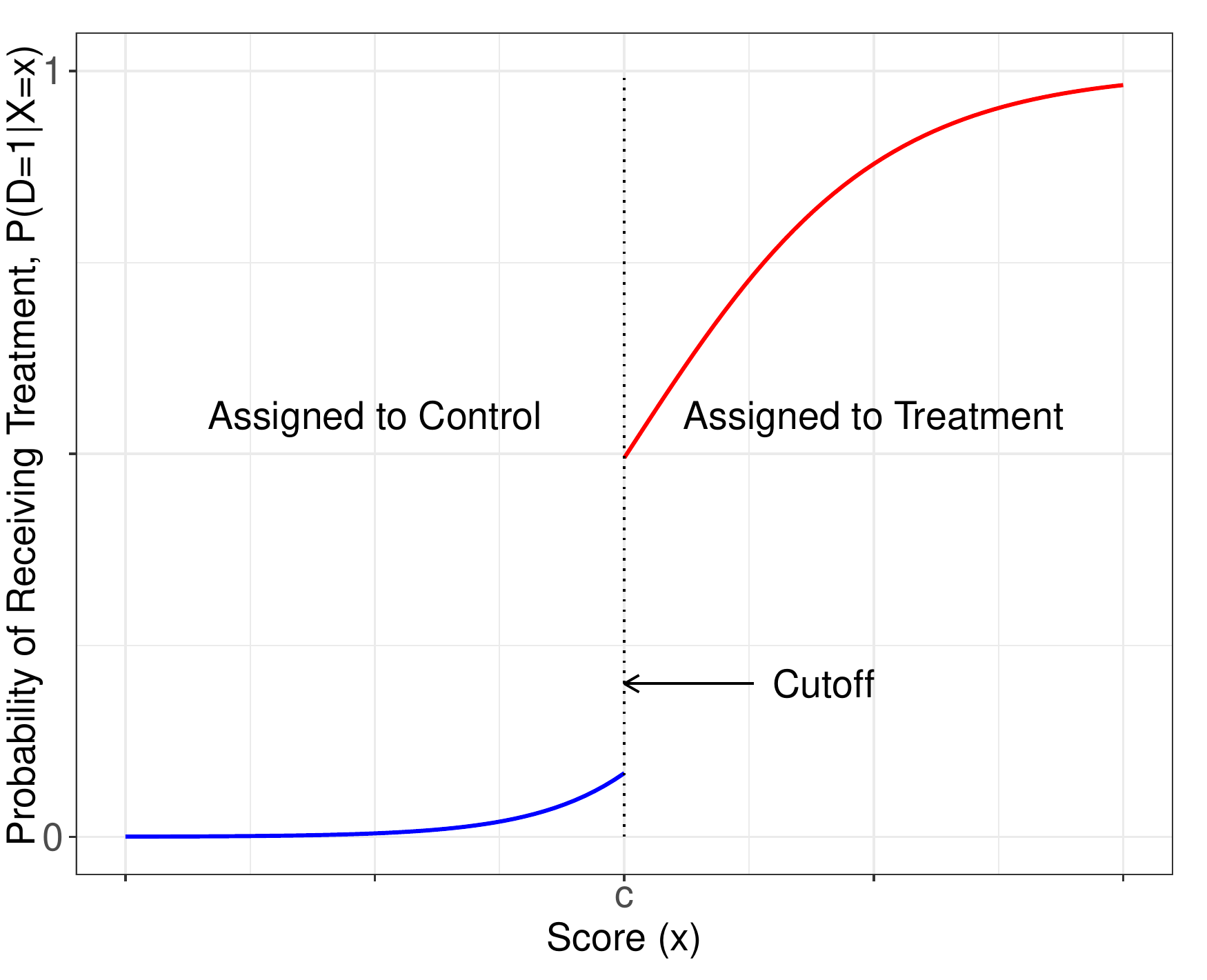}
		\caption{Fuzzy RD}\label{fuzzy}
	\end{subfigure}
	\caption{Conditional Probability of Receiving Treatment in Sharp vs. Fuzzy RD Designs}\label{fig:tr}
\end{figure}

We illustrate the difference between the Sharp and Fuzzy RD designs in Figure \ref{fig:tr}, where we plot the conditional probability of receiving treatment given the score, $\P(\D_i=1 | X_i = x)$, for different values of the running variable $X_i$. As shown in Figure \ref{fig:tr}(\subref{sharp}), in a Sharp RD design the probability of receiving treatment changes exactly from zero to one at the cutoff. In contrast, in a Fuzzy RD design, the change in the probability of being treated at the cutoff is always less than one. Figure \ref{fig:tr}(\subref{fuzzy}) illustrates a Fuzzy RD design with so-called two-sided non-compliance: near the cutoff, some control units receive the treatment, and some treated units fail to receive the treatment.

The treatment received $\D_i$, also known as the \textit{treatment take-up}, has two potential values: $\D_{i}(1)$ is the treatment received by $i$ when this unit is assigned to the treatment condition (i.e, when $X_i \geq \C$ and $T_i =1$) and $\D_{i}(0)$ is the treatment received when this unit is assigned to the control condition (i.e, when $X_i < \C$ and $T_i =0$), with $\D_{i}(1),\D_{i}(0) \in \left\{0,1\right\}$. For example, if unit $i$ receives the treatment when assigned to the control condition, we write $D_{i}(0)=1$, and if this unit complies with the control assignment, we write $D_{i}(0)=0$. The observed treatment taken is $\D_i = \T_i \cdot \D_{i}(1) + (1-\T_i) \cdot \D_{i}(0)$ and the fundamental problem of causal inference now extends to the treatment received in addition to the outcome: for every unit, we observe either $\D_{i}(1)$ or $\D_{i}(0)$, but never both. The quantities $\D_{i}(1)$ and $\D_{i}(0)$ are thus the potential decisions to comply with the treatment assignment; for brevity, we refer to them as \textit{potential treatments}.

Given the possibility of noncompliance, we generalize the notation for the potential outcomes to $Y_i(T_i, D_i(T_i))$, which now includes both the treatment assigned ($T_i$) and the treatment received ($D_i$) as arguments. Because $T_i$ and $D_i$ are both binary, we now have four potential outcomes instead of two. The potential outcome when unit $i$ is assigned to treatment is $Y_i(1, D_i(1))=D_i(1)Y_i(1,1) + (1-D_i(1)) Y_i(1,0)$, which results in $Y_i(1,1)$ or $Y_i(1,0)$ depending on whether $D_i(1)$ is equal to 1 or 0. Similarly, the potential outcome when $i$ is assigned to control is $Y_i(0, D_i(0)) = D_i(0)Y_i(0,1) + (1-D_i(0)) Y_i(0,0) $. The observed outcome is now $Y_i = T_i Y_i(1, D_i(1)) + (1-T_i) Y_i(0, D_i(0))$.

In the Fuzzy RD design, researchers are usually interested in the effects of both assigning the treatment and receiving the treatment on the outcome of interest. Since it is always the case that all units below the cutoff are assigned to control and all units above it are assigned to treatment, the analysis of the effect of assigning the treatment follows standard Sharp RD design methods. In contrast, studying the effect of receiving the treatment requires modifications and different assumptions. We devote this section to discussing both types of effects, organizing our discussion around the same topics previously discussed in \textit{Foundations} and in the last section: estimation of effects, inference, falsification, graphical illustration, and interpretation. As in the Sharp RD case, the analysis of Fuzzy RD designs can be based on a continuity-based approach or a local randomization approach, depending on the assumptions invoked. After introducing our empirical example, we discuss and illustrate both approaches together in Section \ref{subsec:FuzzyAnalysis}.

\subsection{The Effect of Financial Aid on Post-Secondary Education Attainment}\label{subsec:FuzzyApplication}

We illustrate with the study by \citet*{LondonoVelezRodriguezSanchez_2020_AEJ} of the effects of a governmental subsidy for post-secondary education in Colombia. The program, Ser Pilo Paga (SPP), funds the full tuition of a four-year or five-year undergraduate program in any government-certified higher education institution (HEI) with high-quality status. Program eligibility depends on both merit and economic need: in order to qualify for the program, students must obtain a high grade in Colombia's national standardized high school exit exam, SABER 11, and they must also come from economically disadvantaged families, measured by a survey-based wealth index known as SISBEN. In both cases, eligibility follows a deterministic rule with fixed cutoffs: students must obtain a SABER 11 score in the top $9$ percent of scores, and they must come from a household with SISBEN index below a region-specific threshold. 
 
The analysis includes only students who took the SABER 11 test in the fall of 2014; this is the first cohort of beneficiaries of the SPP program. Because program eligibility is based on whether observed scores exceed fixed cutoffs, the SPP program is a clear example of an RD design. However, it differs from the setup discussed in \textit{Foundations} and in the prior section in this Element in two ways. First, because some eligible students did not receive the SPP subsidy, there is imperfect compliance with the treatment assignment, making this a Fuzzy RD design. Second, program eligibility is determined by two scores as opposed to one, which makes this a multi-dimensional RD design in general.
 
For the purposes of this section, we transform this two-dimensional RD design into a one-dimensional RD design by considering only the subset of students whose SABER 11 score is above the merit cutoff. For this subsample, program eligibility obeys a one-dimensional Fuzzy RD design where the score is the SISBEN wealth index. (We re-analyze this application using the full sample and considering both scores simultaneously in Section \ref{sec:multiRD}, where we discuss multi-dimensional RD designs.) In this one-dimensional Fuzzy RD design, the unit of analysis is a student, the running variable is the student's SISBEN wealth index, the treatment is receipt of the SPP subsidy, and the cutoff varies according to the student's area of residence ($40.75$ in rural areas, $57.21$ in the fourteen main metropolitan areas, and 56.32 in other urban areas). The SISBEN wealth index is continuous and ranges between 0 (poorest) and 100 (richest); it is constructed based on a household survey that measures housing quality, ownership of durable goods, pubic utility services, and other indicators of wealth. The main outcome of interest is enrollment in a high-quality HEI.
 
In our replication dataset there are 23,132 total observations, corresponding to students whose SABER 11 score was above the cutoff and whose household received welfare benefits. The running variable $X_1$ is the difference between the student's SISBEN wealth index and her corresponding cutoff, ranging from $-43.84$ to $56.23$ in the sample; the cutoff is thus normalized to zero. The treatment assignment ($T_i$) is an indicator equal to one when the running variable is below zero, which indicates that the student is eligible to receive the SPP subsidy. The treatment received ($D_i$) is an indicator equal to one if the student actually received the subsidy, regardless of their SISBEN score value. Approximately $66.7\%$ of the students in this sample are eligible to receive the SPP program, but only $40\%$ actually receive it. As we will see, the $40\%$ of students who do receive the program does not include any students with SISBEN score below the eligibility cutoff, which makes this an example of a Fuzzy RD design with one-sided non-compliance.

The main outcome of interest is an indicator equal to one if the student enrolled in a HEI immediately after receiving the subsidy ($Y_i$). We also use six predetermined covariates for the falsification analysis: an indicator equal to one if the student identifies as female (\texttt{icfes\_female}), the student's age (\texttt{icfes\_age}), an indicator equal to one if the student identifies as an ethnic minority (\texttt{icfes\_urm}), the residential stratum of the student's household (\texttt{icfes\_stratum}), an indicator equal to one if the student attends a private high school (\texttt{icfes\_privatehs}), and the student's family size (\texttt{icfes\_famsize}). 

\subsection{Estimation and Inference Methods}
\label{subsec:FuzzyAnalysis}

The presence of non-compliance complicates the study of RD treatment effects. As it occurs in randomized experiments, complications arise because some units may strategically decide to take or refuse the treatment based on their expected gains from receiving the treatment. This introduces confounding between potential outcomes and compliance decisions that, in the absence of additional assumptions, prevents us from learning causal treatment effects for all units. Thus, in settings with non-compliance, it is common to shift the focus to different parameters that can still be recovered under reasonable assumptions and, despite being less general than the average treatment effect, are still of potential interest. 

In order to define such parameters, we start by applying the Sharp RD estimation strategy to the observed outcome in a Fuzzy RD design, and considering what kind of effect is recovered in this case. As we discussed in \textit{Foundations}, a continuity-based analysis of a Sharp RD proceeds by separately (and locally) estimating the average observed outcome given the score, $\E[Y_i | X_i = x]$, for observations above and below the cutoff, and then taking the limit of those averages as $x$ approaches the cutoff $\C$. When we apply this same strategy in the Fuzzy RD context, we estimate the following parameter,
\begin{align*}
\tau_Y &\equiv \lim_{x\downarrow \C}\E[Y_i|X_i=x]-\lim_{x\uparrow \C}\E[Y_i|X_i=x]\\
       &= \lim_{x\downarrow \C}\E[Y_i(1,D_i(1))|X_i=x]-\lim_{x\uparrow \C}\E[Y_i(0,D_i(0))|X_i=x],
\end{align*}
where the equality follows from the more general definition of the observed outcome given above and thus requires no special assumptions.

Analogously, applying the Sharp RD estimation strategy in the local randomization framework yields
 \begin{align*}
 \theta_Y &\equiv 	\frac{1}{N_\W} \sum_{i:X_i\in\W} \E_\W\Big[ \frac{T_i Y_i}{\P_\W[T_i=1]} \Big] 
 	- \frac{1}{N_\W} \sum_{i:X_i\in\W} \E_\W\Big[ \frac{(1-T_i) Y_i}{1-\P_\W[T_i=1]} \Big] \\
 	 & = 	 \frac{1}{N_\W} \sum_{i:X_i\in\W} \E_\W\Big[ \frac{T_i Y_i(1, D_i(1))}{\P_\W[T_i=1]} \Big] 
 	- \frac{1}{N_\W} \sum_{i:X_i\in\W} \E_\W\Big[ \frac{(1-T_i) Y_i(0, D_i(0))}{1-\P_\W[T_i=1]} \Big].
 \end{align*}
In words, $\tau_Y$ and $\theta_Y$ are the parameters that are estimated in a Fuzzy RD design when we compare the average outcome of observations just below the cutoff to the average outcome of observations just above the cutoff using, respectively, a continuity-based approach that takes the limit to the cutoff or a local randomization approach that compares observations in the small window $\W$.

A natural approach to the analysis of a Fuzzy RD design is to investigate different assumptions under which these quantities yield parameters that are of interest to the researcher. There are two main strategies. One is to focus on assumptions that allow us to interpret $\tau_Y$ and $\theta_Y$ as the effect of \textit{assigning} the treatment on the outcome. The other is to focus on assumptions that allow us to learn about the effect of \textit{receiving} the treatment on the outcome, at least for some subpopulation of units. Whether one or the other strategy is preferable depends on the specific application and goals of the researcher.

\subsubsection{Intention-to-treat Effects}

We start by considering the first strategy, where the focus is on learning about the effects of the treatment assignment, not the treatment received. Following the experimental literature, we call the effects of assigning the treatment on any outcome of interest \textit{intention-to-treat} (ITT) effects.

To obtain ITT parameters in the continuity-based framework, we generalize the conditions discussed in \textit{Foundations}, and assume that the regression functions $\E[Y_i(1,D_i(1)) | X_i = x]$ and $\E[Y_i(0,D_i(0)) | X_i = x]$ are smooth near the cutoff $\C$. In other words, seeing $T_i$ as the intervention of interest, we ask that the regression functions for both values of this variable be continuous in the score at the cutoff. This assumption implicitly restricts how compliance decisions change at the cutoff: for example, if $D_i(1)$, seen as a function of the score $x$, changes discontinuously at $x=\C$ for some units and that leads to discontinuity of $\E[Y_i(1,D_i(1)) | X_i = x]$ at $c$, the assumption would not hold. Under continuity, we have
\begin{align*}
\tau_Y = \tau_{\mathtt{ITT}}, \qquad \tau_{\mathtt{ITT}} \equiv \E[Y_i(1,D_i(1)) - Y_i(0, D_i(0)) | X_i = \C],
\end{align*}
and the estimated jump in the average observed outcome at the cutoff recovers the average effect of $T_i$ on $Y_i$ at $\C$, which we denote $\tau_{\mathtt{ITT}}$.

In the local randomization framework, adapting the assumptions discussed in Section \ref{sec:localrand}, we continue to require that the treatment assignment mechanism be known and unconfounded in $\W$, but now we also require that the augmented potential outcomes $Y_i(1,D_i(1))$ and $Y_i(0,D_i(0))$ not be functions of $X_i$ inside $\W$. Under these assumptions, we have
 \begin{align*}
\theta_Y = \theta_{\mathtt{ITT}}, \qquad \theta_{\mathtt{ITT}} \equiv \frac{1}{N_{\W}} \sum_{i:X_i \in W} \EW[Y_i(1,D_i(1)) - Y_i(0, D_i(0))],
\end{align*}
and thus the estimated difference in the average observed outcomes inside the window recovers the average ITT effect of $T_i$ on $Y_i$ in $\W$, which we call $\theta_{\mathtt{ITT}}$.

The perfect compliance Sharp RD setting can now be understood as a particular case of the fuzzy RD design where $\P[D_i(0)=0|X_i=x]=1$ for $x<c$ (no units with score below the cutoff receive the treatment) and $\P[D_i(1)=1|X_i=x]=1$ for $x\geq c$ (all units with score above the cutoff receive the treatment), the treatment assignment rule reduces to the sharp rule, $D_i = T_i=\I(X_i\geq c)$, the four potential outcomes reduce to $Y_i(1, 1)=Y_i(1)$ and $Y_i(0, 0)=Y_i(0)$, and the ITT parameters become $\tau_{\mathtt{ITT}} = \E[Y_i(1) - Y_i(0) | X_i = \C]$ and $\theta_{\mathtt{ITT}} = \EW[Y_i(1) - Y_i(0) | X_i \in \W ]$. Thus, when compliance is perfect, the ITT effects of the treatment assignment on the outcome reduce to the Sharp RD effects of the treatment received.

In addition to investigating the effects on the outcome, an ITT analysis of a Fuzzy RD design should include a study of how the RD assignment rule affects the probability of receiving the treatment. The effect of the treatment assignment on the treatment received reveals information about compliance and the effectiveness of the RD rule in inducing individuals to take the treatment. We define parameters analogous to $\tau_{\mathtt{Y}}$ and $\theta_{\mathtt{Y}}$, but this time treating $D_i$ as the outcome. Applying a Sharp RD estimation strategy that compares observations just above and below the cutoff, we define
\[\tau_{\mathtt{D}} \equiv \lim_{x\downarrow \C}\E[D_i|X_i=x]-\lim_{x\uparrow \C}\E[D_i|X_i=x]\]
and
\[\theta_{\mathtt{D}} \equiv \frac{1}{N_\W} \sum_{i:X_i\in\W} \E_\W\Big[ \frac{T_i D_i}{\P_\W[T_i=1]} \Big] 
	- \frac{1}{N_\W} \sum_{i:X_i\in\W} \E_\W\Big[ \frac{(1-T_i) D_i}{1-\P_\W[T_i=1]} \Big],\]
for the continuity-based and local randomization frameworks, respectively. Since $D_i$ is binary, $\tau_{\mathtt{D}}$ and $\theta_{\mathtt{D}}$ capture the difference in the probability of receiving the treatment between units assigned to treatment and units assigned to control, at the cutoff or in the window. 

In order to interpret these parameters as the causal effect of $T_i$ on $D_i$, we must extend the above assumptions. In the continuity-based case, we assume continuity at $\C$ of $\E[D_i(1) | X_i = x]$ and $\E[D_i(0) | X_i = x]$, seen as functions of $x$. In the local randomization case, we require that the potential treatments, $\D_{i}(1)$ and $\D_{i}(0)$, satisfy the local randomization assumptions in $\W$. Under these assumptions, we have
\[\tau_{\mathtt{D}} = \tau_{\mathtt{FS}}, \qquad \tau_{\mathtt{FS}} \equiv \E[D_i(1) - D_i(0)|X_i=\C]\]
and
\[\theta_{\mathtt{D}} = \theta_{\mathtt{FS}}, \qquad \theta_{\mathtt{FS}} \equiv \frac{1}{N_{\W}} \sum_{i:X_i \in W} \EW[D_i(1) - D_i(0)].\]
The parameters $\tau_{\mathtt{FS}}$ and $\theta_{\mathtt{FS}}$ thus capture the effect of assigning the treatment on receiving the treatment for units with scores near or at the cutoff. Following the instrumental variables (IV) literature, we call them \textit{first-stage} effects. 

In sum, to study ITT effects in a Fuzzy RD design we must augment the continuity and local randomization assumptions appropriately to cover the regression functions of the augmented potential outcomes $Y_i(t,d)$, and the additional potential treatments $D_i(t)$. In the continuity-based framework, we require continuity of the regression functions of the potential outcomes, $\E[Y_i(1, D_i(1)) | X_i = x]$ and $\E[Y_i(0, D_i(0)) | X_i = x]$, and the potential treatments, $\E[D_i(1) | X_i = x]$ and $\E[D_i(0) | X_i = x]$. In the local randomization framework, the exclusion restriction requires that both the potential outcomes and the potential treatments be unaffected by the score within $\W$. Informally, these extended continuity and local randomization assumptions require that near the cutoff, the outcomes of units with scores below the cutoff be similar to the outcomes that the units with scores above the cutoff would have had if they had been assigned to the control condition instead of the treatment. As in any Sharp RD design, if any important variable other than the treatment assignment changes abruptly at the cutoff, these assumptions will fail to hold.

Once we generalize the continuity and local randomization conditions to accommodate non-compliance, estimation, inference, and validation for the ITT parameters $\theta_{\mathtt{ITT}}$, $\theta_{\mathtt{FS}}$, $\tau_{\mathtt{ITT}}$, and $\tau_{\mathtt{FS}}$ proceed by applying the methods of analysis for Sharp RD designs. More precisely, local randomization and continuity-based Sharp RD methods are deployed where $X_i$ remains the RD score, $T_i = \I(X_i \geq \C)$ is seen as the ``treatment'' of interest, and now both $Y_i$ and $D_i$ are viewed as outcomes for the analysis. In the continuity-based framework, the ITT parameters can be estimated with the difference in the intercepts of local polynomials of the observed outcome on the score, fit separately for observations above and below the cutoff,
\[\hat{\tau}_{\mathtt{ITT}} = \lim_{x\downarrow \C}\widehat{\E}[Y_i|X_i=x]-\lim_{x\uparrow \C}\widehat{\E}[Y_i|X_i=x]\]
and
\[\hat{\tau}_{\mathtt{FS}} = \lim_{x\downarrow \C}\widehat{\E}[D_i|X_i=x]-\lim_{x\uparrow \C}\widehat{\E}[\D_i|X_i=x], 
\]
with bandwidth selection and inference methods as discussed in Sections 4.2 and 4.3 in \textit{Foundations}.

In the local randomization framework, $\theta_{\mathtt{ITT}}$ and $\theta_{\mathtt{FS}}$ can be estimated by calculating sample difference-in-means between units above and below the cutoff for units with scores in $\W$:
\[\widehat{\theta}_{\mathtt{ITT}} = \bar{Y}_{\W,+} - \bar{Y}_{\W,-} \qquad\text{and}\qquad
  \widehat{\theta}_{\mathtt{FS}} = \bar{D}_{\W,+} - \bar{D}_{\W,-},
\]
where
\[\bar{Y}_{\W,+} = \frac{1}{N_{\W,+}} \sum_{i:X_i\in\W} \omega_i T_i Y_i,\qquad
  \bar{Y}_{\W,-} = \frac{1}{N_{\W,-}} \sum_{i:X_i\in\W} \omega_i (1-T_i) Y_i
\]
and
\[\bar{D}_{\W,+}  = \frac{1}{N_{\W,+}} \sum_{i:X_i\in\W} \omega_i T_i D_i,\qquad
  \bar{D}_{\W,-} = \frac{1}{N_{\W,-}} \sum_{i:X_i\in\W} \omega_i (1-T_i) D_i,
\]
with the weights appropriately selected as discussed in Section \ref{subsec:locrand-estimation}, and after the window $\W$ has been selected, preferably based on pre-treatment covariates as discussed in Section \ref{subsec:chooseW}. Inference can similarly proceed based on the methods we discussed in Section \ref{sec:localrand}, using either Fisherian or super-population methods, depending on whether the potential outcomes are seen as fixed or stochastic.

Under the appropriate assumptions, $\theta_{\mathtt{ITT}}$ and $\tau_{\mathtt{ITT}}$ capture the average overall effect of assigning the treatment on the outcome, not of receiving the treatment. In some applications, this effect will be of primary interest. For example, when households with income below a cutoff are eligible to receive a cash transfer, households whose income would have been above but near the cutoff might decrease their labor supply so that they become eligible for the program. In such a case, the effects of program eligibility on outcomes of interest are essential to characterizing the impact of the program in the population.

\subsubsection{Treatment Effects for Subpopulations}

In some Fuzzy RD applications, researchers are interested in learning about the effect of receiving the treatment itself rather than the effect of the assignment. In these cases, it is common to consider other parameters which, under different assumptions, can provide information about the treatment received, at least for a subpopulation of units.

When interest is on the effect of the treatment received, it is common to focus on the parameters
\[\tau_{\mathtt{FRD}} \equiv \frac{\tau_\mathtt{Y}}{\tau_\mathtt{D}} \qquad\text{and}\qquad
  \theta_{\mathtt{FRD}} \equiv \frac{\theta_\mathtt{Y}}{\theta_\mathtt{D}}
\]
for the continuity-based and local randomization frameworks, respectively. We call these parameters the \textit{Fuzzy RD parameters}, and discuss conditions under which they can be interpreted as the average effect of the treatment for some subpopulations. 

Under the augmented continuity and local randomization assumptions discussed above for the ITT effects, these Fuzzy parameters will be equal to the ratio of the effects of the treatment assignment on the outcome and the treatment received, $\tau_{\mathtt{FRD}} = \frac{\tau_\mathtt{ITT}}{\tau_{\mathtt{FS}}}$ and $\theta_{\mathtt{FRD}} = \frac{\theta_\mathtt{ITT}}{\theta_{\mathtt{FS}}}$. This interpretation of the Fuzzy RD parameters as the ratio of two ITT effects is analogous to results in the IV literature. However, our upcoming discussion does not assume that these conditions hold.

Putting aside conditions to recover ITT effects, we now explore assumptions under which the Fuzzy RD parameters $\theta_{\mathtt{FRD}}$ and $\tau_{\mathtt{FRD}}$ can be directly interpreted as treatment effects. The first assumption we discuss is analogous to the exclusion restriction in IV settings: the treatment assignment must affect the potential outcomes and potential treatments only via the treatment received, but not directly. In other words, given a particular value of the treatment received, $D_i=d$, the potential outcomes (or their distributions) should not be affected by the value of $T_i$, at least not near the cutoff.

This exclusion restriction is already imposed in the local randomization framework when we assume that the potential outcomes and potential treatments cannot be a function of the score within $\W$. Since the assignment $T_i=\I{(X_i\leq \C)}$ is a function of $X_i$, assuming that $Y_i(T_i, 1)$ and $Y_i(T_i, 0)$ (or their distributions) are not functions of $X_i$ within $\W$ implies assuming that, given a value of the treatment received, $D_i=d$, the potential outcomes (or their distributions) do not depend on $T_i$. In contrast, in the continuity-based framework, it is still possible for $X_i$ to affect the potential outcomes directly because all parameters are defined at the same point $X_i=\C$, which makes any direct effects of $X_i$ irrelevant. In this case, the exclusion restriction requires, at a minimum, that conditional on $D_i=d$, there is no discontinuity in the regression functions at the cutoff---that is, that $\E[Y_i(T_i,0) |X_i=x]$ and $\E[Y_i(T_i,1) |X_i=x]$ are continuous in $x$ at $\C$, ensuring that the effect of the treatment assignment at the cutoff is entirely driven by the treatment received.

Assuming that the treatment assignment has no effect on the outcome except via the treatment received is not sufficient to recover the effects of the treatment in the Fuzzy RD design. The reason is that the decision to comply or not with the treatment assignment is still unrestricted and does not allow us to decompose $\tau_{\mathtt{Y}}$ and $\theta_{\mathtt{Y}}$ in a way that leads to treatment effects of interest, which is the typical strategy to obtain treatment effects in IV settings. We thus need an additional assumption.

One possibility is to assume that the potential outcomes are unrelated to or independent of the potential treatments for values of the score in $\W$ or near the cutoff. In this case, the Fuzzy RD parameters $\tau_\mathtt{FRD}$ and $\theta_\mathtt{FRD}$ simplify to $\tau_\mathtt{SRD}$ and $\theta_\mathtt{SRD}$, respectively, thus capturing the same effects as in the Sharp RD setup. This assumption is rarely invoked because it requires that an individual's decision to comply or not with the treatment assignment be unrelated to the anticipated gains from taking the treatment, which is implausible in most applications. 

An alternative assumption restricts the type of non-compliance that may occur in a more flexible way, allowing for strategic compliance decisions based on anticipated treatment gains. To explain this assumption, we informally define four different groups of units according to their compliance decisions. \textit{Compliers} are those units whose treatment received coincides with their treatment assigned; \textit{Never-takers} are those who always refuse the treatment regardless of their assignment; \textit{Always-takers} are those who always take the treatment regardless of their assignment; and \textit{Defiers} are those who receive the opposite treatment to the one they are assigned. These definitions are formalized differently depending on whether the local randomization or the continuity-based framework is used and usually applies to units with scores near the cutoff.

Under the assumption that there are no defiers inside $\W$ or at (or near) $\C$, known as \textit{monotonicity}, $\theta_{\mathtt{FRD}}$ and $\tau_{\mathtt{FRD}}$ recover the effect of the treatment received for the subpopulation of units that comply with the treatment assignment. We let $B_i$ be a binary variable that denotes compliance status, with $B_i=1$ if unit $i$ is a complier and $B_i=0$ otherwise. Adding monotonicity, it can be shown that $\tau_\mathtt{FRD} = \E[Y_i(1,1)-Y_i(0,0)|X_i= \C, B_i=1 ]$ and $\theta_{\mathtt{FRD}} = \frac{1}{N_\W}\sum_{i: B_i=1, X_i \in W}\EW[Y_i(1,1) - Y_i(0,0)| B_i = 1]$. In other words, under the additional assumption of monotonicity, the Fuzzy RD parameters recover the average effect of the treatment at the cutoff for compliers. Following the IV literature, these effects are sometimes called Local Average Treatment Effects (LATE) or Complier Average Treatment Effects (CATE). 

In the local randomization framework, the formal definitions of complier strata and monotonicity are analogous to those in the IV literature, restricted to units whose scores are in $\W$. In contrast, in the continuity-based framework, the formalization of monotonicity and related conditions is less straightforward, and several alternatives have been proposed. The technical details are beyond the scope of our practical guide, but we offer references at the end of this section for the interested reader. The general conclusion is that by adding a monotonicity assumption to the continuity conditions, the Fuzzy RD parameters can be interpreted as the average treatment effect at the cutoff for the compliers.

Naturally, estimation and inference for the Fuzzy RD parameters $\theta_{\mathtt{FRD}}$ and $\tau_{\mathtt{FRD}}$ proceed in the same manner regardless of what particular assumptions are invoked to interpret them. Estimation proceeds by simply using local polynomials or difference-in-means to estimate the numerator and denominator, and taking their ratio, which entails taking the ratio of the estimators defined above,
\[\widehat{\tau}_{\mathtt{FRD}} = \frac{\widehat{\tau}_\mathtt{Y}}{\widehat{\tau}_\mathtt{D}} = \frac{\lim_{x\downarrow \C}\widehat{\E}[Y_i|X_i=x]-\lim_{x\uparrow \C}\widehat{\E}[Y_i|X_i=x]}{\lim_{x\downarrow \C}\widehat{\E}[D_i|X_i=x]-\lim_{x\uparrow \C}\widehat{\E}[\D_i|X_i=x]}\]
and
\[\widehat{\theta}_{\mathtt{FRD}} = \frac{\widehat{\theta}_\mathtt{Y}}{\widehat{\theta}_\mathtt{D}} = \frac{\bar{Y}_{\W,+} - \bar{Y}_{\W,-}}{\bar{D}_{\W,+} - \bar{D}_{\W,-}}.\]

Inference methods are analogous to those used in the Sharp RD design, with some modifications. In the local randomization approach, inferences in the super-population framework rely on standard IV large sample approximations (based on the Delta method) to the sampling distribution of the ratio of the two effects, applied to observations with scores inside $\W$. In the Fisherian framework, inferences are implemented as before by permuting the vector of treatment assignments $T_i$ according to the assumed distribution. Under the sharp null hypothesis of no treatment effect for any unit, no modifications are needed; under the null hypothesis that the effect is the same $\gamma$ for every unit, implementation relies on testing the sharp null on the adjusted observed outcomes $Y_i - D_i \gamma$. In all cases, the implementation permutes the treatment assignment, not the treatment received.

In the continuity-based framework, inferences can be based on robust local polynomial methods. As discussed in \textit{Foundations} for the Sharp RD case, these methods use polynomials to approximate the unknown regression functions in a neighborhood or bandwidth around the cutoff, and in general contain an error because the approximation is not exact. This error of approximation (also known as \textit{bias}) is controlled by the bandwidth and affects the large sample distribution of the test statistics of interest. For this reason, it is recommended to employ robust bias-correction methods for inference when local polynomial methods are used for RD analysis, particularly when the bandwidth is chosen to be mean-squared-error optimal, as is common (and recommended) in practice. Analogously to the Sharp RD case, the resulting bias-corrected robust confidence interval is not centered around $\hat{\tau}_{\mathtt{FRD}}$, but rather around 	$\hat{\tau}_{\mathtt{FRD}}$ minus its estimated bias. All conceptual issues are the same as those discussed in \textit{Foundations} for the case of Sharp RD (Sections 4.2 and 4.3). Similarly, covariate adjustment for the ITT parameters follows directly the procedures discussed in \textit{Foundations} and Section \ref{sec:localrand}, depending on the framework adopted. As discussed in \citet{Cattaneo-Keele-Titiunik_2023_HandbookCh}, covariate adjustment can be used for efficiency gains without altering the parameter of interest, but not to ``fix'' RD designs where predetermined covariates are imbalanced at or near the cutoff.

The main points of our discussion so far can be summarized as follows:
\begin{itemize}
	\item In the Fuzzy RD design, some units fail to comply with the treatment they are assigned, which introduces a distinction between the treatment assignment and the treatment received. Researchers must decide whether they are interested in the effect of the treatment assignment, the effect of the treatment received, or both.
			
	\item When interest is on the effect of the treatment assignment, estimation and inference proceed analogously to the Sharp RD case. The effect of the treatment assignment ($T_i$) on both the outcome ($Y_i$) and the treatment received ($D_i$) can be studied in a straightforward manner using Sharp RD methods where $T_i$ is seen as the treatment of interest and the outcomes are $Y_i$ and $D_i$. The local randomization and continuity assumptions required are the same as those discussed in Section \ref{sec:localrand} and \textit{Foundations}, respectively, with appropriate extensions. These are called the intention-to-treat effects and capture the effect of assigning the treatment, not the effect of receiving the treatment.
	
	\item When interest is on the effect of the treatment received, it is common to focus on the Fuzzy RD parameter, which is the ratio (at or near the cutoff) of the difference in average outcomes and the difference in the probability of receiving treatment between units above and below the cutoff. Under appropriate assumptions, this is equal to the effect of the treatment received for all or a subset of units near or at the cutoff. For example, under monotonicity, it is equal to the effect of the treatment received near the cutoff for compliers; under local independence, it is equal to the effect of the treatment received for all units near the cutoff. Moreover, under appropriate assumptions, the Fuzzy RD parameter can be interpreted as the ratio (at or near the cutoff) of the ITT effect of the treatment assignment on the outcome and the ITT effect of the treatment assignment on the treatment received.
		
\end{itemize}

\subsubsection{Bandwidth and Window Selection}

In the continuity-based framework, the bandwidth selection for estimation of the ITT parameters proceeds exactly as explained in \textit{Foundations} (Section 4.2.2). The bandwidth for $\tau_{\mathtt{FRD}}$, however, requires further consideration. Because this parameter is a ratio, the question arises of whether researchers should use a different bandwidth for the denominator and numerator, or the same bandwidth for both. If the focus is on ITT effects, the numerator and denominator parameters will be of independent interest; in this case, the researcher should estimate these parameters separately, selecting a separate optimal bandwidth for each. But if the researcher is also (or only) interested in $\tau_{\mathtt{FRD}}$, using different bandwidths for the numerator and denominator has the disadvantage that the estimators will be constructed based on a different set of observations (naturally, one group will be a subset of the other). 

Thus, when interest is on the ratio parameter $\tau_{\mathtt{FRD}}$, it is sensible to use the same bandwidth to estimate both the numerator and the denominator. This adds transparency to the analysis, as researchers can clearly explain which observations are included in the calculations. This is the approach adopted by \texttt{rdrobust}, where the default is to choose a single mean-squared-error (MSE) optimal bandwidth by minimizing the MSE of a linear approximation of the ratio estimator $\hat\tau_{\mathtt{FRD}}$; the single MSE objective function leads to a single bandwidth for estimation of the Fuzzy RD parameter, avoiding different bandwidths for numerator and denominator.

This issue does not arise in the local randomization framework, because the local randomization assumptions are invoked in a single window that applies to all outcomes. The window selection should be implemented by assessing whether pre-determined covariates are balanced in nested windows around the cutoff, exactly as discussed in Section \ref{sec:localrand} for the Sharp RD case. In other words, window selection in the Fuzzy RD design is unchanged.

\subsubsection{Weak Assignment}

Because the Fuzzy RD parameter is the ratio of two parameters, it will be undefined if the denominator is zero. Thus, the study of this parameter requires the additional assumption that the denominator is non-zero. This assumption is equivalent to the notion of ``relevant instrument'' in the IV literature and can be studied empirically with a test of the null hypothesis that $\tau_{\mathtt{D}}$ or $\theta_{\mathtt{D}}$ are zero, which should be the first step in the analysis of any Fuzzy RD design. If the p-value associated with this hypothesis is smaller than conventional thresholds, the evidence allows us to conclude that the effect is non-zero.

However, the assumption of a non-zero first-stage effect is not enough. When the RD rule has a non-zero but very small effect on the probability of receiving the treatment, the standard Gaussian approximations to the distributions of the RD test statistics (i.e., those based on difference-in-means in $\W$ or limits of local polynomial estimators at the cutoff) are not reliable, and statistical inferences based on those approximations will be invalid. In the IV literature, this is a well-understood problem known as \textit{weak instruments} or, more generally, \textit{weak identification}, which may persist even if the number of observations is very large. 

Following standard results, it is possible to construct inference procedures that are robust to the weak instrument problem. For example, in the context of local randomization methods, Fisherian inference methods continue to be valid in the sense that randomization-based tests of the sharp null hypothesis are valid even if the treatment assignment is a weak instrument for the treatment received. Furthermore, all standard super-population approaches based on large sample approximations under weak-IV asymptotics can be deployed for units with scores within $\W$. Similar approaches can be implemented within the continuity-based framework based on local polynomial methods. The applicability of these results is often limited, however, because confidence intervals for the Fuzzy RD treatment effects often become long and practically uninformative when compliance with the treatment assignment is very weak.

Although the IV literature has developed methods to obtain valid inferences in the presence of weak instruments and some of those methods can be extended to the RD context, our practical recommendation is for researchers to avoid Fuzzy RD designs where the RD treatment assignment has a weak and small effect on the treatment received. A weak effect implies that the RD assignment rule failed to induce a large change in the probability of taking the treatment. When this occurs, any attempt to learn about the effects of the treatment received on the outcome will be severely limited. Our recommendation is, therefore, to always investigate the first stage effect ($\tau_{\mathtt{D}}$ or $\theta_{\mathtt{D}}$) first, and only proceed with the analysis of the ratio parameter if the estimated first stage is strong. Analogously to IV settings, strength can be measured empirically by the size of the F-statistic in the first-stage regression. The rule of thumb in the IV literature is to conclude that an instrument is weak if the F-statistic is less than 10; in the RD context, this threshold is likely to be too low, and recommendations are a minimum F-statistic of 20 or more. If the first-stage effect is weak, researchers should report only ITT effects and interpret these parameters carefully in light of the very weak relationship between treatment assignment and treatment received. An analysis that reveals a weak first-stage effect, however, can be very valuable, as it will likely offer important lessons about the design of the program and the strategic compliance decisions of individuals near the cutoff. In fact, in this case, the ITT effect can be a useful test of the exclusion restriction.

\subsubsection{Validation and Falsification}

As our discussion has emphasized, the methods of analysis for Fuzzy RD designs largely resemble those of Sharp RD designs, either exactly (in the case of ITT parameters) or conceptually (in the case of Fuzzy parameters). Nonetheless, the presence of non-compliance leads to some specific issues that may be important for implementation. The strategies for validation and falsification in the Fuzzy RD design are largely the same as those discussed in \textit{Foundations} (Section 5) and in Section \ref{sec:localrand} for the Sharp RD design: density test, treatment effects on covariates and placebo outcomes, artificial cutoffs, and sensitivity to the local neighborhood. In order to avoid repetition, we only focus on the modifications that are required to accommodate imperfect compliance. 

For implementation of the density test and estimation of effects on predetermined covariates and placebo outcomes, researchers should focus on the intention-to-treat effects. Because the goal of these falsification analyses is to assess whether the observations just above the cutoff are similar to the observations just below the cutoff, the effects of interest are those of $T_i$ (the treatment assignment) on the covariates and placebo outcomes. Similarly, because the goal of the density test is to assess whether the number of observations above the cutoff is similar to the number of observations below the cutoff, the relevant density test is one that compares the number of observations with $T_i=1$ and $T_i=0$ near the cutoff, not those with $D_i=1$ and $D_i=0$.

\subsection{Fuzzy RD in Practice}

We illustrate how to analyze a Fuzzy RD design using the SPP application introduced in Section \ref{subsec:FuzzyApplication}. We first implement a continuity-based analysis based on local polynomial methods. Since we illustrated Sharp RD methods in \textit{Foundations}, we omit details except when they pertain to issues that arise specifically due to non-compliance. We then illustrate the use of local randomization methods.

\begin{figure}[H]
	\centering
	\includegraphics[scale=0.4]{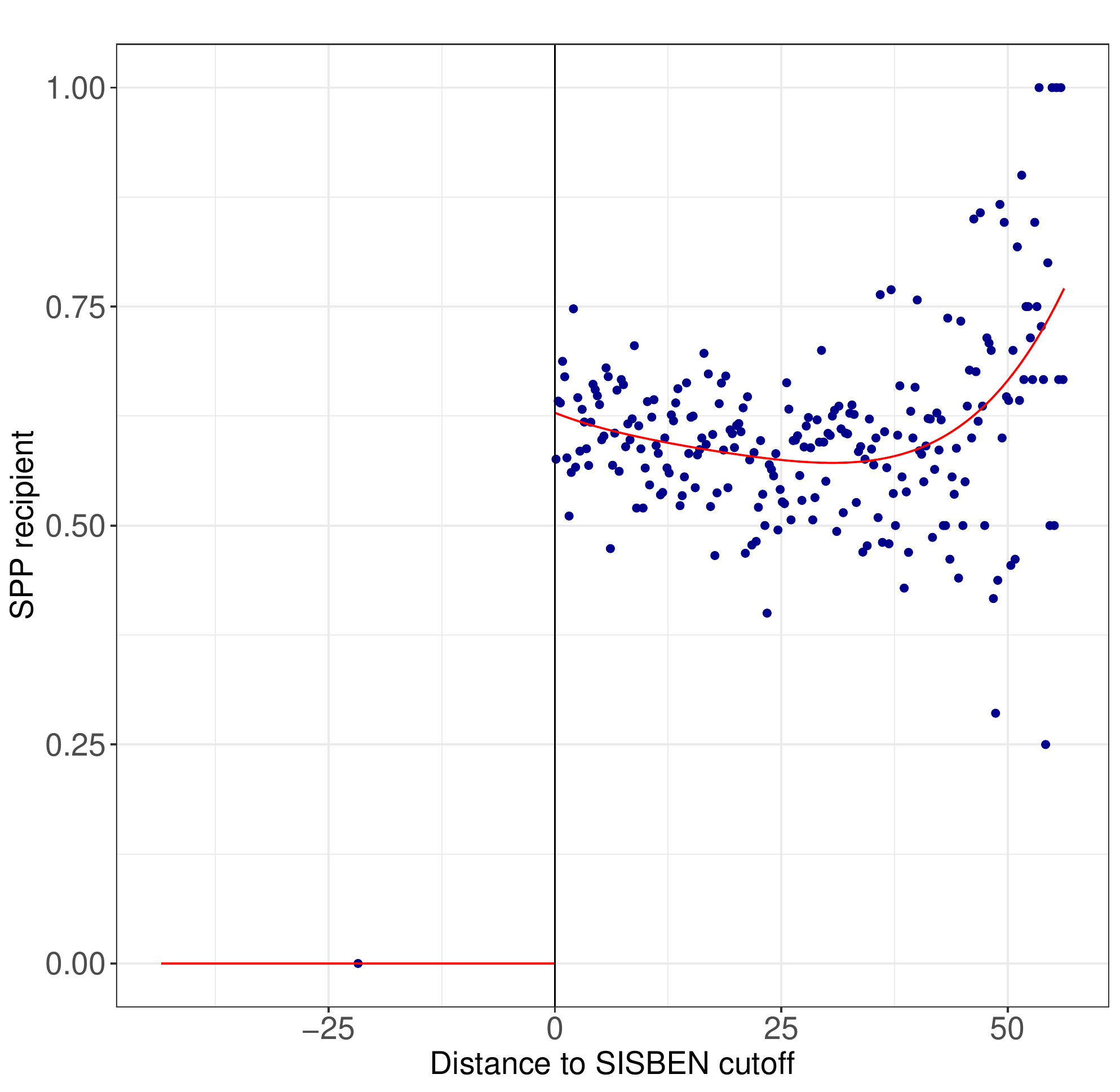}
	\caption{RD Plot: First Stage---SPP data}\label{fig:fuzzy-LRS-firststage}
\end{figure}

We start by investigating the ITT (sharp) RD effects of assigning SPP eligibility. The first step is to analyze the first-stage relationship between the eligibility to receive SPP funding ($\mathtt{T}$) and the actual receipt of SPP funds ($\mathtt{D}$). Figure \ref{fig:fuzzy-LRS-firststage} shows the corresponding RD plot with default choices, as discussed in \textit{Foundations}. The figure shows that this application has one-sided non-compliance: no student whose SISBEN score is below the cutoff receives SPP funding, but some of the students whose scores are above the cutoff fail to receive funding despite being eligible. The estimated probability of receiving the treatment of SPP funding thus jumps from zero to around 0.60 at the cutoff. 

A formal continuity-based analysis of the first stage can be conducted with local polynomial methods. Using the command \texttt{rdrobust}, we first choose a MSE-optimal bandwidth and then fit two linear polynomials of $Y_i$ on $X_i$ within this bandwidth, separately for observations above and below the cutoff. The RD effect is the difference between the estimated intercepts in both regressions, and we build confidence intervals using robust bias-corrected inference---for details, see Section 3 in \textit{Foundations}. Because compliance is one-sided and all students with SISBEN score below the cutoff fail to receive SPP funding, there is no need to fit a polynomial with observations below the cutoff: all those observations have $D_i=0$ and thus the intercept is zero. It follows that there is no need to select a bandwidth below the cutoff either; in fact, the bandwidth selector is undefined when the observations are constant. 

We use \texttt{rdrobust} to estimate the first-stage relationship between SPP eligibility and the actual receipt of funding.
\labelsnippet{fuzzyA}
\rsnip{Vol-2-R_LRS_rdrobust_firststage.txt}{\Rlink{\thesection}{\thefuzzyA}}
\statasnip{Vol-2-STATA_LRS_rdrobust_firststage}{\Slink{\thesection}{\thefuzzyA}}

The lack of variability below the cutoff leads \texttt{rdrobust} to behave as follows: (i) it prints a warning (not shown) stating that there is not enough variability in the data, (ii) it sets the bandwidth below the cutoff equal to the smallest value of the score, and (iii) it implements the bandwidth selector above the cutoff as usual. Although the output shows that $18.511$ is the bandwidth on both sides, this value corresponds to the MSE-optimal bandwidth above the cutoff. Re-running the command allowing for different bandwidths above and below the cutoff (option \texttt{rdbwselect=msetwo}) reports the bandwidth below the cutoff to be $43.480$, which is the minimum value of the SPP score in the data.

The output shows that the first-stage effect is $0.625$, consistent with the jump observed in Figure \ref{fig:fuzzy-LRS-firststage}. The effect is highly statistically significant, with a tight $95\%$ robust confidence interval between $0.595$ and $0.652$. This is evidence of a very strong effect of eligibility on receiving SPP funding, showing that approximately $62\%$ of those who are barely eligible to receive SPP funding do in fact receive it, compared with $0\%$ of those who are barely ineligible---an infinite change!

\begin{figure}[ht]
	\centering
	\includegraphics[scale=0.4]{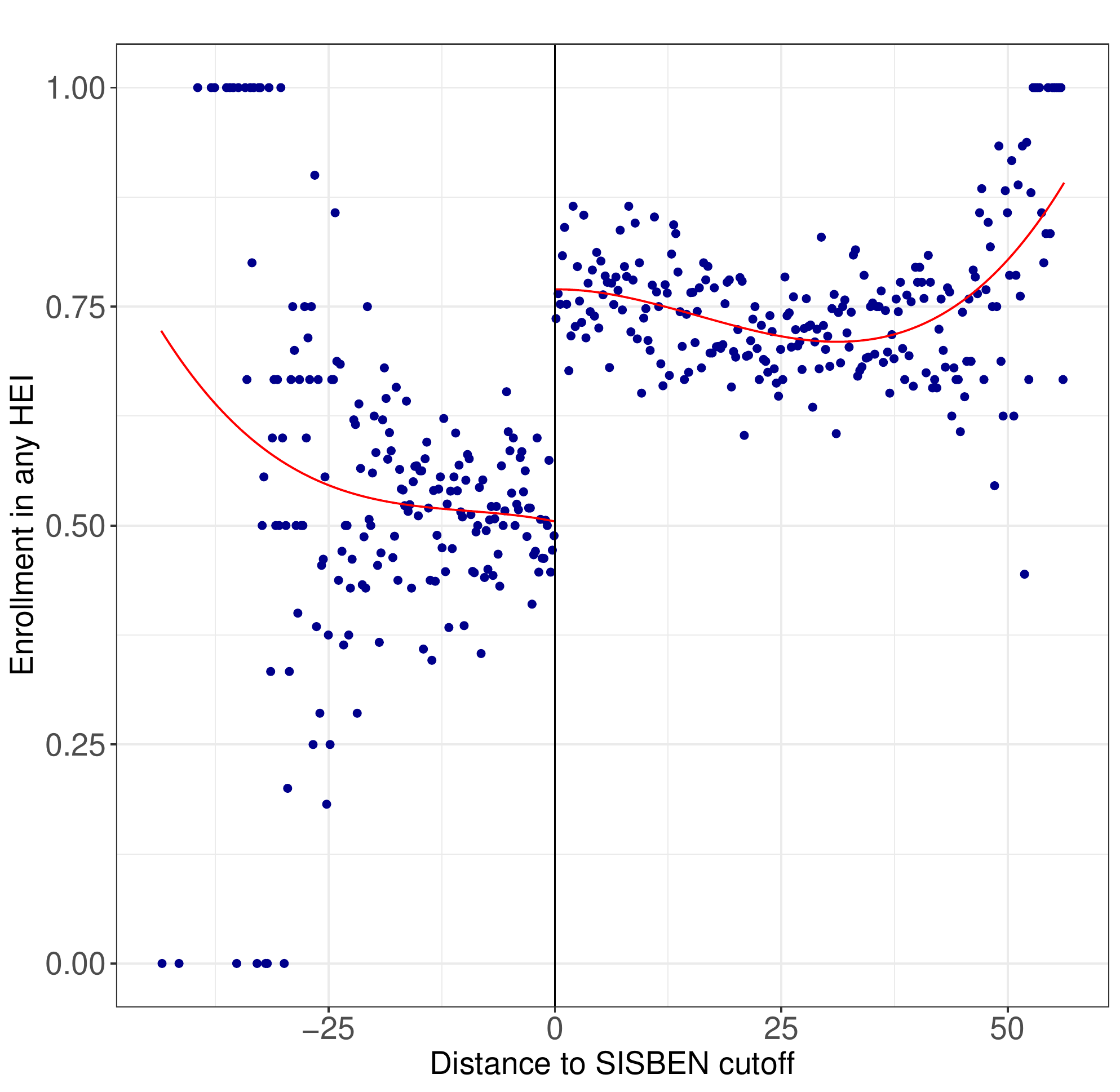}
	\caption{RD Plot: Intention-to-treat ($p=3$)---SPP data}\label{fig:fuzzy-LRS-reducedform}
\end{figure}

We continue with the analysis of the ITT effect of SPP eligibility on the outcome of interest, attending an HEI immediately after becoming eligible ($\mathtt{Y}$). The RD plot of this effect is shown in Figure \ref{fig:fuzzy-LRS-reducedform}, and we conduct formal estimation and inference with \texttt{rdrobust}.

\labelsnippet{fuzzyB}
\rsnip{Vol-2-R_LRS_rdrobust_reducedform.txt}{\Rlink{\thesection}{\thefuzzyB}}
\statasnip{Vol-2-STATA_LRS_rdrobust_reducedform}{\Slink{\thesection}{\thefuzzyB}}

The plot and the formal analysis both show a large effect at the cutoff: students whose SISBEN scores are barely above the cutoff and are thus barely eligible to receive SPP funding are about $27$ percentage points more likely to enroll in a HEI, jumping from near $50\%$ enrollment just below the cutoff to about $77\%$ just above the cutoff.

In order to point-estimate the fuzzy RD parameter, $\tau_{\mathtt{FRD}}$, we could simply take the ratio between the two ITT effects estimated above, $0.269/0.625=0.4304$. If we follow this approach, each effect is estimated using a different bandwidth, chosen to be MSE-optimal for each individual effect. The difference in bandwidths is considerable, $18.511$ for the denominator versus $9.041$ for the numerator, implying that many more observations are used to estimate the denominator than the numerator. Although this poses no problems theoretically, applied researchers may prefer to use the same observations to estimate both quantities; this can be achieved with the \texttt{fuzzy} option.

\labelsnippet{fuzzyC}
\rsnip{Vol-2-R_LRS_rdrobust_fuzzy.txt}{\Rlink{\thesection}{\thefuzzyC}}
\statasnip{Vol-2-STATA_LRS_rdrobust_fuzzy}{\Slink{\thesection}{\thefuzzyC}}

When the \texttt{fuzzy} option is specified, \texttt{rdrobust} first computes a single optimal bandwidth and then uses this bandwidth to estimate the denominator, the numerator, and the ratio. When non-compliance is one-sided as in our SPP example, \texttt{rdrobust} uses the optimal bandwidth for estimation of the Sharp ITT effect $\tau_{\mathtt{Y}}$. When compliance is imperfect on both sides of the cutoff, it chooses a bandwidth that is optimal for point-estimation of the linearized ratio $\tau_{\mathtt{Y}}/\tau_{\mathtt{D}}$. In both cases, the result is a single optimal bandwidth that is used to estimate all effects. An additional advantage of using the \texttt{fuzzy} option is that it reports robust bias-corrected confidence intervals for the fuzzy RD effect $\tau_{\mathtt{Y}}/\tau_{\mathtt{D}}$.
 
The estimated fuzzy RD effect using the \texttt{fuzzy} option is $0.434$, very similar to the ratio of 0.430 obtained above using different bandwidths (the difference occurs because the first-stage point estimate changes from 0.625 within the 18.511 bandwidth to 0.619 within the 9.041 bandwidth). As discussed before, the interpretation of this parameter depends on the assumptions invoked. Under appropriate continuity and monotonicity conditions, it is showing that receiving SPP funding resulted in an increase at the cutoff of roughly $43$ percentage points in the probability of enrolling in a HEI for the subset of students who are compliers.

We now present the analysis based on local randomization methods. The first step is to select the local randomization window $\W$ using the four covariates presented above: \texttt{icfes\_female}, \texttt{icfes\_age}, \texttt{icfes\_urm}, \texttt{icfes\_stratum}, \texttt{icfes\_privatehs}, and \texttt{icfes\_famsize}.

\labelsnippet{fuzzyG}
\rsnip{Vol-2-R_LRS_rdwinselect.txt}{\Rlink{\thesection}{\thefuzzyG}}
\statasnip{Vol-2-STATA_LRS_rdwinselect}{\Slink{\thesection}{\thefuzzyG}}

The minimum p-value is above $0.200$ in the first six windows, dropping to $0.111$ in the seventh window. The chosen window is therefore $[-0.13, 0.13]$, which has a total of $63$ control and $56$ treated observations. Once the window is selected, we use \texttt{rdrandinf} to estimate the first-stage parameter, to assess whether the RD eligibility rule (\texttt{T}) did in fact have the effect of changing the probability of receiving SPP funding (\texttt{D}) for students with scores within this window. We perform this analysis by selecting (\texttt{D}) as the outcome of interest in our call to \texttt{rdrandinf}.
\labelsnippet{fuzzyH}
\rsnip{Vol-2-R_LRS_rdrandinf_firststage.txt}{\Rlink{\thesection}{\thefuzzyH}}
\statasnip{Vol-2-STATA_LRS_rdrandinf_firststage}{\Slink{\thesection}{\thefuzzyH}}

Consistent with the continuity-based analysis, the output shows a very strong first stage: within the window, $57.1\%$ of students above the cutoff received SPP, and no students below the cutoff received it (from \texttt{Mean of outcome} row). This leads to a difference-in-means of $57.1\%$, as shown in the main output row, an extremely large effect that is statistically different from zero according to both Fisherian and large sample tests. Using both continuity-based and local randomization methods, the evidence is clear that SPP eligibility induces a large take-up of the program near the cutoff. The local randomization first-stage point estimate is very similar to the value of $0.625$ that we estimated above using continuity-based methods. This similarity shows that the first-stage effect in this application is remarkably robust, as the conclusions from the empirical analysis are similar whether we use continuity-based methods with approximately $8,000$ observations in a bandwidth of $\pm 18.5$ or local randomization methods with just $130$ observations in a $\pm 0.13$ window.
 
We continue by considering the ITT effect of being eligible to receive SPP ($T_i$) on our outcome of interest, HEI enrollment. We estimate it inside the chosen window using \texttt{rdrandinf}, this time using \texttt{Y} as the outcome of interest. 

\labelsnippet{fuzzyI}
\rsnip{Vol-2-R_LRS_rdrandinf_reducedform.txt}{\Rlink{\thesection}{\thefuzzyI}}
\statasnip{Vol-2-STATA_LRS_rdrandinf_reducedform}{\Slink{\thesection}{\thefuzzyI}}
 
The results estimate the effect of being barely eligible to receive SPP on HEI enollment within $[-0.13,0.13]$, the ITT parameter $\theta_{\mathtt{ITT}}$. This effect is estimated to be $0.171$, with a Fisherian p-value of $0.064$ and a large sample p-value of $0.056$. This estimated effect is lower than the $0.269$ effect estimated with continuity-based methods, and the p-values are larger. The increase in p-values is expected, as the number of effective observations decreases from over $8,000$ to just $130$. The decrease in the point estimate is considerable. Nevertheless, the overall conclusion is the same: becoming just eligible to receive SPP funding increases enrollment in HEI. 
 
In order to study the ratio parameter $\theta_{\mathtt{FRD}}$, we call \texttt{rdrandinf} using the option \texttt{fuzzy = c(D, "tsls")}, where the first argument (\texttt{D}) is the indicator for treatment received and the second argument requests the two-stage least-squares (TSLS) statistic---the estimate of the ratio $\theta_{\mathtt{Y}}/\theta_{\mathtt{D}}$.

\labelsnippet{fuzzyJ}
\rsnip{Vol-2-R_LRS_rdrandinf_TSLS.txt}{\Rlink{\thesection}{\thefuzzyJ}}
\statasnip{Vol-2-STATA_LRS_rdrandinf_TSLS}{\Slink{\thesection}{\thefuzzyJ}}

When the \texttt{tsls} option is chosen, available inference results are only based on large sample approximations, and the finite sample p-value is not returned. The column labeled \texttt{T} reports the test statistic, which is $0.299$. This number is the ratio between the two ITT effects reported above: the effect of SPP eligibility on HEI enrollment ($\theta_{\mathtt{Y}}$), $0.171$, over the effect of SPP eligibility on SPP funding ($\theta_{\mathtt{D}}$), $0.571$. The effect is estimated to be 0.299 with p-value $0.038$. This point estimate is smaller than the continuity-based estimate of 0.434, a difference mostly due to the smaller $\theta_{\mathtt{Y}}$ estimate ($0.171$ versus $0.269$).

We now briefly illustrate falsification in local randomization and continuity frameworks. As discussed above, in a Fuzzy RD design, the main falsification analyses should be implemented based on the treatment assignment and not on the treatment received; in other words, falsification should follow the same procedures as in the Sharp RD design. 

For the density tests, we use the \texttt{rddensity} command, which tests the (continuity-based) hypothesis that the densities of assigned-to-treatment ($T=1$) and assigned-to-control ($T=0$) observations are the same at the cutoff. This command now also reports results from a binomial test in several windows around the cutoff, which allows researchers to easily implement the density test in a local randomization framework.

\labelsnippet{fuzzyK}
\rsnip{Vol-2-R_LRS_rddensity.txt}{\Rlink{\thesection}{\thefuzzyK}}
\statasnip{Vol-2-STATA_LRS_rddensity}{\Slink{\thesection}{\thefuzzyK}}

The results (not shown) indicate that, in both frameworks, the null hypothesis fails to be rejected (with p-values of 0.4243 and 0.5825) and there is therefore no evidence of `sorting' around the cutoff based on this measure.

We then formally estimate ITT effects on pre-determined covariates in the continuity-based and local randomization frameworks using, respectively, the commands \texttt{rdrobust} and \texttt{rdrandinf}. For example, for the covariate \texttt{icfes\_female}, we estimate

\labelsnippet{fuzzyL}
\rsnip{Vol-2-R_LRS_rdrobust_reducedform_examplecovariate.txt}{\Rlink{\thesection}{\thefuzzyL}}
\statasnip{Vol-2-STATA_LRS_rdrobust_reducedform_examplecovariate}{\Slink{\thesection}{\thefuzzyL}}

with continuity-based methods, and 

\labelsnippet{fuzzyM}
\rsnip{Vol-2-R_LRS_rdrandinf_ITT_examplecovariate.txt}{\Rlink{\thesection}{\thefuzzyM}}
\statasnip{Vol-2-STATA_LRS_rdrandinf_ITT_examplecovariate}{\Slink{\thesection}{\thefuzzyM}}

with local randomization methods, where we omit the outputs to conserve space.

The ITT falsification effects for all the pre-determined covariates are reported in Table \ref{tab:LRS_sharprdrobust_covariates} using continuity-based methods and Table \ref{tab:LRS_rdrandinf_covariates} using local randomization methods. The goal of these falsification analyses is inference, not point estimation, because we know that the effect of the treatment assignment on any pre-determined covariate is zero. For this reason, we implement the local polynomial analysis with a bandwidth that optimizes the coverage-error (CER) of the confidence intervals.

\begin{table}[ht]
	\centering
	\resizebox{\textwidth}{!}{\begin{tabular}{lcccc}
			\toprule
			\multicolumn{1}{c}{\multirow{2}{*}{Variable}} & CER-Optimal &  \multicolumn{2}{c}{Robust Inference} & Number of \\
			\cmidrule{3-4}
			& Bandwidth   & p-value & 95\% CI                             & Observations\\
			\midrule
			=1 if female in exam day (icfes\_female) & $4.91$ & $0.52$ & $[-0.09, 0.05]$ & $4,276$ \\
Age in exam day (icfes\_age) & $5.50$ & $0.36$ & $[-0.16, 0.46]$ & $4,746$ \\
=1 if self-identifies as ethnic minority in exam day (icfes\_urm) & $6.68$ & $0.49$ & $[-0.04, 0.02]$ & $5,775$ \\
Household residential stratum (icfes\_stratum) & $6.56$ & $0.47$ & $[-0.05, 0.12]$ & $5,678$ \\
Family size in exam day (icfes\_famsize) & $5.51$ & $0.16$ & $[-0.29, 0.05]$ & $4,783$ \\ \bottomrule
 \\
	\end{tabular}}
	\caption{Sharp Continuity-Based Analysis for Covariates---SPP data}\label{tab:LRS_sharprdrobust_covariates}
\end{table}

\begin{table}[ht]
	\centering
	\resizebox{\textwidth}{!}{\begin{tabular}{lccccc}
			\toprule
			\multicolumn{1}{c}{\multirow{2}{*}{Variable}} & Mean of  & Mean of & Diff-in-Means & Fisherian  & Number of \\
			& Controls & Treated & Statistic     & p-value    & Observations\\
			\midrule
			=1 if female in exam day (icfes\_female) & $0.52$ & $0.39$ & $-0.13$ & $0.21$ & $119$ \\
Age in exam day (icfes\_age) & $16.41$ & $16.89$ & $0.48$ & $0.56$ & $119$ \\
=1 if self-identifies as ethnic minority in exam day (icfes\_urm) & $0.02$ & $0.02$ & $0.00$ & $1.00$ & $119$ \\
Household residential stratum (icfes\_stratum) & $2.19$ & $2.27$ & $0.08$ & $0.67$ & $119$ \\
Family size in exam day (icfes\_famsize) & $4.06$ & $4.02$ & $-0.05$ & $0.90$ & $119$ \\ \bottomrule
 \\
	\end{tabular}}
	\caption{Local Randomization Analysis (ITT) for Covariates---SPP data}\label{tab:LRS_rdrandinf_covariates}
\end{table}

Both frameworks lead to the same conclusion: there is no evidence that the treatment assignment is correlated with the covariates at or near the cutoff. In other words, students who are barely eligible to receive SPP funding are similar to students who are barely ineligible in terms of age, sex, minority status, household stratum, and family size. This kind of evidence suggests that the continuity and local randomization assumptions are plausible in this application.

\subsection{Further Reading}

Identification in Fuzzy RD designs was first discussed in \citet*{Hahn-Todd-vanderKlaauw_2001_ECMA}, and later in a sequence of papers including \citet{Dong_2018_OxfordBull}, \citet*{Cattaneo-Keele-Titiunik-VazquezBare_2016_JOP} and \citet*{Arai-Hsu-Kitagawa-Mourifie-Wan_2022_QE}. Estimation and inference methods are discussed in \citet*{Calonico-Cattaneo-Titiunik_2014_ECMA}, \citet*{Calonico-Cattaneo-Farrell-Titiunik_2019_RESTAT} and \citet*{Calonico-Cattaneo-Farrell_2020_ECTJ} within the continuity-based framework, and in \citet*{Cattaneo-Frandsen-Titiunik_2015_JCI} and \citet*{Cattaneo-Titiunik-VazquezBare_2017_JPAM} within the local randomization framework. Weak instrument issues in the RD context are discussed in \citet*{Feir-Lemieux-Marmer_2016_JBES}; see also \citet{Cattaneo-Titiunik_2022_ARE} for more references. For a review of weak instrument methods in IV regression, see \cite{Andrews-Stock-Sun_2019_ARE}.

\clearpage
\section{RD Designs with Discrete Running Variables}
\label{sec:discrete}
\setcounter{figuras}{1}
\setcounter{snippet}{1}
\setcounter{tablas}{1}

The canonical continuity-based RD design assumes that the score that determines treatment assignment is a continuous random variable. A random variable is continuous when it can take an uncountable number of values. For example, a share such as a party's proportion of the vote is continuous because it can take any value in the $[0,1]$ interval. In practical terms, when the score is continuous, all the observations in the dataset have distinct score values---i.e., there are no ties. In contrast, a discrete score such as date of birth can only take a finite number of values; as a result, a random sample of a discrete running variable will exhibit ``mass points''---that is, many observations share the same value of the score. 

When the RD score is not a continuous random variable, the continuity-based local polynomial methods we discussed in \textit{Foundations} are not directly applicable. This is practically important because many RD applications have a discrete score. The key issue when deciding how to analyze an RD design with a discrete score is the number of distinct mass points. Local polynomial methods will behave essentially as if each mass point is a single observation; therefore, if the score is discrete but the number of mass points is sufficiently large (and ideally close to the cutoff), then using local polynomial methods may still be appropriate under reasonable assumptions. In contrast, if the number of mass points is small (and sparse, away from the cutoff), then local polynomial methods will not be directly applicable in the absence of more restrictive assumptions. In this case, analyzing the RD design using the local randomization approach is a natural alternative. When the score is discrete, the local randomization approach has the advantage that the window selection procedure is often no longer needed, as the smallest window is well-defined and typically has enough observations. In both cases, the discreteness of the running variable leads to unavoidable extrapolation to the cutoff, a point that must be taken into account when interpreting the continuity-based and the local randomization RD treatment effects.

We devote the rest of this section to discuss an empirical RD example with a discrete running variable in order to illustrate how identification, estimation, and inference can be modified when the dataset contains multiple observations with the same value of the RD score. We illustrate how to count the number of mass points, how to implement a continuity-based analysis using both the raw and collapsed data, and how to implement a local randomization analysis in the smallest (or a small) window around the cutoff.

\subsection{The Effect of Academic Probation on Future Academic Achievement}

We analyze the study by \citet*{Lindo-Sanders-Oreopoulos_2010_AEJ}, who used an RD design to investigate the impact of placing students on academic probation on their future academic performance. Our choice of an education example is intentional. The origins of the RD design can be traced to the education literature, and RD methods continue to be used extensively in education because interventions such as scholarships or remedial programs are often assigned on the basis of a test score and a fixed approving threshold. Moreover, despite being continuous in principle, it is common for test scores and grades to be discrete in practice. 

This application analyzes a policy at a Canadian university that places students on academic probation when their grade point average (GPA) falls below a threshold. The treatment involves setting a standard for the student's future academic performance: a student who is placed on probation in a given term must improve her GPA in the next term according to campus-specific standards, or face suspension. Thus, in this RD design, the unit of analysis is the student, the score is the student's GPA, the treatment of interest is placing the student on probation, and the cutoff is the GPA value that triggers probation placement. Students come from three different campuses. In campuses 1 and 2, the cutoff is 1.5; in campus 3 the cutoff is 1.6. In their original analysis, \citet*{Lindo-Sanders-Oreopoulos_2010_AEJ} normalized the score, centering each student's GPA at the appropriate cutoff, and pooling the observations from the three campuses in a single dataset. (This approach is standard in Multi-Cutoff RD design settings, as we discuss in Section \ref{sec:multiRD}.) The resulting running variable is, therefore, the difference between the student's GPA and the cutoff; this variable ranges from $-1.6$ to $2.8$, with negative values indicating that the student was placed on probation, and positive values indicating that the student was not placed on probation, with a cutoff of zero. 

 There are $40,582$ student-level observations coming from the $1996$--$2005$ period. The outcome we analyze is the GPA obtained by the student in the term immediately after he was placed on probation (\texttt{Next Term GPA}). Naturally, this variable is only observed for students who decide to continue at the university; thus, the effects of probation on this outcome must be interpreted with caution, as the decision to leave the university may itself be affected by the treatment. (The original study contains an analysis of the effects of the treatment on dropout rates; we omit this outcome to simplify our illustration.) We also investigate some predetermined covariates: the percentile of the student's average GPA in standard high school classes (\texttt{hsgrade\_pct}), the total number of credits for which the student enrolled in the first year (\texttt{totcredits\_year1}), the student's age at entry (\texttt{age}), an indicator for whether the student is male (\texttt{male}), and an indicator for whether the student was born in North America (\texttt{bpl\_north\_america}).

\subsection{Counting the Number of Mass Points in the RD Score}

The crucial issue in the practical analysis of RD designs with discrete scores is the number of mass points (i.e., unique values) that actually occur in the dataset. When this number is large, it may be possible to apply continuity-based methods for RD analysis, after changing the interpretation of the treatment effect of interest and imposing additional assumptions enabling extrapolation. In contrast, when the number of unique score values is either moderately small or very small, a local randomization approach may be more appropriate. With few mass points, local or global polynomial fitting will be useful only as an exploratory strategy but not as a formal method of analysis, unless the researcher is willing to impose strong parametric assumptions. Therefore, the first step in the analysis of an RD design with a discrete running variable is to analyze the empirical distribution of the score and determine (i) the total number of observations, (ii) the total number of mass points, and (iii) the total number of observations per mass point. We illustrate this step with the academic probation application. 

\begin{figure}[ht]
    \centering
	\begin{subfigure}{0.48\textwidth}
		\centering
		\includegraphics[scale=0.45]{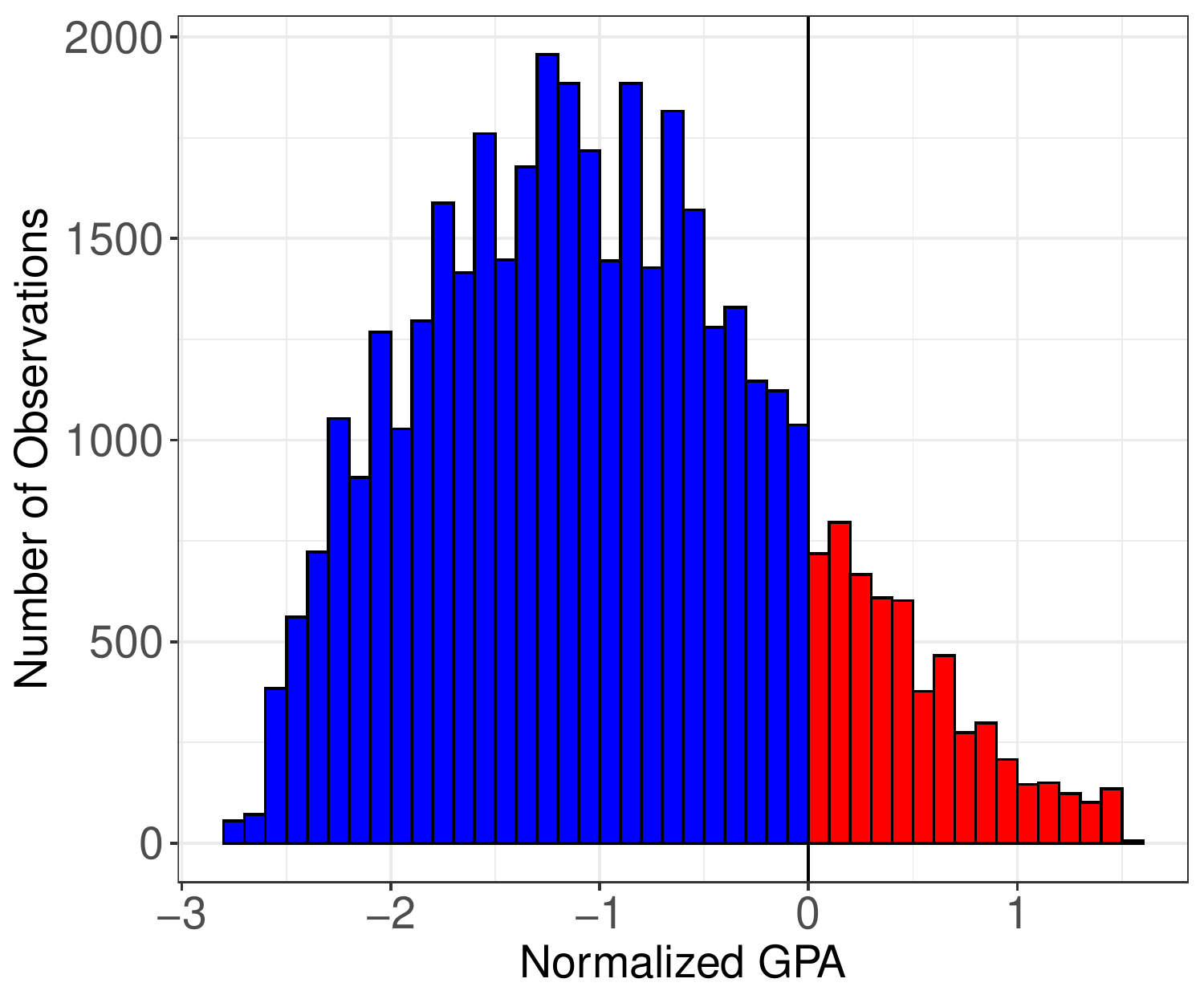}
		\caption{Histogram}\label{fig:LSOhist}		
	\end{subfigure}
	\begin{subfigure}{0.48\textwidth}
		\centering
		\includegraphics[scale=0.45]{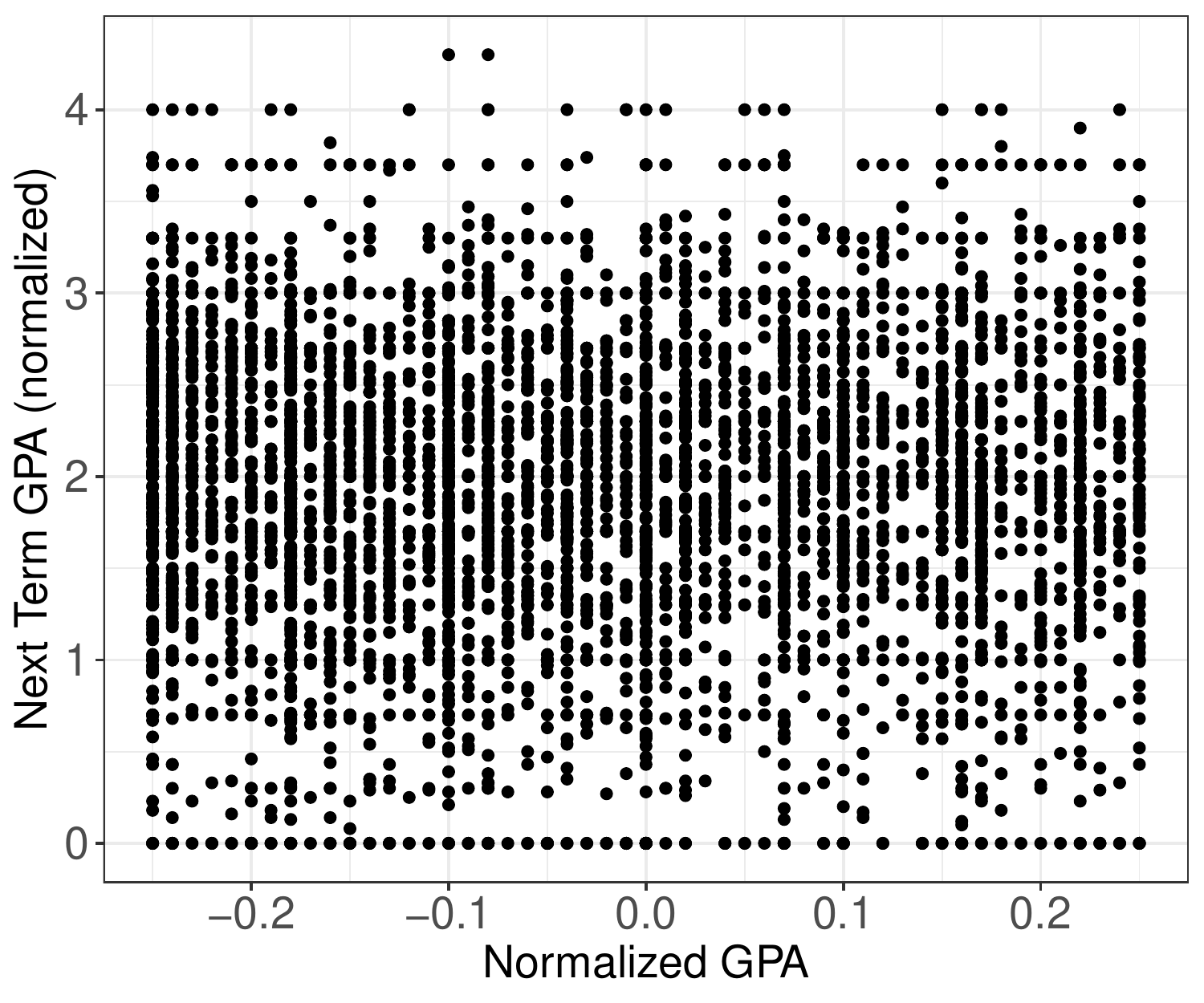}
		\caption{Scatter plot}\label{fig:LSOscatter}
	\end{subfigure}
	\caption{Distribution of Score and Outcome---Academic Probation Data}\label{fig:LSOdiscrv}
\end{figure}

Since only students who have a GPA below a threshold are placed on probation, the treatment is administered to students whose GPA is to the left of the cutoff. It is customary to define the RD treatment indicator as equal to one for units whose score is greater than the cutoff. To conform to this convention, we multiply the original running variable (the distance between GPA and the campus cutoff) by $-1$, so that students placed on probation are now above the cutoff. For example, a student who has $X_i=-0.2$ in the original score is placed on probation because her GPA is $0.2$ units below the threshold. The value of the transformed running variable for this treated student is $\tilde{X}_i=0.2$. Moreover, since we define the treatment as $\I(\tilde{X}_i \geq 0)$, this student will now be placed above the cutoff. The only caveat is that we must shift slightly those students whose original GPA is exactly equal to the cutoff (and thus are not placed on probation), since for these students the original normalized running variable is exactly zero and thus multiplying by $-1$ does not alter their score. In the scale of the transformed variable, we need these students to be below zero to continue to assign them to the control (i.e., non-probation) condition. We manually change the score of students who are exactly at zero to $X_i=-0.000005$ so that the rule $\I(\tilde{X}_i \geq 0)$ correctly identifies treated and control students. A histogram of the transformed running variable is shown in Figure \ref{fig:LSOhist}.

We first check how many total observations we have in the dataset, that is, how many observations have a non-missing value of the score.

\labelsnippet{countX}
\rsnip{Vol-2-R_lso_countX.txt}{\Rlink{\thesection}{\thecountX}}
\statasnip{Vol-2-STATA_lso_countX}{\Slink{\thesection}{\thecountX}}

The total sample size in this application is large: $40,582$ observations. However, because the running variable is discrete, the crucial step is to calculate how many mass points we have.

\labelsnippet{uniqueX}
\rsnip{Vol-2-R_lso_uniqueX.txt}{\Rlink{\thesection}{\theuniqueX}}
\statasnip{Vol-2-STATA_lso_uniqueX}{\Slink{\thesection}{\theuniqueX}}

The $40,582$ total observations in the dataset take only $429$ distinct values. This means that, on average, there are roughly $95$ observations per value. To have a better idea of the density of observations near the cutoff, Table \ref{tab:educ_masspoints} shows the number of observations for the six mass points closest to the cutoff; this table also illustrates how the score is transformed. Since the original score ranges between $-1.6$ and $2.8$, our transformed score ranges from $-2.8$ to $1.6$. Both the original and the transformed running variables are discrete, because the GPA increases in increments of 0.01 units and there are many students with the same GPA value. For example, there are $72$ students who are $0.02$ GPA units above the cutoff. Of these $72$ students, $41+5= 46$ have a GPA of $1.52$ (because the cutoff in Campuses 1 and 2 is $1.5$), and $26$ students have a GPA of $1.62$ (because the cutoff in Campus 3 is $1.6$). The same phenomenon of multiple observations with the same value of the score occurs at all other values of the score; for example, there are $208$ students who have a value of zero in the original score (and $-0.000005$ in our transformed score). 

\begin{table}[ht]
	\centering
	\resizebox{\textwidth}{!}{\begin{tabular}{ccccccc}
			\toprule
			Original & Transformed & Treatment  &  \multicolumn{4}{c}{Number of Observations}\\
			\cmidrule{4-7}    
			Score    & Score       & Status     & All Campuses & Campus 1 & Campus 2 & Campus 3 \\
			\midrule
			$\displaystyle \vdots$ & $\displaystyle \vdots$ & $\displaystyle \vdots$ & $\displaystyle \vdots$ & $\displaystyle \vdots$ & $\displaystyle \vdots$ & $\displaystyle \vdots$ \\
			$0.02$ & $-0.02$ & Control & $72$ & $41$ & $5$ & $26$ \\
$0.01$ & $-0.01$ & Control & $65$ & $23$ & $13$ & $29$ \\
$0.00$ & $-0.000005$ & Control & $208$ & $94$ & $50$ & $64$ \\
$-0.01$ & $0.01$ & Treated & $67$ & $28$ & $10$ & $29$ \\
$-0.02$ & $0.02$ & Treated & $122$ & $52$ & $29$ & $41$ \\
$-0.03$ & $0.03$ & Treated & $47$ & $16$ & $4$ & $27$ \\

			$\displaystyle \vdots$ & $\displaystyle \vdots$ & $\displaystyle \vdots$ & $\displaystyle \vdots$ & $\displaystyle \vdots$ & $\displaystyle \vdots$ & $\displaystyle \vdots$ \\
			\bottomrule
	\end{tabular}}
	\caption{Observations at Closest Mass Points---Academic Probation data}\label{tab:educ_masspoints}
\end{table}

\subsection{Using the Continuity-Based Approach when the Number of Mass Points is Large}

When the number of mass points in the discrete score is sufficiently large, we can use the continuity-based approach to RD analysis that we discussed extensively in \textit{Foundations}. The academic probation application illustrates a case in which a continuity-based analysis might be possible, since the total number of mass points is $429$, a moderate value. Because there are mass points, extrapolation between these points is unavoidable; however, in practical terms, this is no different from analyzing a dataset from any continuous score RD design with a sample of size $429$.

We start with a falsification analysis, doing a continuity-based density test and a continuity-based analysis of the effect of the treatment on predetermined covariates. First, we use \texttt{rddensity} to test whether the density of the score is continuous at the cutoff.

\labelsnippet{rddensityA}
\rsnip{Vol-2-R_lso_falsification_rddensity.txt}{\Rlink{\thesection}{\therddensityA}}
\statasnip{Vol-2-STATA_lso_falsification_rddensity}{\Slink{\thesection}{\therddensityA}}

The p-value is $0.082$, and we fail to reject the hypothesis that the density of the score changes discontinuously at the cutoff point at the conventional 5\% level. However, the p-value is below 10\%, suggesting a possible density imbalance. This is consistent with the local randomization binomial density test that we discuss below, and with the jump from 208 to 67 observations in the mass points closest to the cutoff shown in Table \ref{tab:educ_masspoints}.  

Next, we employ local polynomial methods to perform falsification analyses on several predetermined covariates. We use \texttt{rdrobust} with the default polynomial of order one. Because the focus is on inference and not on point estimation, we select the bandwidth to be coverage-error (CER) optimal. (For further discussion, see Section 5 in \textit{Foundations}.)

Table \ref{tab:RDcovariates-LSO} presents a summary of the results for the five predetermined covariates presented above. The results indicate that the probation treatment has no effect on the covariates, with the exception of \texttt{totcredits\_year1}, which has an associated p-value of $0.001$, rejecting the hypothesis of no effect at standard levels. The point estimate of the effect on this covariate (not shown, estimated with MSE-optimal bandwidth) is small: treated students take an additional $0.08$ credits in the first year, but the average value of \texttt{totcredits\_year1} in the overall sample is $4.43$, with a standard deviation of roughly $0.5$. This covariate imbalance might require further investigation if there were reasons to believe that small differences in prior credits could affect future academic performance---but such investigation is beyond the scope of our illustration.

\begin{table}[ht]
	\centering
	\resizebox{\textwidth}{!}{\begin{tabular}{lccccc}
			\toprule
			\multicolumn{1}{c}{\multirow{2}{*}{Variable}} & CER-Optimal & RD        & \multicolumn{2}{c}{Robust Inference} & Number of \\
            \cmidrule(rl){4-5}
			& Bandwidth   & Estimator & p-value & 95\% CI                             & Observations\\
			\midrule
			High school grade percentile & $0.30$ & $1.43$ & $0.28$ & $[-1.25, 4.29]$ & $5,769$ \\
Credits attempted in first year & $0.17$ & $0.13$ & $0.00$ & $[0.05, 0.22]$ & $3,276$ \\
Age at entry & $0.25$ & $0.00$ & $0.94$ & $[-0.10, 0.10]$ & $4,590$ \\
Male & $0.31$ & $-0.02$ & $0.49$ & $[-0.08, 0.04]$ & $5,885$ \\
Born in North America & $0.26$ & $0.01$ & $0.69$ & $[-0.03, 0.05]$ & $4,875$ \\ \bottomrule

	\end{tabular}}
	\caption{RD Effects on Predetermined Covariates---Academic Probation data}\label{tab:RDcovariates-LSO}
\end{table}

\begin{figure}[H]
	\centering
	\includegraphics[scale=0.5]{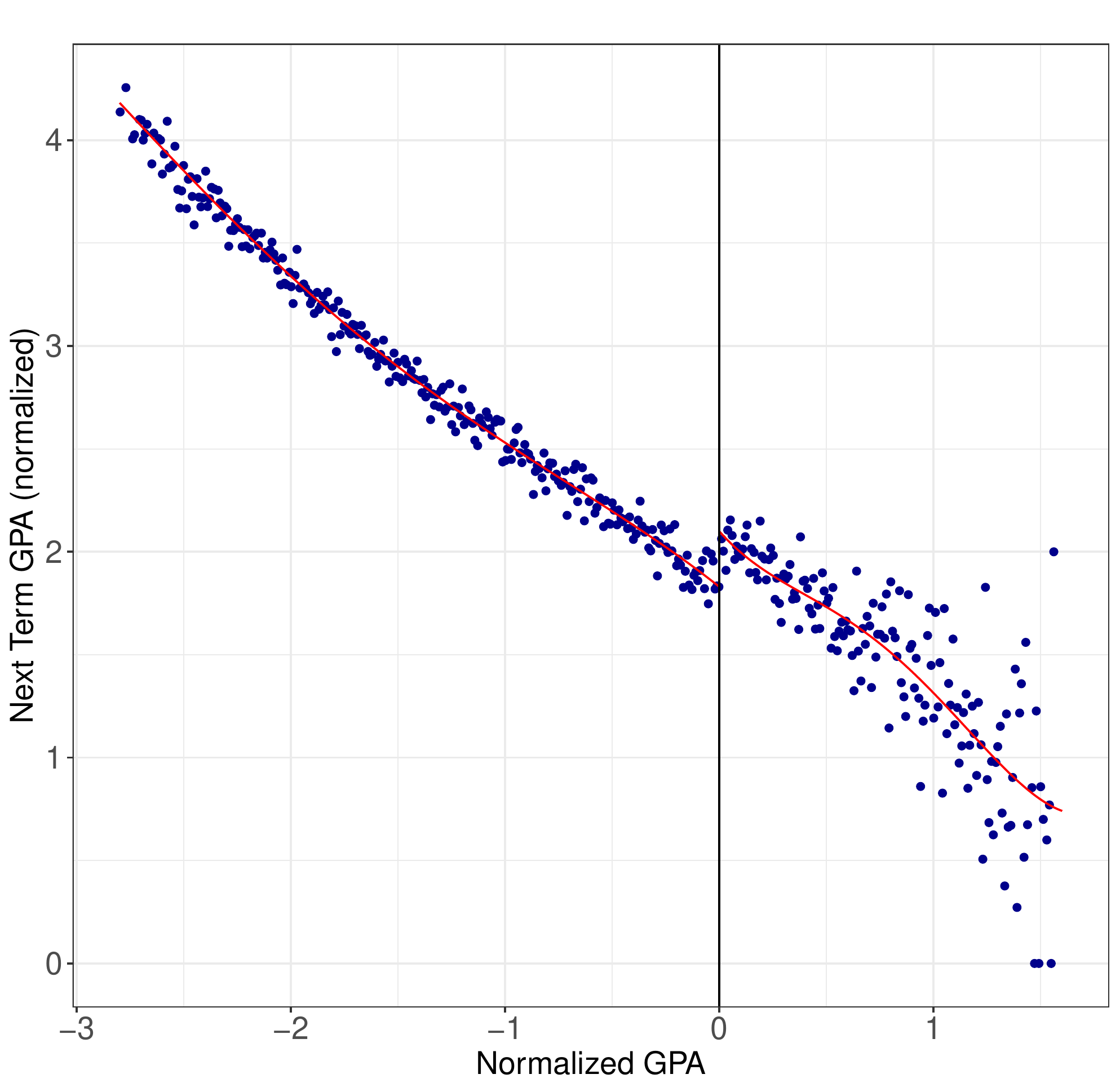}
	\caption{RD Plot: \texttt{nextGPA}---Academic Probation data}\label{fig:RDplotLSO-NextGPA}
\end{figure}

Next, we analyze the effect of being placed on probation on the outcome of interest, \texttt{nextGPA}, the GPA in the following academic term. We first visualize the effect with an RD plot, shown in Figure \ref{fig:RDplotLSO-NextGPA}.

The RD plot suggests a negative relationship between the running variable and the outcome: students who have a low GPA in the current term (and thus have a higher value of the running variable) tend to also have a low GPA in the following term. The plot also shows that students with scores just above the cutoff (who are just placed on probation) tend to have a higher GPA in the following term relative to students who are just below the cutoff and just avoided probation. These results are confirmed when we use a local linear polynomial and robust bias-corrected inference to provide a formal statistical analysis of the RD effect.

\labelsnippet{rdrobustC}
\rsnip{Vol-2-R_lso3_rdrobust_triangular_mserd_p1_rhofree_regterm1.txt}{\Rlink{\thesection}{\therdrobustC}}
\statasnip{Vol-2-STATA_lso3_rdrobust_triangular_mserd_p1_rhofree_regterm1}{\Slink{\thesection}{\therdrobustC}}

As shown, students who are just placed on probation improve their GPA in the following term by approximately $0.224$ additional points, relative to students who just miss probation. The robust p-value is less than $0.00005$, and the robust $95\%$ confidence interval ranges from $0.126$ to $0.304$. Thus, the evidence indicates that, conditional on not leaving the university, being placed on academic probation translates into an increase in future GPA. The point estimate of $0.224$---obtained with \texttt{rdrobust} within a MSE-optimal bandwidth of $0.470$---is very similar to the effect of $0.23$ grade points found in the original study (which employed an ad-hoc bandwidth of $0.6$).

To better understand this treatment effect, we may be interested in knowing the point estimate for the controls and treated students separately. To see this information, we explore the information returned by \texttt{rdrobust}.

\labelsnippet{rdrobustD}
\rsnip{Vol-2-R_lso3_rdrobust_triangular_mserd_p1_rhofree_regterm1_namescoefsout_all.txt}{\Rlink{\thesection}{\therdrobustD}}
\statasnip{Vol-2-STATA_lso3_rdrobust_triangular_mserd_p1_rhofree_regterm1_namescoefsout_all}{\Slink{\thesection}{\therdrobustD}}

This output shows the estimated intercept and slope from the two local regressions estimated separately to the right (\texttt{beta\_Y\_p\_r}) and left (\texttt{beta\_Y\_p\_l}) of the cutoff. At the cutoff, the average GPA in the following term for control students who just avoid probation is $1.844$, while the average future GPA for treated students who are just placed on probation is $2.068$. The increase is the estimated RD effect reported above, $2.068-1.844 = 0.224$. This represents approximately a $11\%$ GPA increase relative to the control group.

In some applications, it may be desirable to cluster the standard errors by every value of the score. We implement this using the \texttt{cluster} option in \texttt{rdrobust}. 

\labelsnippet{rdrobustE}
\rsnip{Vol-2-R_lso3_rdrobust_triangular_mserd_p1_rhofree_regterm1_cluster.txt}{\Rlink{\thesection}{\therdrobustE}}
\statasnip{Vol-2-STATA_lso3_rdrobust_triangular_mserd_p1_rhofree_regterm1_cluster}{\Slink{\thesection}{\therdrobustE}}

The conclusions remain essentially unaltered, as the $95\%$ robust confidence interval changes only slightly from $[0.126, 0.304]$ to $[0.140, 0.284]$. The point estimate moves slightly from $0.224$ to $0.221$ because the MSE-optimal bandwidth with clustering shrinks to 0.428 from 0.470, and the bias bandwidth also decreases (output not shown). 

\subsubsection{Interpreting Continuity-Based RD Analysis with Mass Points}

Provided that the number of mass points in the score is reasonably large, it is possible to analyze an RD design with a discrete score using the tools from the continuity-based approach. However, it is important to understand how to correctly interpret the results from such analysis. We discuss the academic probation application further, with the goal of clarifying these issues.

When there are mass points in the running variable, local polynomial methods for RD analysis behave essentially as if we had as many observations as mass points. In other words, when applied to an RD design with a discrete score, the effective number of observations used by continuity-based methods is the number of mass points or distinct values, not the total number of observations. In practical terms, this means that fitting a local polynomial to the raw data with mass points is roughly equivalent to fitting a local polynomial to a ``collapsed'' version of the data where we aggregate the original observations by the discrete score values, calculating the average outcome for all observations that share the same score value. The total number of observations in the collapsed dataset is equal to the number of mass points in the running variable.

More formally, suppose that the score variable takes on values $\{\mathsf{x}_{K_{-}}$, \dots, $\mathsf{x}_{-2}$, $\mathsf{x}_{-1}$, $\mathsf{x}_0=\C$, $\mathsf{x}_1$, $\mathsf{x}_2$, \dots, $\mathsf{x}_{K_{+}}\}$, where $K_-$ denotes the number of unique values below the cutoff, $K_+$ denotes the number of unique values above the cutoff, and the cutoff $\C$ is assumed to be one of the possible values of the score. The process of collapsing the data reduces the original observations to the pairs $(\mathsf{x}_{K_{-}},\bar{Y}_{K_{-}})$, \dots, $(\mathsf{x}_{-1} ,\bar{Y}_{-1})$, $(\C ,\bar{Y}_{\C})$, $(\mathsf{x}_{1} ,\bar{Y}_{1})$, \dots, $(\mathsf{x}_{K_{+}},\bar{Y}_{K_{+}})$, where $\bar{Y}_{k} = \frac{1}{\#\{i: X_i = \mathsf{x}_k\}} \sum_{i=1}^n Y_i \cdot \I( X_i = \mathsf{x}_k)$ and $\#\{A\}$ denotes the number of elements in the set $A$. The number of observations in this collapsed dataset is $K_{-}+K_{+}+1$. Once the collapsed dataset is constructed, the continuity-based analysis can proceed as described in \textit{Foundations}. The exception is the RD plot; instead of using data-driven criteria to choose the number of bins, the most natural way of constructing the RD plot is to graph the sample average of the outcome for each score value against the unique score values.

To illustrate this procedure with the academic probation application, we calculate the average outcome for each of the $429$ mass points in the score value. The resulting dataset has $429$ observations, where each observation consists of a score-outcome pair: every score value is paired with the average outcome across all students in the original dataset whose score is equal to that value. We then use \texttt{rdrobust} to estimate the RD effect with a local polynomial.

\labelsnippet{rdrobustF}
\rsnip{Vol-2-R_lso3_rdrobust_collapsed.txt}{\Rlink{\thesection}{\therdrobustF}}
\statasnip{Vol-2-STATA_lso3_rdrobust_collapsed}{\Slink{\thesection}{\therdrobustF}}

The estimated effect is $0.246$, with a robust p-value less than $0.00005$. This is similar to the $0.224$ point estimate obtained with the raw dataset. The two estimates are very similar, even though the former is calculated using $429$ observations, while the latter is calculated using $40,582$ observations. The difference stems from the fact that different bandwidths are estimated in each case. Moreover, the inference conclusions from both analyses are consistent, as the robust $95\%$ confidence interval using the raw data is $[0.126, 0.304]$, while the robust confidence interval using the collapsed data is $[0.166, 0.316]$, indicating that plausible values of the effect are in roughly the same positive range in both cases.

This analysis shows that the seemingly large number of observations in the raw dataset is effectively much smaller, and that the behavior of the continuity-based results is governed by the average behavior of the data at every mass point. Thus, a natural point of departure for researchers who wish to study an RD design with a discrete score and many mass points is to collapse the data and estimate the effects on the aggregate results. As a second step, these aggregate results can be compared to the results using the raw, uncollapsed data---in most cases, both sets of results should lead to the same conclusions.

While the mechanics of local polynomial fitting using a discrete running variable are clear, the actual relevance and interpretation of the treatment effect may change. As we discuss below, researchers may want to change the parameter of interest when the score is discrete; if they do not, then parametric extrapolation will be unavoidable to achieve point identification. Because the score is discrete, it is not possible to nonparametrically point identify the continuity-based RD treatment effect at the cutoff, $\tau_\mathtt{SRD}=\E[Y_i(1)|X_i=\C]-\E[Y_i(0)|X_i=\C]$, because the lack of denseness of $X_i$ near the cutoff makes it impossible to appeal to limit arguments and large sample approximations. Put differently, if the researcher insists on retaining the same parameter of interest as in the canonical RD design, then extrapolation via additional parametric assumptions from the closest mass points above and below the cutoff to the cutoff point will be needed, no matter how large the sample size is.

Since parametric extrapolation is unavoidable when the running variable is discrete and the parameter $\tau_\mathtt{SRD}$ is still of interest, a simple local linear extrapolation towards the cutoff may be a reasonable strategy. This extrapolation approach will always operate in the background when continuity-based methods are used to analyze an RD design with discrete score. However, if the number of mass points is very small, bandwidth selection methods will not be appropriate; in this case, the researcher may conduct linear parametric extrapolation globally, fitting the polynomial using all the observations (i.e., employing the few unique values of the score). This runs counter to the local nature of the RD parameter, but it is essentially the only possibility for implementation if the goal is to estimate the canonical continuity-based RD parameter and the number of mass points is small.

\subsection{Local Randomization RD Analysis with Discrete Score}

A natural alternative for the analysis of an RD design with a discrete running variable is to use the local randomization approach discussed in Section \ref{sec:localrand}, which effectively changes the parameter of interest from the RD treatment effect at the cutoff ($\tau_\mathtt{SRD}$) to the RD treatment effect in the neighborhood $\W$ around the cutoff where local randomization is assumed to hold ($\theta_\mathtt{SRD}$). A key advantage of this alternative framework is that, unlike the continuity-based approach, it can be used even when there are very few mass points in the running variable; indeed, it can be used with as few as two mass points (one on each side of the cutoff).

Consider a hypothetical example where the score takes five values, $X_i \in \{-3,-2,-1,0,1,2\}$, the RD cutoff is $\C=0$, and the treatment assignment is therefore $T_i=\I(X_i\geq 0)$. In this case, the continuity-based RD treatment effect at the cutoff is $\tau_\mathtt{SRD}=\E[Y_i(1)|X_i=0]-\E[Y_i(0)|X_i=0]$, which is not identifiable nonparametrically because the score can never get close enough to 0 for untreated observations (the closest score value that an untreated observation can have is $-1$). However, if the local randomization assumptions hold in the window $\W=[-1,0]$, we can define the local randomization parameter $\theta_\mathtt{SRD}= \E[Y_i(1) -Y_i(0) | X_i \in [-1,0]]$, which is nonparametrically identifiable under the conditions discussed in Section \ref{sec:localrand} (we are assuming random potential outcomes for simplicity). These conditions imply that changing the score from -1 to 0 does not change the average potential outcome under control, $\E[Y_i(0)|X_i=-1] = \E[Y_i(0)|X_i=0]$---that is, the average control outcome at the score value just below the cutoff is the same as the average control outcome we would have observed if we had increased the score by one but had not changed the units' status to treated.

More generally, if $ X_i \in \{\mathsf{x}_{K_{-}}, \dots, \mathsf{x}_{-2},\mathsf{x}_{-1},\C,\mathsf{x}_1,\mathsf{x}_2,\dots,\mathsf{x}_{K_{+}}\}$, a natural local randomization RD treatment effect parameter is $\theta_\mathtt{SRD}=\frac{1}{N_{\W}}\sum_{i: X_i \in \W} \EW[Y_i(1) - Y_i(0)]$ for $\W=[\mathsf{x}_{-1},\C]$, that is, the average difference between treated and control potential outcomes for observations with scores in the smallest possible window around the cutoff, which ranges from the smallest score value that leads to treatment assignment to the largest score value that leads to control assignment. In applications with a large number of observations per cutoff, there will be enough observations with $X_i=\C$ and $X_i=\mathsf{x}_{-1}$ so that $\bar{Y}_\C$ and $\bar{Y}_{-1}$ can be used as consistent estimators of $\frac{1}{N_{\W}}\sum_{i: X_i \in \W} \EW[Y_i(1)]$ and $\frac{1}{N_{\W}}\sum_{i: X_i \in \W} \EW[Y_i(0)]$, respectively. In this case, window selection is not necessary, because the smallest possible window is $\W=[\mathsf{x}_{-1},\C]$, and the number of observations permits estimation of the effect in this window. Because this is the window where extrapolation is smallest, if the effect can be estimated in this window, it is not necessary to consider other windows.

The plausibility of the local randomization assumptions in $\W=[\mathsf{x}_{-1},\C]$ may depend on the scale of measurement of the running variable. For example, if the running variable is the date of birth measured in days and individuals become eligible to vote when they turn $18$ years old, $\E[Y_i(0)|X_i=\mathsf{x}_{-1}]$ represents the average control outcome the day before turning $18$. Since age is measured in days for most social science purposes, we do not expect that the $23$ hours and $59$ minutes of additional age will significantly affect average potential outcomes such as political knowledge, and thus we expect $\E[Y_i(0)|X_i=\mathsf{x}_{-1}]$ and $\E[Y_i(0)|X_i=\C]$ to be largely similar. In contrast, in other applications, the extrapolation may be significant and have stronger conceptual consequences. For example, if the policy is receiving social security benefits at age $65$, the running variable is measured in years, and the outcome is overall health, the difference between $\E[Y_i(0)|X_i=\mathsf{x}_{-1}]$ and $\E[Y_i(0)|X_i=\C]$ may be considerable if one extra year of age at $64$ is enough to affect overall average health.

Our discussion shows that, when the score is discrete, the local randomization approach for RD analysis does not typically require choosing a window in most applications, because the researcher knows the exact location of the minimum window around the cutoff: $\W=[\mathsf{x}_{-1},\C]$. Crucially, if local randomization holds, then it must hold for the \textit{smallest} window in the absence of design failures such as manipulation of the running variable. As shown in Table \ref{tab:educ_masspoints}, in the academic probation application the original score has a mass point at zero where all observations are control (because they reach the minimum GPA required to avoid probation), and the mass point immediately below it occurs at $-0.01$, where all students are placed on probation because they fall short of the threshold to avoid probation. Thus, the smallest window around the cutoff in the scale of the original score is $\W=[0.00,-0.01]$. In the scale of the transformed score, the minimum window is $\W=[-0.000005,0.01]$.

Regardless of the scale used, the important point is that the minimum window around the cutoff in a local randomization analysis of an RD design with a discrete score is precisely the interval between the two consecutive mass points where the treatment status changes from zero to one. The particular values taken by the score are irrelevant, as the analysis will proceed to assume that the treated and control groups were assigned to treatment as-if randomly, and will typically make the exclusion restriction assumption that the particular value of the score has no direct impact on the outcome of interest. Moreover, the location of the cutoff is no longer meaningful, as any cutoff value between the minimum value of the score on the treated side and the maximum value of the score on the control side will produce identical treatment and control groups. Once the researcher finds the treated and control observations located at the two mass points around the cutoff, the local randomization analysis can proceed as explained in Section \ref{sec:localrand}. 

In applications where the smallest window contains too few observations, estimation of $\theta_\mathtt{SRD}$ for $\W=[\mathsf{x}_{-1},\C]$ will not be possible. In such cases, researchers can use the covariate-based window selection procedure discussed in Section \ref{sec:localrand} to select a larger window with enough observations where pre-treatment covariates are balanced, enlarging the window one mass point at a time in each direction.

Finally, in some applications it may be of interest to redefine the parameter to $\theta_\mathtt{SDS}=\E[Y_i(1)|X_i=\C]-\E[Y_i(0)|X_i=\mathsf{x}_{-1}]$ (where SDS refers to \textit{sharp discrete score} and we are again using random potential outcomes for simplicity). This parameter is always nonparametrically identifiable and it explicitly indicates the inability to identify $\E[Y_i(0)|X_i=\C]$ without additional assumptions. Without invoking the local randomization assumptions, however, the interpretation of this parameter can only be descriptive, not causal.

We proceed to illustrate the local randomization analysis using the academic probation example. We first conduct a falsification analysis to determine whether the assumption of local randomization in the window $[-0.00005, 0.1]$ seems consistent with the empirical evidence. We conduct a binomial hypothesis test to test whether the density of observations in this window is consistent with the density that would have been observed in a series of unbiased coin flips. 

\labelsnippet{binomialB}
\rsnip{Vol-2-R_lso_falsification_binomial_byhand.txt}{\Rlink{\thesection}{\thebinomialB}}
\statasnip{Vol-2-STATA_lso_falsification_binomial_byhand}{\Slink{\thesection}{\thebinomialB}}

As shown previously in Table \ref{tab:educ_masspoints}, there are $208$ control observations immediately below the cutoff and $67$ above the cutoff. In other words, there are $208$ students who get exactly the minimum GPA needed to avoid probation, and $67$ students who get the maximum possible GPA that still places them on probation. The number of control observations is roughly three times higher than the number of treated observations, a ratio that is inconsistent with the assumption that the probability of treatment assignment in this window was $1/2$---the p-value of the Binomial test (not shown) is indistinguishable from zero.

Although these results alone do not imply that the local randomization RD assumptions are violated, the imbalance in the number of observations is consistent with what one would expect if students were actively avoiding probation (presumably an undesirable outcome). The results raise some concern that students may have been aware of the probation cutoff, and may have tried to appeal their final GPA in order to avoid being placed on probation. This may justify seeking additional qualitative evidence; for example, it would be helpful to know whether there is a record of requests for GPA modifications, and whether those requests were more likely among students whose GPAs would have just placed them on probation.

It is also possible that, given the typical number of courses taken by students, the typical grades given in each course, and the number of credits associated with each course, GPA values are not all equally likely. In particular, if obtaining a GPA of 1.50 or 1.60 were more likely than obtaining a GPA of 1.49 or 1.59, we would expect to see fewer students just assigned to the probation treatment than just assigned to the control. Investigating this issue is beyond the scope of our practical guide, but we note that if the probabilities of obtaining different GPA values could be approximated or estimated, one could implement the local randomization framework using these probabilities instead of assuming that both GPA values in the window around the cutoff are equally likely. This might also resolve imbalances in pre-determined covariates, which would be expected to occur if the wrong probabilities are used.

Strictly speaking, an imbalanced number of observations would not pose any problems if the types of students in the treated and control groups were on average similar. To establish whether treated and control students at the cutoff are similar in terms of observable characteristics, we use \texttt{rdrandinf} to estimate the RD effect of probation on the predetermined covariates introduced above. A summary of the results is reported in Table \ref{tab:RDcovariates-LSO-LR}. Treated and control students seem indistinguishable in terms of prior high school grade, total number of credits, age, sex, and place of birth. The minimum p-value is 0.138, which is slightly smaller than our recommended value of 0.15, but still considerably larger than conventional levels. 

\begin{table}[ht]
	\centering
	\resizebox{\textwidth}{!}{\begin{tabular}{lccccc}
			\toprule
			\multicolumn{1}{c}{\multirow{2}{*}{Variable}} & Mean of  & Mean of & Diff-in-Means & Fisherian  & Number of \\
			& Controls & Treated & Statistic     & p-value    & Observations\\
			\midrule
			High school grade percentile & $29.13$ & $33.13$ & $4.01$ & $0.17$ & $275$ \\
Credits attempted in first year & $4.23$ & $4.33$ & $0.10$ & $0.14$ & $275$ \\
Age at entry & $18.79$ & $18.70$ & $-0.09$ & $0.36$ & $275$ \\
Male & $0.38$ & $0.42$ & $0.03$ & $0.67$ & $275$ \\
Born in North America & $0.88$ & $0.84$ & $-0.04$ & $0.40$ & $275$ \\ \bottomrule
 
	\end{tabular}}
	\caption{RD Effects on Predetermined Covariates---Academic Probation data}\label{tab:RDcovariates-LSO-LR}
\end{table}

In order to compare the smallest window around the cutoff including only two mass points to slightly larger windows, we employ the window selector discussed in Section \ref{sec:localrand}. This selector considers a sequence of nested windows, starting with the smallest, and in each window conducts balance tests for each covariate specified. We use the command \texttt{rdwinselect} with the default randomization inference method for the difference in means test statistic:
\labelsnippet{rdwinselectF}
\rsnip{Vol-2-R_lso_rdwinselect_consecutive_windows.txt}{\Rlink{\thesection}{\therdwinselectF}}
\statasnip{Vol-2-STATA_lso_rdwinselect_consecutive_windows}{\Slink{\thesection}{\therdwinselectF}}

The empirical results show that the minimum p-value in the smallest window is 0.138, as we had seen in Table \ref{tab:RDcovariates-LSO-LR}. The results also show that as soon as we consider the next largest window, the minimum p-value drops to less than 0.00005, suggesting the treated and control students are not comparable in larger windows around the cutoff. Given this, we only report the outcome analysis in the smallest window.

We investigate the local randomization RD treatment effect on the main outcome of interest using \texttt{rdrandinf}.

\labelsnippet{rdrandinfJ}
\rsnip{Vol-2-R_lso3_rdrandinf_adhocsmall_p0.txt}{\Rlink{\thesection}{\therdrandinfJ}}
\statasnip{Vol-2-STATA_lso3_rdrandinf_adhocsmall_p0}{\Slink{\thesection}{\therdrandinfJ}}

The difference-in-means between the $208$ control students and the $67$ treated students in the smallest window around the cutoff is $0.234$ grade points, remarkably similar to the continuity-based local polynomial RD effects of $0.224$ and $0.246$ that we found using the raw and aggregated data, respectively. Moreover, we can reject the null hypothesis of no effect at $6\%$ level using both the Fisherian and the large-sample inference approaches. This shows that the results for next term GPA are generally robust: we found similar results using the $208+67=275$ observations closest to the cutoff in a local randomization analysis, the total $40,582$ observations using a continuity-based analysis, and the $429$ collapsed observations in a continuity-based analysis.

\subsection{Further Reading}

\cite{Lee-Card_2008_JoE} discuss alternative local polynomial methods in the continuity-based RD framework when the running variable is discrete. \cite{Dong_2015_JAE} and \citet{Barreca-Lindo-Waddell_2016_EI} discuss issues of rounding and heaping in the running variable. \citet[Section 6.2]{Cattaneo-Frandsen-Titiunik_2015_JCI} discuss explicitly the connections between discrete scores and the local randomization approach; see also \citet{Cattaneo-Titiunik-VazquezBare_2017_JPAM}. \citet{Cattaneo-Titiunik_2022_ARE} review other methods and extensions.

\clearpage
\section{Multi-Dimensional RD Designs}
\label{sec:multiRD}
\setcounter{figuras}{1}
\setcounter{snippet}{1}
\setcounter{tablas}{1}

The standard RD design assumes that the treatment is assigned on the basis of a score $X_i$ and a cutoff $\C$ according to the rule $T_i=\I(X_i \geq \C)$, where both the score and the cutoff are scalars (i.e., one-dimensional variables). In contrast, multi-dimensional RD designs occur when the treatment is assigned on the basis of more than one score or more than one cutoff---or both. 

In the Multi-Cutoff RD design, the treatment is assigned on the basis of a scalar score, but different groups of units face different cutoff values. A common instance occurs when a federal program is administered by sub-national agencies, and each agency chooses a different cutoff value to determine program eligibility. For example, in order to target the most disadvantaged households in a given area, the Mexican conditional cash transfer program Progresa determined program eligibility based on a household-level poverty index. In rural areas, the cutoff that determined program eligibility varied geographically, with seven different cutoffs used in seven different regions.

In the Multi-Score RD design, the treatment is assigned on the basis of two or more scores, where typically a different cutoff is used for each score, and the treatment is assigned to a unit only if all scores simultaneously exceed their respective cutoffs. For example, in education settings, it is common to award scholarships to students who score above a cutoff in both a mathematics exam and a language exam. This leads to two running variables---the student's grade in the mathematics exam and her grade in the language exam---and two (possibly different) cutoffs. Another common example of a Multi-Score RD design is the Geographic RD design, where treatment eligibility is determined based on the location of the units relative to a geographic boundary. This type of RD settings are also known as \textit{Boundary Discontinuity Designs}.

We discuss both types of multi-dimensional RD designs. We start with the Multi-Cutoff RD design in the next section and continue with the Multi-Score RD design. We discuss interpretation and analysis and use different empirical examples to illustrate how to implement estimation, inference, and falsification using both the continuity-based and the local randomization frameworks.

\subsection{Multi-Cutoff RD Design}

To formalize the Multi-Cutoff RD design we assume that the cutoff is a random variable $C_i$ taking on $J$ distinct values $\mathcal{C} = \{c_1, c_2, \ldots, c_J\}$, instead of a single known constant as in the standard RD design. (Although the setup could be generalized to a continuously distributed $C_i$, we do not consider this possibility because most practical applications focus on finitely many cutoffs.) The treatment assignment is extended to $\T_i = \I(X_i \geq C_i)$ for $C_i\in\mathcal{C}$, and we also define the treatment assignment rule for each cutoff value $\T_i(c) = \I(X_i \geq c)$ so that $T_i=T_i(C_i)$. The single-cutoff RD design is a particular case of this setup when $\mathcal{C}=\{\C\}$ and thus $\P[C_i=\C]=1$. In the Multi-Cutoff RD design, we have $\P[C_i=c]\in(0,1)$ for each $c\in\mathcal{C}$.

We continue to employ standard potential outcomes notation. Because the particular cutoff to which a unit is exposed may affect the potential outcomes, treatment effects can display heterogeneity. In this sense, the cutoff $c\in\mathcal{C}$ may be understood as indexing a subpopulation type ($C_i=c$), with different types having possibly different potential outcome distributions.

\subsubsection{Cumulative versus Non-cumulative Cutoffs}

\begin{figure}[ht]
    \centering
	\begin{subfigure}{0.48\textwidth}
		\centering
		\includegraphics[scale=\kDF]{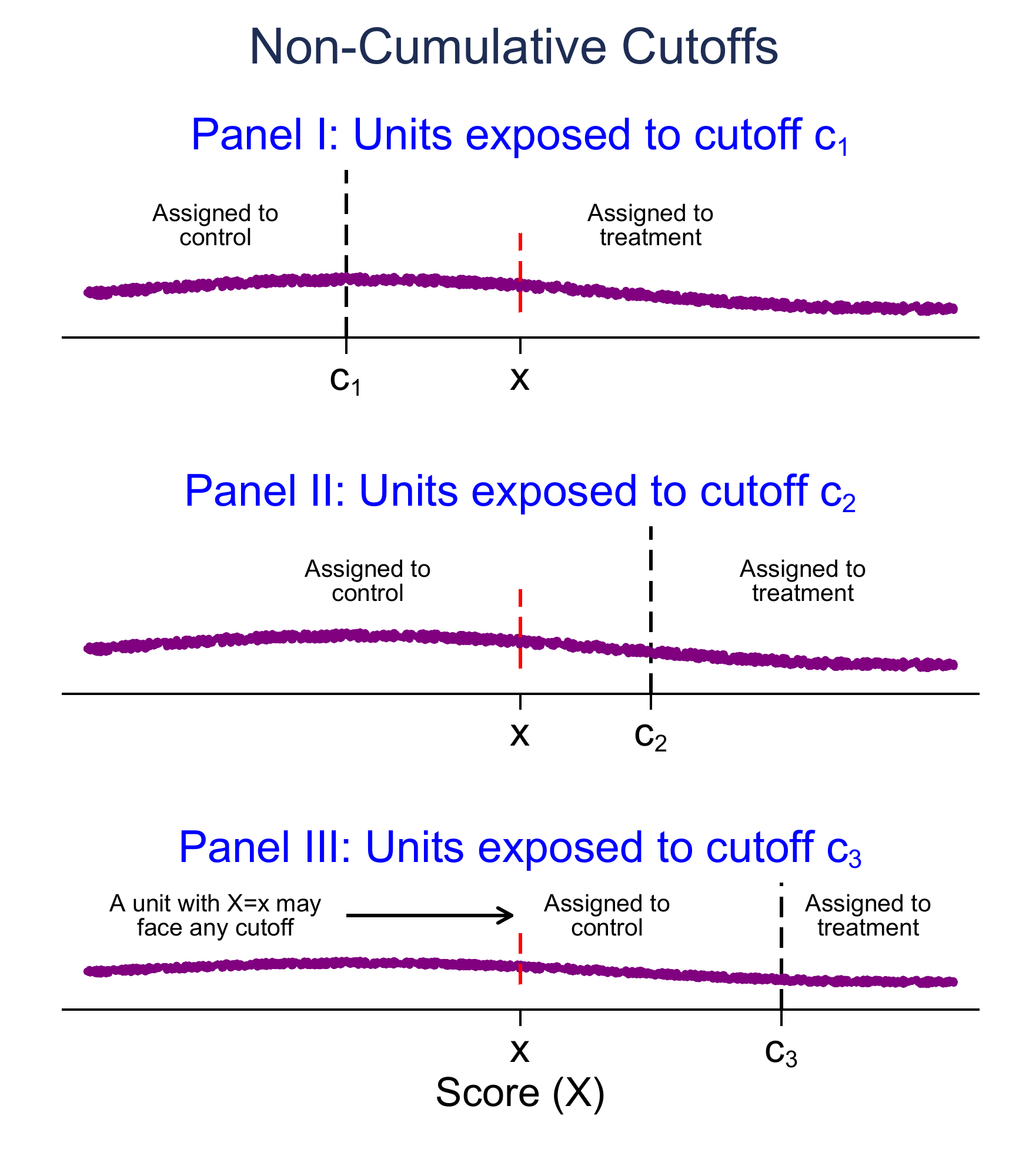}
		\caption{Non-cumulative Cutoffs}\label{fig:NoncumCutoffs}
	\end{subfigure}
	\begin{subfigure}{0.48\textwidth}
		\centering
		\includegraphics[scale=\kDF]{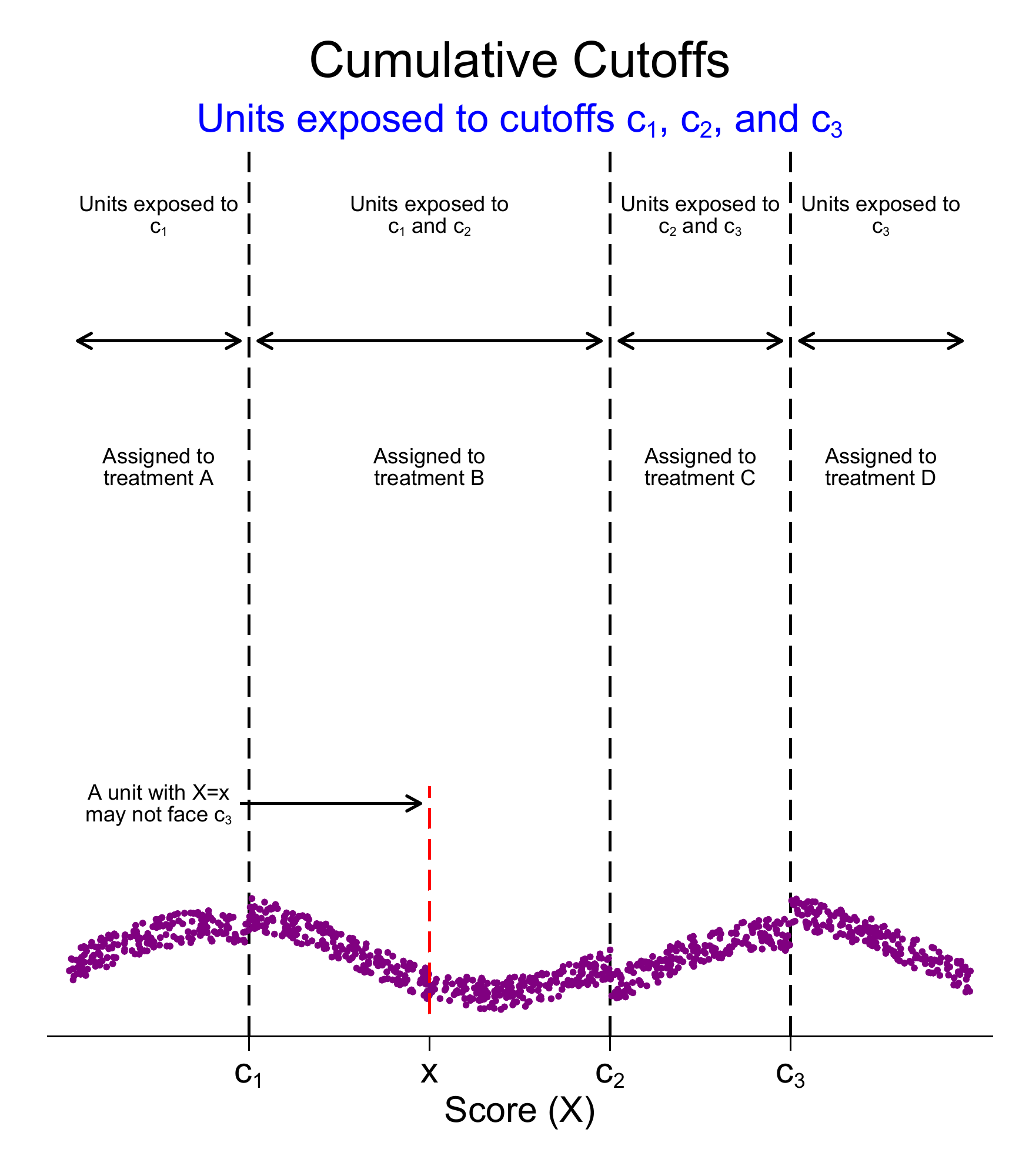}
		\caption{Cumulative Cutoffs}\label{fig:CumCutoffs}
	\end{subfigure}
	\caption{Non-cumulative vs. Cumulative Cutoffs in Multi-Cutoff RD Design}\label{fig:CutoffTaxonomy}
\end{figure}

An important practical issue in the Multi-Cutoff RD design is the relationship between the multiple cutoffs and the score induced by the treatment assignment mechanism. If a unit with score $X_i=x$ can be exposed to any cutoff $c \in \mathcal{C}$, we say that the cutoffs are non-cumulative. Figure \ref{fig:NoncumCutoffs} shows a hypothetical Multi-Cutoff RD design with three different non-cumulative cutoffs, $c_1$, $c_2$, and $c_3$, where a particular value $x$ is shown for the three subpopulations exposed to each of the three cutoffs. Panels I, II, and III show that a unit with $X_i=x$ could be exposed to any one of the three cutoff values. Although the process that determines whether a unit faces $c_1$, $c_2$ or $c_3$ may be related to $X_i$, the support of the score is common across the three subpopulations.

In contrast, when cutoffs are cumulative, a unit's score value restricts the number of cutoffs to which the unit can be exposed. This case arises most frequently when different doses of a treatment are given for different ranges of the running variable, making the cutoff faced by each unit a deterministic function of the unit's score. In Figure \ref{fig:CumCutoffs}, units with $X_i < c_1$ receive treatment A, units with $c_1\leq X_i < c_2$ receive treatment B, units with $c_2 \leq X_i < c_3$ receive treatment C, and units with $c_3 \leq X_i$ receive treatment D. Thus, a unit's score value is sufficient to know which cutoff (or pair of cutoffs) the unit faces.

There are three important practical consequences of this distinction. First, in Multi-Cutoff RD designs with non-cumulative cutoffs, it is common for all units to receive the same treatment regardless of which cutoff they are exposed to. In contrast, when cutoffs are cumulative, it is common for the treatment to vary by cutoff. For example, in the SPP program introduced in Section \ref{sec:FuzzyRD}, all students in Colombia receive the same subsidy, but the cutoff for eligibility varies by geographic region. This is a case of a non-cumulative Multi-Cutoff RD design. Alternatively, for example, municipalities may receive a different amount of federal transfers depending on the municipality's population, or patients may receive different medicine dosages depending on the result of some continuously distributed laboratory result. With cumulative cutoffs, every time a different cutoff is crossed, the treatment received typically increases or decreases---but it could also change altogether. Researchers interested in an overall effect may need to redefine the treatment of interest. In the federal transfers example, we can redefine the treatment as receiving higher transfers regardless of the particular amount, and treat all units exposed to different cutoffs as receiving the same treatment. From this perspective, the presence of multiple cutoffs can imply observable heterogeneity in the treatment. 

Second, the cumulative rule implies a lack of common support in the value of the running
variable for units facing different cutoffs. For example, in Figure \ref{fig:CumCutoffs}, a unit with $X_i= x$ for $c_1<x <c_2$ can only be exposed to cutoffs $c_1$ or $c_2$ but not $c_3$, and all units exposed to the highest cutoff $c_3$ must have score higher than or equal to $c_2$. In general, with cumulative cutoffs, the subpopulations of units exposed to different cutoffs will have systematically different values of the running variable. If the score is related to the potential outcomes, as is common, the type of units exposed to one cutoff may thus be different from the type of units exposed to a different cutoff, which may lead to important heterogeneity in treatment effects, and even lack of meaningful comparability across treatment effects.

Finally, the subpopulations exposed to the different cutoffs are well-defined in the non-cumulative case but are ambiguously defined in the cumulative case. When cutoffs are non-cumulative as in the scenario in Figure \ref{fig:NoncumCutoffs}, every unit is exposed to exactly one cutoff, and the subpopulations exposed to each cutoff $c_1, c_2, \ldots, c_J$ are defined straightforwardly by selecting units with $C_i=c_1$, $C_i=c_2$, $\ldots$, $C_i=c_J$. In contrast, when cutoffs are cumulative, the same unit may be exposed to two cutoffs. For example, in Figure \ref{fig:CumCutoffs}, the unit with $X_i = x$ is above the cutoff $c_1$ and below the cutoff $c_2$. Thus, a cutoff-specific analysis of $c_1$ may include this unit as a treated unit, while a cutoff-specific analysis of $c_2$ may include the same unit as a control. This would lead the estimated cutoff-specific effects to be correlated with each other, in addition to altering the interpretation of the treatment effects. To avoid this, researchers can calculate some midpoint between $c_1$ and $c_2$ such as $c_{21}=\frac{c_2-c_1}{2}$ or $c_{21}=\text{median}(X_i:c_1\leq X_i<c_2)$, and use units with $X_i\leq c_{21}$ in the analysis of the effect at $c_1$, and units with $X_i > c_{21}$ in the analysis of the effect at $c_2$. 

\subsubsection{Multi-Cutoff Treatment Effects}

One direct consequence of the multi-dimensionality of the Multi-Cutoff RD design is that we can define multiple average treatment effects of interest. We discuss three kinds of parameters: cutoff-specific effects that capture the average treatment effect for units exposed to one of the multiple cutoffs, a pooled effect that combines all observations into a single RD parameter by normalizing the score, and extrapolated RD effects for values of the score in between the cutoffs. 

\noindent\underline{Cutoff-specific Treatment Effects}

When cutoffs are non-cumulative, the \textit{cutoff-specific} RD treatment effects are defined analogously to the standard one-dimensional RD effects. For example, focusing on sharp treatment effects for $c\in\mathcal{C}$,
\[\tau_{\mathtt{SRD}}(c) \equiv \E[Y_i(1)-Y_i(0)|X_{i}=c, C_i =c]\]
within the continuity-based framework, and
\[\theta_{\mathtt{SRD}}(c) \equiv \frac{1}{N_{\W_c}}\sum_{i: X_i \in \W_c} \E_{\W_c}\left[Y_i(1)-Y_i(0)\right],\]
within the local randomization framework, where $\W_c = [c-w, c+w]$ is a window around the specific cutoff $c$, $N_{\W_c}$ denotes the number of units with $X_i \in \W_c$, and $\E_{\W_c}$ is defined as in Section \ref{sec:localrand} for each $\W_c$. 

\begin{figure}[ht]
	\centering
	\includegraphics[scale=\kSF]{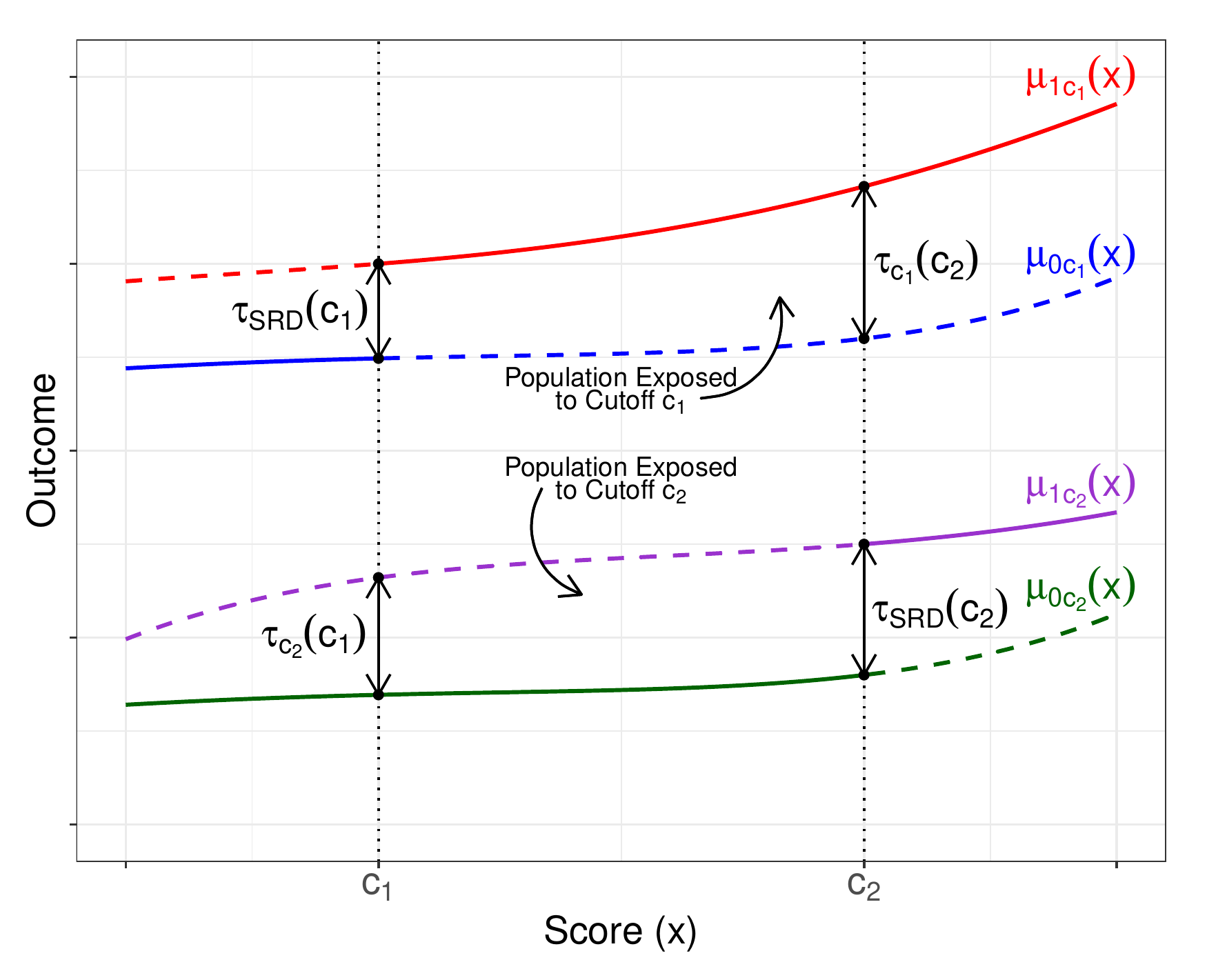}
	\caption{Multi-Cutoff RD Design with Two Non-cumulative Cutoffs}\label{fig:RDMultiCutoff}
\end{figure}

The interpretation of these treatment effects follows closely the interpretation of the effect in the standard single-cutoff RD design: it is the average change in outcomes that would be observed near the cutoff if we changed the status of all units exposed to cutoff $c$ from control to treated. Because each of the effects $\tau_{\mathtt{SRD}}(c)$ and $\theta_{\mathtt{SRD}}(c)$ focuses on the subpopulation of units exposed to the cutoff $c$, a cutoff-specific analysis allows researchers to explore the heterogeneity of the treatment effect across the subpopulations exposed to different cutoffs. Figure \ref{fig:RDMultiCutoff} illustrates a hypothetical continuity-based Multi-Cutoff RD design with two non-cumulative cutoffs. The figure shows the two cutoff-specific effects, $\tau_\mathtt{SRD}(c_1)$ and $\tau_\mathtt{SRD}(c_2)$, as well as two infeasible effects $\tau_{c_2}(c_1)$ (effect at cutoff $c_1$ for the subpopulation exposed to cutoff $c_2$) and $\tau_{c_1}(c_2)$ (effect at cutoff $c_2$ for the subpopulation exposed to cutoff $c_1$). (The notation $\tau_{c}(x)$ will be explained when we discuss extrapolation effects below.)

Estimation, inference, and falsification for $\tau_{\mathtt{SRD}}(c)$ and $\theta_{\mathtt{SRD}}(c)$ can be implemented using the one-dimensional continuity-based methods discussed in \textit{Foundations} and the one-dimensional local randomization methods discussed in Section \ref{sec:localrand}, respectively. These methods are implemented by considering each subsample defined by all units $i$ with $C_i=c$, for $c\in\mathcal{C}$, and analyzing each subsample separately. These effects can then be collected for further analysis and interpretation under additional conditions.

Restricting the analysis to units with cutoff equal to $c$, a continuity-based analysis can be based on the standard single-cutoff identification result for the Sharp RD design, 
\[\tau_\mathtt{SRD}(c) = \lim_{x\downarrow c}\E[Y_{i}|X_{i}=x, C_i = c]-\lim_{x\uparrow c}\E[Y_{i}|X_{i}=x, C_i=c],\]
and a local randomization analysis can be based on the analogous result 
\[\theta_{\mathtt{SRD}}(c) = \frac{1}{N_{\W_c}}\sum_{i: X_i \in \W_c} \E_{\W_c}\left[\frac{T_iY_i}{\P_{\W_c}[T_i=1]}\right] - \frac{1}{N_{\W_c}}\sum_{i: X_i \in \W_c} \E_{\W_c}\left[\frac{(1-T_i)Y_i}{1-\P_{\W_c}[T_i=1]}\right],
\]
where $N_{\W_c}$, $\P_{\W_c}$ and $\E_{\W_c}$ are defined as in Section \ref{sec:localrand} for each $\W_c$, $c\in\mathcal{C}$. Plug-in estimators and related inference procedures proceed as discussed in previous sections.

Cutoff-specific treatment effects can also be defined when the cutoffs are cumulative. Denoting the different values or doses of the treatment as $t_j$, $j=0,1,\dots,J$, the treatment level variable is $L_i=\sum_{j=1}^J (t_j-t_{j-1}) \I(X_i\geq c_j)\in\{t_1,t_2,\ldots, t_J\}$ with $t_{-1}=0$. This assignment rule is an RD assignment because for each cutoff the assignment of treatment still obeys the rule $T_i(c)=\I(X_i\geq c)$. It follows that $\tau_\mathtt{SRD}(c)$ and $\theta_{\mathtt{SRD}}(c)$ continue to have the same interpretation as before with the caveat that now each treatment effect is defined relative to the previous treatment level in a cumulative way. As discussed above, observations that are exposed to two different cutoffs can be used to estimate two different but consecutive treatment effects. For example, a unit with score $ c_{j} <X_i < c_{j+1}$ will receive treatment dosage $t_j$ and could be used both as a treatment unit when estimating $\tau_\mathtt{SRD}(c_j)$ and as a control unit when estimating $\tau_\mathtt{SRD}(c_{j+1})$. As a result, cutoff-specific estimators may not be independent, although the dependence disappears if the bandwidths or window lengths are chosen so that they do not overlap across cutoffs, which implies that units contribute to estimation and inference for only one cutoff. Once the observations have been assigned to each cutoff under analysis, local polynomial or local randomization methods can be applied cutoff by cutoff.

\noindent\underline{Normalized-and-pooled Treatment Effects}

It is also common to define a single treatment effect for all units after normalizing the score. We define the normalized score $\tilde{X}_i = X_i-C_i$, pool all observations using the normalized score instead of the original score, and use zero as the cutoff for all observations. The treatment assignment indicator is therefore $\T_i = \I(X_i - C_i \geq 0) = \I(\tilde{X}_i \geq 0)$ for all units. This normalizing-and-pooling strategy combines all observations exposed to different cutoffs into a single parameter, called the \textit{normalized-and-pooled} RD treatment effect. For example, for the continuity-based Sharp RD case, this treatment effect is
\[\tau^\mathtt{P}_\mathtt{SRD} = \lim_{x\downarrow0} \E[Y_i | \tilde{X}_i=x] - \lim_{x\uparrow0} \E[Y_i | \tilde{X}_i=x].
\]
Like in the cutoff-specific case, estimation and inference for $\tau^\mathtt{P}_\mathtt{SRD}$ can proceed in the same way as in the standard Sharp RD design with a single cutoff, using the methods discussed in \textit{Foundations} with $\tilde{X}_{i}$ as the score and $c=0$ as the cutoff. The local randomization parameter can be defined analogously using the notation introduced in Section \ref{sec:localrand}:
\[\theta^\mathtt{P}_\mathtt{SRD} =\frac{1}{N_{\tilde{\W}}} \sum_{i:\tilde{X}_i\in\tilde{\W}} \E_{\tilde{\W}}\Big[\frac{T_i Y_i}{\P_{\tilde{\W}}[T_i=1]} \Big] 
	- \frac{1}{N_{\tilde{\W}}} \sum_{i:\tilde{X}_i\in\tilde{\W}} \E_{\tilde{\W}}\Big[ \frac{(1-T_i) Y_i}{1-\P_{\tilde{\W}}[T_i=1]}\Big],
\]
where $\tilde{\W} = [-w,w]$ for $w>0$, $N_{\tilde{\W}}$ denotes the number of observations with normalized score $\tilde{X}_i$ within $\tilde{\W}$, and $\P_{\tilde{\W}}[\cdot]$ and $\E_{\tilde{\W}}[\cdot]$ denote probability and expectations computed conditionally for those units with normalized score $\tilde{X}_i$ within $\tilde{\W}$.

Under regularity conditions, these pooled treatment effects are equal to a weighted average of the corresponding cutoff-specific RD treatment effects. For example, in the continuity-based framework,
\[\tau^\mathtt{P}_\mathtt{SRD} = \sum_{c\in\mathcal{C}}\tau_{\mathtt{SRD}}(c) \omega(c), \qquad
  \omega(c) \equiv \P[C_i=c|\tilde{X}_i=0] = \frac{f_{X|C}(c|c)\P[C_i=c]}{\sum_{c\in\mathcal{C}} f_{X|C}(c|c)\P[C_i=c]}
\]
with $f_{X|C}(x|c)$ denoting the conditional density of the score given the cutoff. In the local randomization framework, a similar representation of $\theta^\mathtt{P}_\mathtt{SRD}$ as a weighted average of $\theta_{\mathtt{SRD}}(c)$, $c\in\mathcal{C}$, can be derived. These results show that the cutoff-specific effects that contribute the most to $\tau^\mathtt{P}_\mathtt{SRD}$ and $\theta^\mathtt{P}_\mathtt{SRD}$ are those whose cutoffs have a relatively high number of observations near them.

\noindent\underline{Extrapolation Treatment Effects}

It is also possible to define \textit{extrapolation} parameters, which capture the effect of the treatment at values of the score other than the cutoff $c\in\mathcal{C}$. These parameters are useful because they allow researchers to learn about the effect of the treatment for units whose score values are not necessarily close to the specific cutoff used for treatment assignment. Because the treatment assignment is still based on the RD rule, the fundamental problem of causal inference makes it impossible to learn about effects arbitrarily far away from the cutoff in the absence of additional assumptions. We discuss one possible assumption that explicitly exploits the presence of multiple cutoffs.

We focus on the continuity-based Sharp RD case for simplicity. Letting the treatment effect include an additional subscript indicating the cutoff to which units are exposed, we define
\begin{equation*}\label{eq:RD-single}
	\tau_c(x) \equiv \E[Y_i(1)-Y_i(0)| X_i=x,C_i=c] = \mu_{1,c}(x)-\mu_{0,c}(x),
\end{equation*}
where $\mu_{1,c}(x) = \E[Y_{i}(1)|X_i=x,C_i=c]$ and $\mu_{0,c}(x) = \E[Y_{i}(0)|X_i=x,C_i=c]$ denote the regression functions of the potential outcomes under treatment and control, respectively, for a given cutoff $c$. This notation separates the cutoff to which each population is exposed ($c$ subindex) from the value of the score being conditioned on ($x$ argument). In the Multi-Cutoff RD design, $\tau_c(x)$ is the average treatment effect that a population exposed to cutoff $c$ would exhibit at the score value $x$. For a fixed cutoff $c$, this parameter captures how the average treatment effect varies for a subpopulation as the score changes. Our previously defined continuity-based Sharp RD parameter $\tau_\mathtt{SRD}(c)$ is thus $\tau_c(c)$.

Suppose we have two cutoffs, $c_1$ and $c_2$, with $c_2>c_1$. This means that there are two subpopulations: units exposed to the low cutoff, $c_1$, and units exposed to the high cutoff, $c_2$. The standard RD effects at each cutoff are $\tau_{c_1}(c_1)$ and $\tau_{c_2}(c_2)$, which are easily estimated with the methods already discussed. In contrast, the problem of extrapolation is to study an effect such as $\tau_{c_1}(x)$, for $c_1<x\leq c_2$, that is, the average effect of the treatment at a score value away from the cutoffs for the subpopulation of units exposed to the low cutoff $c_1$. The main challenge to identification is that all units exposed to the cutoff $c_1$ are treated for values of the score above $c_1$. That is, for $x\in(c_1,c_2]$, the treatment response function $\mu_{1,c_1}(x)$ and the control response function $\mu_{0,c_2}(x)$ are estimable, while the needed control response of the population exposed to $c_1$, $\mu_{0,c_1}(x)$, is not. 
 
\begin{figure}[ht]
	\centering
	\includegraphics[scale=\kSF]{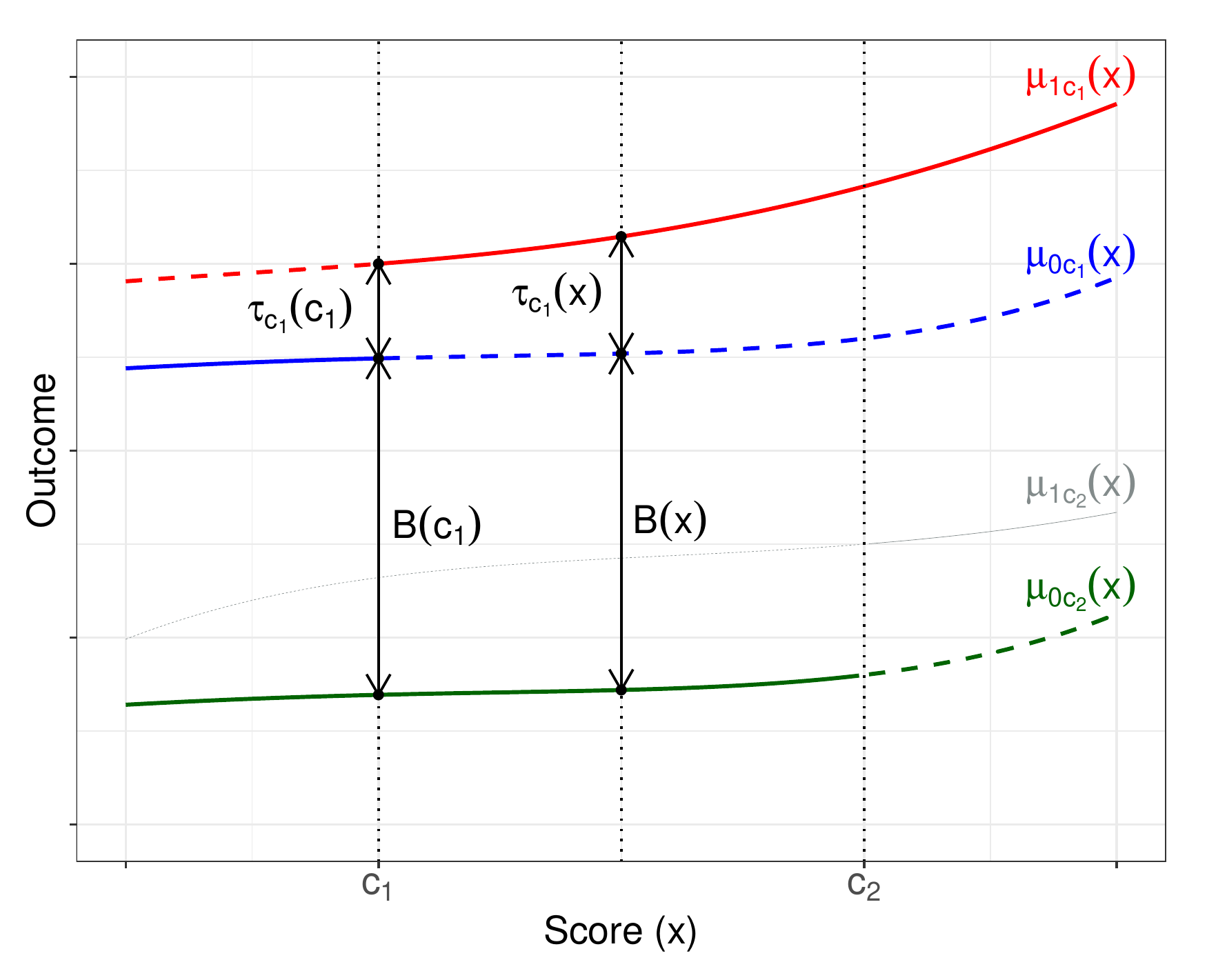}
	\caption{Multi-Cutoff RD Extrapolation with Constant Bias ($B(c_1)=B(x)$).\label{fig:constantbias}}
\end{figure}

A natural approach to identify and estimate $\tau_{c_1}(x)=\mu_{1,c_1}(x)-\mu_{0,c_1}(x)$ for $c_1<x\leq c_2$ is to use the control group of the subpopulation of units exposed to cutoff $c_2$ to learn about $\mu_{0,c_1}(x)$ under appropriate conditions. A na\"ive approach would assume $\mu_{0,c_2}(x) = \mu_{0,c_1}(x)$, that is, the control response of the units exposed to $c_1$ at $x$ would have been the same as the control response of the units exposed to $c_2$ at $x$. This assumption would ignore any systematic differences between both subpopulations; it would be valid if, for example, the cutoffs were randomly assigned to units. An alternative assumption is that any pre-existing differences between the two subpopulations are constant for all values of the score. Then, we can use the ``bias'' or difference $B(c_1) \equiv \mu_{0,c_2}(c_1) - \mu_{0,c_1}(c_1)$, which is estimable from the data, to calculate the difference $B(x) \equiv \mu_{0,c_2}(x) - \mu_{0,c_1}(x)$ for $c_1<x\leq c_2$, because such difference is assumed constant, $B(c_1) = B(x)$. The assumption, which is illustrated in Figure \ref{fig:constantbias}, is analogous to the ``parallel trends'' assumption in the difference-in-differences design. With this assumption, we obtain the identification result
 \begin{equation*}
\tau_{c_1}(x)=\mu_{1,c_1}(x) - \left\{ \mu_{0,c_2}(x) + \mu_{0,c_1}(c_1) - \mu_{0,c_2}(c_1)\right\}, \qquad
x\in(c_1,c_2],
\end{equation*}
where estimation and inference for the four conditional expectations can be conducted with the local polynomial methods discussed in \textit{Foundations} at boundary and interior points (for example, $c_1$ is a boundary point for estimation of $\mu_{0,c_1}(c_1) $ but an interior point for estimation of $\mu_{0,c_2}(c_1)$). An analogous assumption can be invoked in a window around the cutoff to identify, estimate, and conduct inference on extrapolation RD treatment effects using local randomization methods.

\subsubsection{The Effect of Financial Aid on Post-Secondary Education Attainment with Multiple Cutoffs}

We illustrate the analysis of Multi-Cutoff RD designs with the \cite{LondonoVelezRodriguezSanchez_2020_AEJ} application. As in Section \ref{sec:FuzzyRD}, we keep the subset of students in the $2014$ cohort whose SABER 11 score is above the merit cutoff, which results in an RD design with a single score---the SISBEN wealth index. Recall that SPP program eligibility was assigned according to three different cutoffs that varied with the student's area of residence: $40.75$ in rural areas, $57.21$ in the fourteen main metropolitan areas, and $56.32$ in the rest of the urban areas. For simplicity of exposition, we ignore compliance issues and focus on intention-to-treat effects, using a Sharp Multi-Cutoff RD design where the running variable is the SISBEN wealth score, the cutoff varies by region, and the treatment is eligibility to receive the SPP program. As before, the outcome of interest is an indicator equal to one if the student enrolled in a HEI immediately after program eligibility.

Table \ref{tab:LRSapplication-Cutoffs} shows the three different cutoff values, the number of observations exposed to each cutoff, and the maximum and minimum SISBEN score for the subpopulation exposed to each cutoff. 

\begin{table}[ht]
    \centering
    \begin{tabular}{lcccc}
    	\toprule
    	\multicolumn{1}{c}{Subpopulation}  & Cutoff  &  \multirow{1}{*}{Sample Size} &  \multirow{1}{*}{Min $X_i$} &  \multirow{1}{*}{Max $X_i$}\\ 
    	\midrule
    	Area 1 (14 metropolitan areas) & 57.21 	& 11,238 &   .98 & 83.15\\
        Area 2 (other urban areas)	   & 56.32	& 10,053 &  1.78 & 91.91\\
    	Area 3 (rural areas)     	   & 40.75  &  1,841 &  2.89 & 84.23\\
    	\bottomrule
    \end{tabular}
    \caption{Subpopulations exposed to different cutoffs---SPP data}\label{tab:LRSapplication-Cutoffs}
\end{table}

\begin{figure}[ht]
	\centering
	\centering
	\begin{subfigure}{0.48\textwidth}
		\centering
		\includegraphics[width=0.93\textwidth]{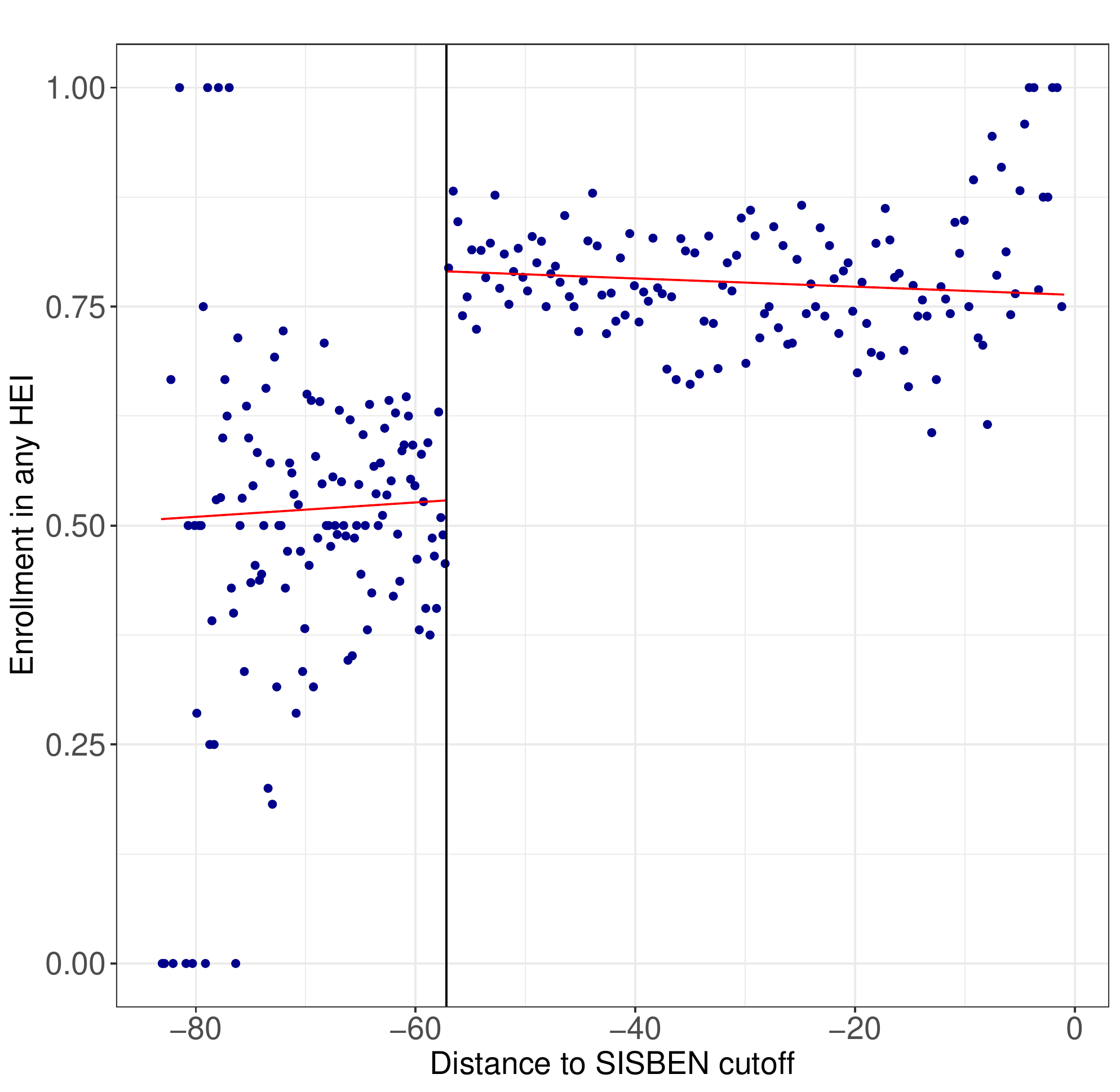}
		\caption{Cutoff $C_i=57.21$ ($J_{-}^\mathtt{MV}$, $J_{+}^\mathtt{MV}$, $p=1$)}\label{RDPLOT-SPP-cutoff=57.21}
	\end{subfigure}
	\begin{subfigure}{0.48\textwidth}
		\centering
		\includegraphics[width=0.9\textwidth]{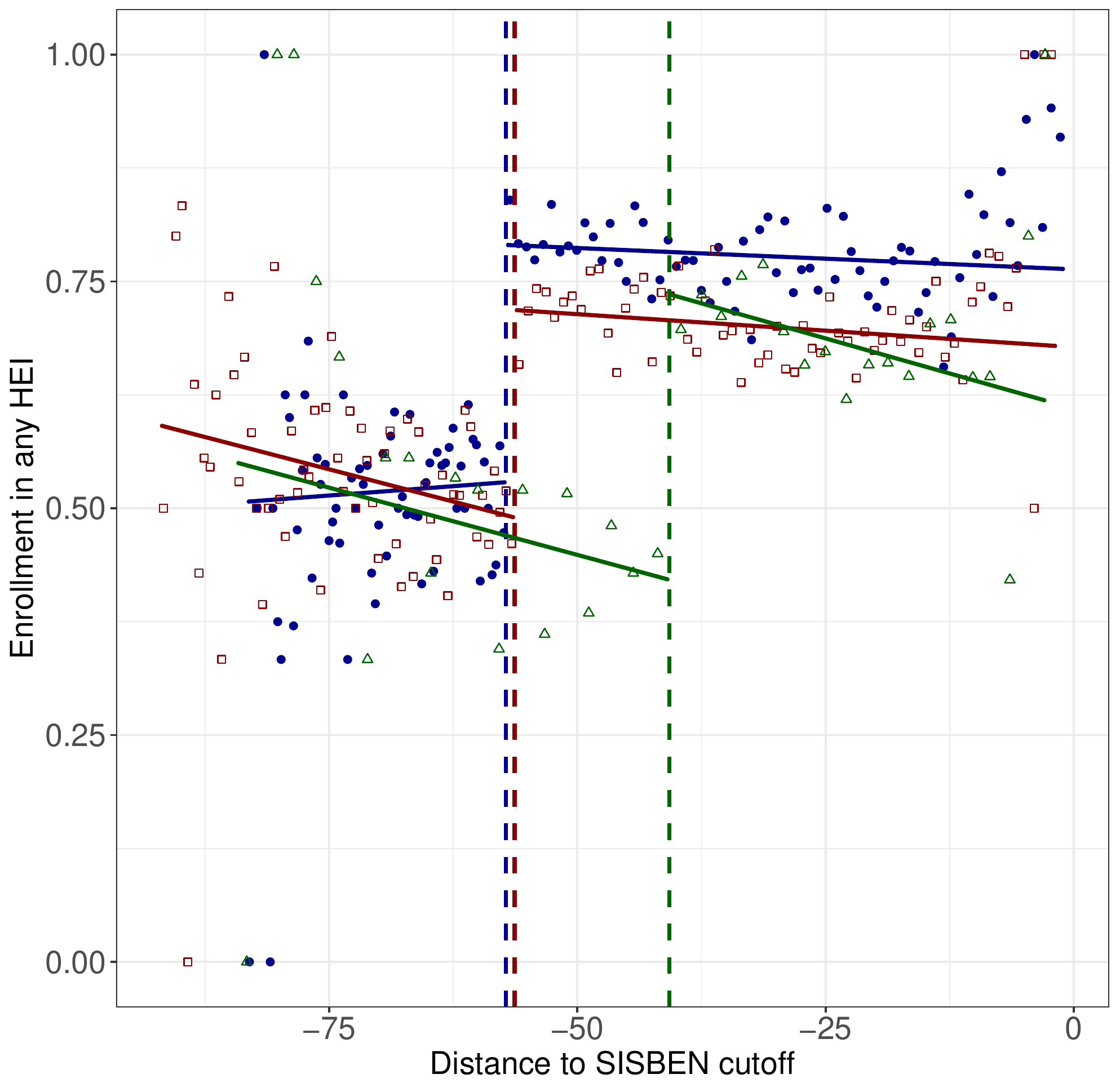}
		\caption{All Cutoffs ($J_{-}^\mathtt{MV}/2$, $J_{+}^\mathtt{MV}/2$, $p=1$)}\label{RDPLOT-SPP-allcutoffs}
	\end{subfigure}
	\caption{RD Plots---SPP data}
	\label{}
\end{figure}

Because SPP eligibility is given to students with wealth \textit{below} the cutoffs, we multiply all cutoffs and all scores by $-1$ to follow the convention that the active treatment is assigned to units above the cutoff. The analysis can be implemented with \texttt{rdrobust} after subsetting the data accordingly, or with \texttt{rdmulti}, which employs \texttt{rdrobust} to perform estimation, inference, and plotting in Multi-Cutoff RD designs. We begin by plotting the data using \texttt{rdplot} for the cutoff $C_i=57.21$ and \texttt{rdmcplot} for all three cutoffs. Figure \ref{RDPLOT-SPP-cutoff=57.21} employs a global linear ($p=1$) polynomial fit to avoid Runge's phenomenon near the cutoff and for comparability with the accompanying Figure \ref{RDPLOT-SPP-allcutoffs}, which reports all cutoffs in a single plot also employing a linear global fit (this figure uses half of the mimicking variance optimal number of bins, $J_{+}^\mathtt{MV}$ and $J_{-}^\mathtt{MV}$, to reduce the overall cluttering and improve visualization).

We now use \texttt{rdrobust} to conduct cutoff-specific and normalizing-and-pooling analyses. We first estimate the RD effect of SPP eligibility on HEI enrollment for the subpopulation of students exposed to the highest cutoff of $57.21$ (SISBEN Area 1), using \texttt{rdrobust} with default specifications (local linear, common MSE bandwidth, and triangular weights) only on the subset of observations exposed to this cutoff.

\labelsnippet{rdmcA}
\rsnip{Vol-2-R_LRS_rdrobust_cutoff1.txt}{\Rlink{\thesection}{\therdmcA}}
\statasnip{Vol-2-STATA_LRS_rdrobust_cutoff1}{\Slink{\thesection}{\therdmcA}}

In this subpopulation, students who are barely eligible to receive the SPP subsidy are $34.6$ percentage points more likely to enroll in a HEI than students who are barely ineligible. The effect is statistically distinguishable from zero, with a robust $95\%$ confidence interval of approximately $[0.269 , 0.452]$. Although this analysis can be repeated for each cutoff using \texttt{rdrobust} for each subpopulation, the analysis can be conducted more succinctly using \texttt{rdmulti}. 

\labelsnippet{rdmcB}
\rsnip{Vol-2-R_LRS_rdmc.txt}{\Rlink{\thesection}{\therdmcB}}
\statasnip{Vol-2-STATA_LRS_rdmc}{\Slink{\thesection}{\therdmcB}}

The first three rows show the cutoff-specific effects, which can be directly reproduced by using \texttt{rdrobust} on the subset of the observations exposed to each individual cutoff. The results in the first row therefore coincide with the \texttt{rdrobust} output for the subpopulation of students in Area 1 just shown. The last two rows show two different versions of the normalized-and-pooled Multi-Cutoff RD effect. The \texttt{Pooled} row displays the normalizing-and-pooling effect, which can also be implemented with \texttt{rdobust} by first creating the normalized score and then using it with a cutoff of zero. We illustrate this next.

\labelsnippet{rdmcC}
\rsnip{Vol-2-R_LRS_rdrobust_pooled_xnorm.txt}{\Rlink{\thesection}{\therdmcC}}
\statasnip{Vol-2-STATA_LRS_rdrobust_pooled_xnorm}{\Slink{\thesection}{\therdmcC}}

The normalizing-and-pooling parameter is a weighted average of the cutoff-specific effects, where each effect receives weight $\omega(c)=\P[C_i=c|\tilde{X}_i=0]$. The \texttt{Weighted} row in the \texttt{rdmulti} output multiplies each cutoff-specific cutoff by the estimated weights, implemented as $\hat{w}(c) = \hat{\P}(C_i = c | \tilde{X}_i= 0) = \sum_{i=1}^n{\I(C_i=c,-h<\tilde{X}_i<h)}/\sum_{i=1}^n\I(-h<\tilde{X}_i<h)$, for bandwidth $h>0$. In other words, in the row labeled \texttt{Weighted}, the weights that are implicitly imposed by normalizing and pooling are directly estimated and then used to calculate the estimated pooled effect by multiplying each cutoff-specific effect by its corresponding weight---this explains why the point estimates in rows \texttt{Weighted} and \texttt{Pooled} are so similar. The point estimate in the \texttt{Weighted} column can thus be obtained by multiplying each cutoff-specific effect by the estimated weights, both of which are returned by \texttt{rdmc}:

\labelsnippet{rdmcD}
\rsnip{Vol-2-R_LRS_rdmc_row_weightedresults.txt}{\Rlink{\thesection}{\therdmcD}}
\statasnip{Vol-2-STATA_LRS_rdmc_row_weightedresults}{\Slink{\thesection}{\therdmcD}}

The weights are approximately $0.384$ for Area 1, $0.534$ for Area 2, and $0.082$ for Area 3. The relatively lower weight for Area 3 is expected, as this comprises all rural areas where the number of observations is much smaller than in the urban areas (see Table \ref{tab:LRSapplication-Cutoffs}). Although the other two areas have similar numbers of total observations, Area 2's estimated weight is larger than Area 1's weight. The difference arises because, compared to Area 1, Area 2 has relatively more students with SISBEN wealth scores near the cutoff. 

Both the cutoff-specific analysis and the normalizing-and-pooling analysis lead to similar conclusions: eligibility for the SPP program leads to a 20 to 30 percentage-point increase in the probability of enrolling in a HEI. Finally, we show how to formally test whether the effects at each specific cutoff are different from each other by manually calculating the t statistic and its corresponding two-sided p-value. 

\labelsnippet{rdmcE}
\rsnip{Vol-2-R_LRS_rdmc_comparing_effects.txt}{\Rlink{\thesection}{\therdmcE}}
\statasnip{Vol-2-STATA_LRS_rdmc_comparing_effects}{\Slink{\thesection}{\therdmcE}}

The SPP eligibility effect is estimated to be roughly $34.6$ percentage points in Area 1; this effect is statistically significantly different from the effect in Area 2, with a p-value of 0.011. (The estimated effect in Area 2 is roughly $20$ percentage points. The difference shown in the output above is 16.3 percentage points, larger than $34.6-20=14.6$, because the point estimates used to construct the z-statistic are the bias-corrected estimates, not the conventional point estimates reported in the printed output.) In contrast, the effects for the smallest two cutoffs (Areas 2 and 3) are indistinguishable from each other.

\subsection{Multi-Score RD Design}

In the Multi-Score RD Design, two or more running variables determine the treatment assignment. To formalize, we assume that each unit's score is a bivariate vector denoted by $\mathbf{X}_i = (X_{1i}, X_{2i})'$. Then, the treatment assignment is $T_i = a(\mathbf{X}_i)$ for some assignment function $a:\mathbb{R}^2\mapsto\{0,1\}$. A common treatment assignment rule is one that requires both scores to be above a cutoff, which leads to $a(\mathbf{X}_i) = \I(X_{1i} > b_1) \cdot \I(X_{2i} > b_2)$ where $b_1$ and $b_2$ denote the cutoffs for each dimension. More complex treatment assignment rules have a varying boundary on the plane, as is common in Geographic RD designs.

\begin{figure}[ht]
    \centering
	\begin{subfigure}{0.48\textwidth}
		\centering
    	\includegraphics[scale=\kDF]{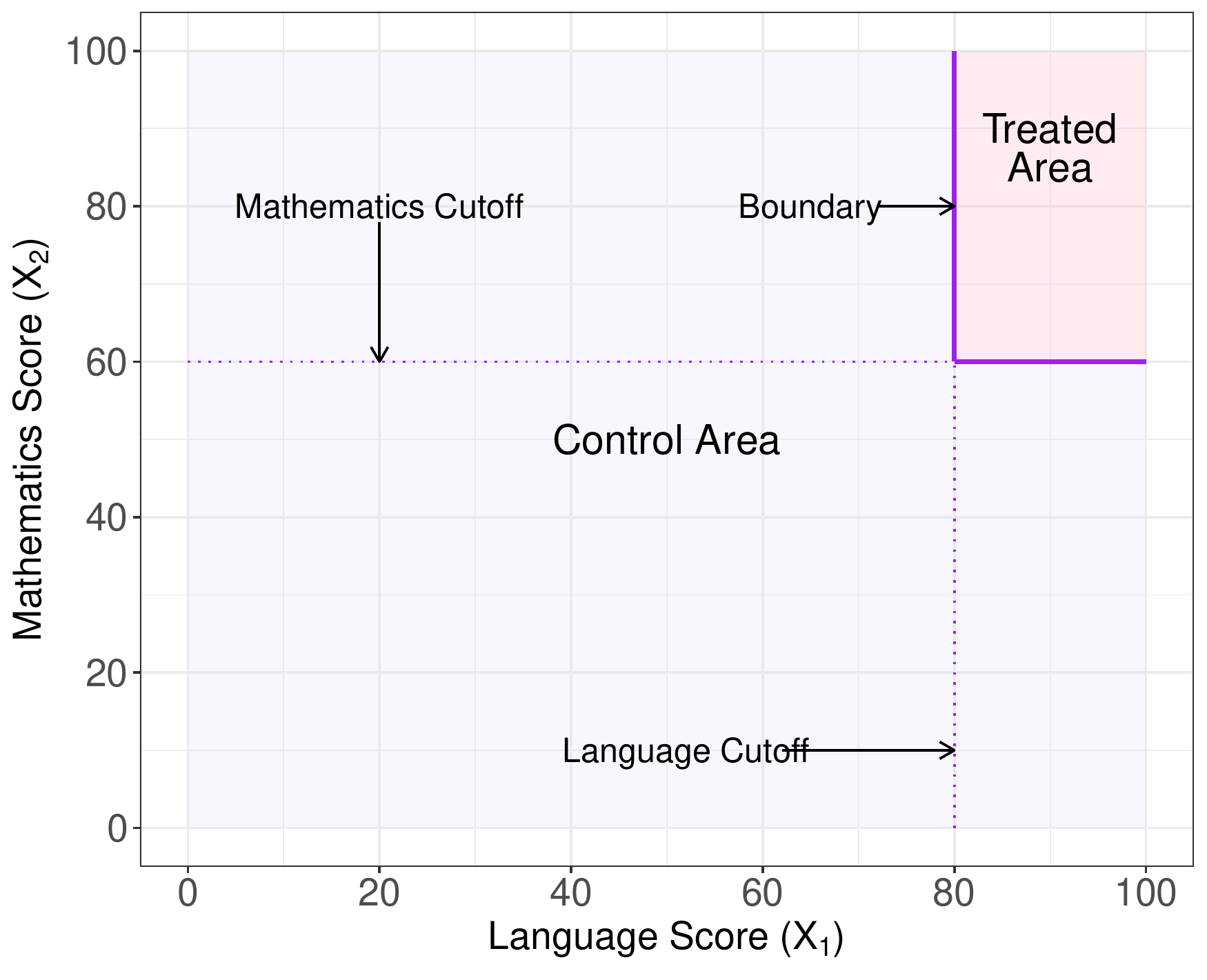}
		\caption{Hard-threshold Assignment}\label{fig:RDMultiScore-educ}		
	\end{subfigure}
	\begin{subfigure}{0.48\textwidth}
		\centering
		\includegraphics[scale=\kDF]{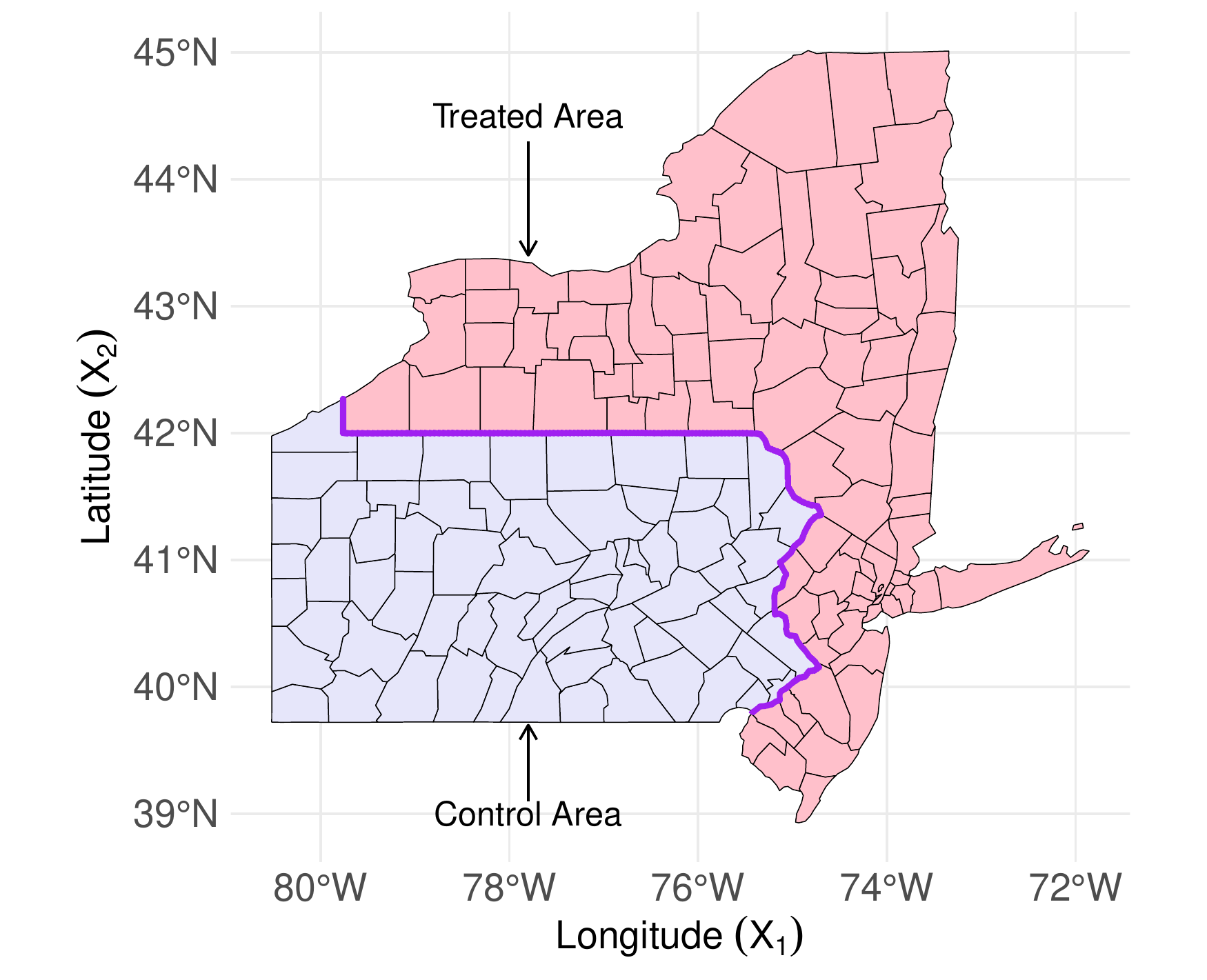}
		\caption{Geographic Assignment}\label{fig:RDMultiScore-geo}
	\end{subfigure}
	\caption{Multi-Score RD Designs: Treated and Control Areas}\label{fig:RDMultiScore}
\end{figure}

To consider a hypothetical education example, we assume that a scholarship is given to all students who score above $60$ in a language exam and above $80$ in a mathematics exam, letting $X_{1i}$ denote the language score and $X_{2i}$ the math score, and $b_1=80$ and $b_2=60$ be the respective cutoffs. According to this treatment assignment rule, a student with score $\mathbf{x}_i = (80,59.9)$ is assigned to the control condition, since $\I(80 \geq 80)\cdot\I(59.9 \geq 60 ) = 1 \cdot 0 = 0$, and misses the treatment only barely---had she scored an additional $1/10$ of a point in the mathematics exam, she would have received the scholarship. Without a doubt, this student is close to the cutoff criteria for receiving the treatment. However, scoring very close to both cutoffs is not the only way for a student to be barely assigned to treatment or control. A student with a perfect $100$ score in language would still be barely assigned to control if he scored $59.9$ in the mathematics exam, and a student with a perfect math score would be barely assigned to control if she got $79.9$ points in the language exam. Thus, with multiple running variables, there is no longer a single cutoff value at which the treatment status of units changes from control to treated; instead, the discontinuity in the treatment assignment occurs along a boundary of points. This is illustrated graphically in Figure \ref{fig:RDMultiScore-educ}.

In Multi-Score RD designs, the boundary where treatment assignment changes discontinuously is $\mathcal{B} = \{\mathbf{x}\in\mathbb{R}^2 : \mathbf{x} \in (\mathtt{bd}(\mathcal{A}_1) \cap \mathtt{bd}(\mathcal{A}_0)) \}$ with $\mathcal{A}_1 = \{\mathbf{x}\in\mathbb{R}^2 : a(\mathbf{x}) = 1\}$ and $\mathcal{A}_0 = \{\mathbf{x}\in\mathbb{R}^2 : a(\mathbf{x}) = 0\}$ denoting the treated and control areas, respectively, and where $\mathtt{bd}(\mathcal{A})$ denotes the frontier or boundary of the set $\mathcal{A}$, defined as the set's closure minus its interior ($\mathtt{bd}(\mathcal{A}) \equiv \mathtt{cl}(\mathcal{A}) \setminus \mathtt{int}(\mathcal{A})$). In the education example depicted in Figure \ref{fig:RDMultiScore-educ}, the assignment boundary takes the simple form $\mathcal{B}=\{\mathbf{x} =(x_1,x_2)' :(x_1 \geq 80 \text{ and } x_2 = 60) \text{ or } (x_1=80 \text{ and } x_2\geq 60)\}$.

An important special case of the Multi-Score RD design is the Geographic RD design, which often arises when adjacent administrative units such as counties, municipalities, or states are assigned opposite treatment status. In this case, the boundary $\mathcal{B}$ at which the treatment assignment changes discontinuously is the border that separates the adjacent administrative units---in other words, $\mathcal{B}$ separates a geographic treated area from a geographic control area. For example, in the 2010 primary election in Colorado in the United States, some counties had all-mail elections where voting could only be conducted by mail and in-person voting was not allowed, while other counties had traditional in-person voting. In areas where the two types of counties are adjacent, the administrative border between the counties induces a discontinuous treatment assignment between in-person and all-mail voting, and a Geographic RD design can be used to estimate the effect of adopting all-mail elections on voter turnout. A hypothetical geographic assignment is shown in Figure \ref{fig:RDMultiScore-geo}.

In the Geographic RD design, the score $\mathbf{X}_i=(X_{1i}, X_{2i})'$ is a vector of two coordinates such as latitude and longitude that determine the exact geographic location of unit $i$. In practice, this score is calculated using Geographic Information Systems (GIS) software, which allows researchers to obtain the coordinates corresponding to the geographic location of each unit in the study, as well as to locate the entire treated and control areas, and all points on the boundary between them. The assignment function $a(\mathbf{x})$ is thus determined by the administrative or otherwise geographic boundary, which is often more complex than the simple function shown in the education example in Figure \ref{fig:RDMultiScore-educ}. In the upcoming sections, we present two empirical illustrations: one has an assignment boundary similar to \ref{fig:RDMultiScore-educ}, while the other one has an assignment boundary similar to \ref{fig:RDMultiScore-geo}.

Analogously to the Multi-Cutoff RD design, the parameters of interest in the Multi-Score RD design change because there is no longer a single cutoff at which the probability of treatment assignment changes discontinuously; instead, the probability of treatment assignment changes discontinuously at an often uncountable collection of locations along the boundary $\mathcal{B}$ induced by the assignment function $a(\mathbf{x})$. One approach is to consider different RD treatment effects for location-specific points on the boundary $\mathcal{B}$, analogous to cutoff-specific effects in the Multi-Cutoff RD design. Another approach is to consider a single treatment effect along the boundary $\mathcal{B}$ by normalizing-and-pooling. We now we discuss both approaches to identification, estimation, and inference when implemented by either considering the multi-dimensional score $\mathbf{X}_i$ directly or reducing the score dimension to scalar score via a distance function. As in the Multi-Cutoff RD design, extrapolation treatment effects can also be defined in the Multi-Score RD design, though we omit them from our discussion to conserve space.

\subsubsection{Multi-Score Treatment Effects}

We assume perfect compliance and no spillovers to simplify the exposition, and thus focus on a Sharp RD setting with multiple scores. The continuity-based parameter of interest in the Multi-Score RD design is a generalization of the standard Sharp RD design parameter, where the average treatment effect is calculated at all points along the boundary $\mathcal{B}$ where the treatment assignment changes discontinuously from zero to one:
\[\tau_\mathtt{SMS}(\mathbf{b}) \equiv \E[Y_i(1) - Y_i(0) | \mathbf{X}_{i} = \mathbf{b}], \qquad \mathbf{b} \in \mathcal{B}.\]
Identification of the Multi-Score RD effect within the continuity-based framework is analogous to the single-score case,
\[\tau_{\mathtt{SMS}}(\mathbf{b})
= \lim_{\mathbf{x}\rightarrow\mathbf{b}: \mathbf{x}\in\mathcal{A}_1} \E[Y_i|\mathbf{X}_{i}=\mathbf{x}]
- \lim_{\mathbf{x}\rightarrow\mathbf{b}: \mathbf{x}\in\mathcal{A}_0} \E[Y_i|\mathbf{X}_{i}=\mathbf{x}],
\qquad \mathbf{b} \in \mathcal{B},
\]
with the difference that now limits are taken along two dimensions. In words, for each two-dimensional cutoff point $\mathbf{b}$ along the boundary $\mathcal{B}$, the treatment effect at that point is identifiable by the limits of the observed bivariate regression functions for the treated and control groups. The important distinction with respect to the one-dimensional score case is that the Multi-Score RD design generates a family or curve of treatment effects $\tau_{\mathtt{SMS}}(\mathbf{b})$, one for each boundary point $\mathbf{b} \in \mathcal{B}$. For instance, in the context of the example in Figure \ref{fig:RDMultiScore}, two different Sharp RD treatment effects are $\tau_{\mathtt{SMS}}(80,70)$ and $\tau_{\mathtt{SMS}}(90,60)$. Treatment effects and related methods within the local randomization framework can also be defined and applied following the analogy with the single-score case.

As we illustrate below, a simple approach for implementation is to choose a grid of points in $\mathcal{B}$ and study treatment effects for each point on the grid, which effectively maps (via discretization of $\mathcal{B}$) the problem to a Multi-Cutoff RD design. In some applications, it may be important to account for joint estimation and inference along the boundary.

Estimation and inference of Multi-Score RD effects can be implemented in two different ways. One is to use the original bivariate score, and modify the continuity-based and local randomization methods to account for the two dimensions. This involves, for example, estimating a local polynomial with two dimensions and selecting a two-dimensional optimal bandwidth. The other is to first reduce the bivariate score to one dimension by using as the score the distance from every unit's location to the particular point on the boundary where the effect is being calculated. The advantage of collapsing the score to one dimension is that all the one-dimensional methods discussed in \textit{Foundations} and the previous sections of this manuscript are directly applicable. However, there are two caveats. First, collapsing the score to one dimension may lead to too much misspecification error when the boundary point is near a kink or otherwise irregular section of the boundary. This may result in inferences that are not uniformly valid---though pointwise validity remains intact. Second, applying MSE-optimal bandwidth selection to the collapsed one-dimensional score results in a bandwidth that is too small in an asymptotic sense, since the optimal one-dimensional bandwidth selection procedures disregard the intrinsic two dimensions underlying the distance-based score. This implementation approach still controls misspecification error, but the implied undersmoothing may lead to efficiency loss.

Our empirical illustration in the rest of this section follows the approach of collapsing the score to one dimension for the analysis of both location-specific and normalized-and-pooled treatment effects. Theory and methods based on the bivariate score are in current development \citep{Cattaneo-Titiunik-Yu_2024_wp}.

\noindent \underline{Location-specific Treatment Effects}

For estimation of treatment effects at specific locations along the boundary $\mathcal{B}$, the bivariate score $\mathbf{X}_i$ is reduced to a scalar score by computing the distance of each unit's multi-dimensional score to the desired location on the boundary. To formalize this approach, suppose $\mathbf{b}=(b_1,b_2)'\in\mathcal{B}$ is the location-specific point. Define
\[\mathfrak{D}_i(\mathbf{b}) = d(\mathbf{X}_i,\mathbf{b}) a(\mathbf{X}_i) - d(\mathbf{X}_i,\mathbf{b}) (1-a(\mathbf{X}_i)),\]
for each unit $i=1,2,\dots, n$, where $d:\mathbb{R}^2\times\mathbb{R}^2\mapsto\mathbb{R}_+$ denotes a distance metric. The choice of distance metric depends on the particular application. For non-geographic applications, this is typically the Euclidean distance, $d(\mathbf{X}_i,\mathbf{b})=\sqrt{(X_{1i} - b_1)^2+ (X_{2i} - b_2)^2}$. For geographic applications, the Euclidean distance may not be appropriate if calculated over a large geographic area, because it fails to account for the approximately spherical shape of Earth. In this case, the geodetic distance (the shortest great-arc distance between points that lie on a spherical surface) or the chordal distance (the distance of the chord joining two points on a sphere), may be more appropriate. In some geographic RD designs, researchers might also be interested in other measures of distance to the boundary such as the driving or walking distance, or the distance along paved roads; these distances will require more geographic information in addition to the unit's geographic coordinates and are typically calculated with GIS software.

Given $(Y_i,\mathfrak{D}_i(\mathbf{b}))$, $i=1,2,\dots, n$, we can identify and estimate the location-specific treatment effect at the point $\mathbf{b}\in\mathcal{B}$ using a one-dimensional RD analysis for all observations together with $X_i=\mathfrak{D}_i(\mathbf{b})$ as the scalar running variable and $c=0$ as the cutoff, employing either the continuity-based or local randomization frameworks. 

In the continuity-based case, 
\[\tau_{\mathtt{SMS}}(\mathbf{b})
    = \lim_{x \downarrow 0} \E[Y_i| \mathfrak{D}_i(\mathbf{b}) = x] 
    - \lim_{x \uparrow 0} \E[Y_i| \mathfrak{D}_i(\mathbf{b}) = x],
\]
and hence the analysis of the location-specific continuity-based treatment effects requires analogous assumptions to the one-dimensional RD design for each boundary point studied and can be implemented using the methods discussed in \textit{Foundations}. 

Similarly, in the local randomization case, 
\[\theta_{\mathtt{SRD}}(\mathbf{b})
   = \frac{1}{N_{\W_\mathbf{b}}}\sum_{i: \mathfrak{D}_i(\mathbf{b}) \in \W_\mathbf{b}} \E_{\W_\mathbf{b}}\left[\frac{T_iY_i}{\P_{\W_\mathbf{b}}[T_i=1]}\right]
   - \frac{1}{N_{\W_\mathbf{b}}}\sum_{i: \mathfrak{D}_i(\mathbf{b}) \in \W_\mathbf{b}} \E_{\W_\mathbf{b}}\left[\frac{(1-T_i)Y_i}{1-\P_{\W_\mathbf{b}}[T_i=1]}\right]
\]
where $N_{\W_\mathbf{b}}$, $\P_{\W_\mathbf{b}}$ and $\E_{\W_\mathbf{b}}$ are defined as in Section \ref{sec:localrand} for each $\W_\mathbf{b}= [-w,w]$, $w>0$, and $\mathbf{b}\in\mathcal{B}$. A local randomization analysis of location-specific effects can be implemented using the methods discussed in Section \ref{sec:localrand}. 

In practice, both approaches are implemented for a finite collection of evaluation points along the boundary $\mathcal{B}$. The choice of the particular boundary points should be generally guided by application-specific considerations, particularly in cases where researchers have substantive reasons to focus on specific sections of the boundary. However, a general principle that can be applied to selecting boundary points for analysis is related to the density of observations. It is common to see sections of the boundary where the number of observations is small, particularly close to the boundary, which will result in excessive extrapolation. In these cases, researchers may wish to analyze boundary points where the density of observations close to the boundary is high for both treated and control groups. Moreover, as in the Multi-Cutoff RD design, reusing observations for multiple boundary points will result in correlated treatment effect estimates; if researchers wish to avoid this, they may restrict each observation to contribute to no more than one analysis.

\noindent \underline{Normalized-and-Pooled Treatment Effects}

It is also possible to analyze all boundary points simultaneously by considering the normalized-and-pooled treatment effect. Instead of performing estimation and inference for multiple treatment effects located at a specific point on the boundary, this approach considers the effects at all boundary points simultaneously by using as the running variable the shortest distance to the boundary and then pooling all observations in a one-dimensional RD analysis. 

Formally, the score for each unit is set to be $X_i=\mathfrak{D}_i = d(\mathbf{X}_i) a(\mathbf{X}_i) - d(\mathbf{X}_i) (1-a(\mathbf{X}_i))$ with $d(\mathbf{X}_i) = \min_{\mathbf{b}\in\mathcal{B}}d(\mathbf{X}_i,\mathbf{b})$, and the cutoff is $c=0$. This approach is analogous to the normalizing-and-pooling approach in the Multi-Cutoff RD design. Because the resulting score is a scalar, estimation and inference use the methods discussed for scalar, single-score RD designs.

Normalizing and pooling is often an effective strategy to summarize the treatment effect, and it is a reasonable first step in the analysis of Multi-Score RD designs. However, by construction, the pooled parameter cannot convey heterogeneity along the boundary. Depending on the application, this may be a limitation. For example, if the boundary is ``long'' and the density of observations along the boundary is different for treated and control groups, treated units close to the boundary may be arbitrarily far from control units close to the boundary, because the normalizing and pooling approach considers only the closest distance to the boundary but not the location along the boundary. Researchers working with long boundaries should complement the normalizing and pooling approach with location-specific effects, to assess whether the conclusions change when the two dimensions are incorporated in the analysis.

\subsubsection{The Effect of Financial Aid on Post-Secondary Education Attainment with Multiple Scores}

\begin{figure}[ht]
	\centering
	\includegraphics[width=0.60\textwidth]{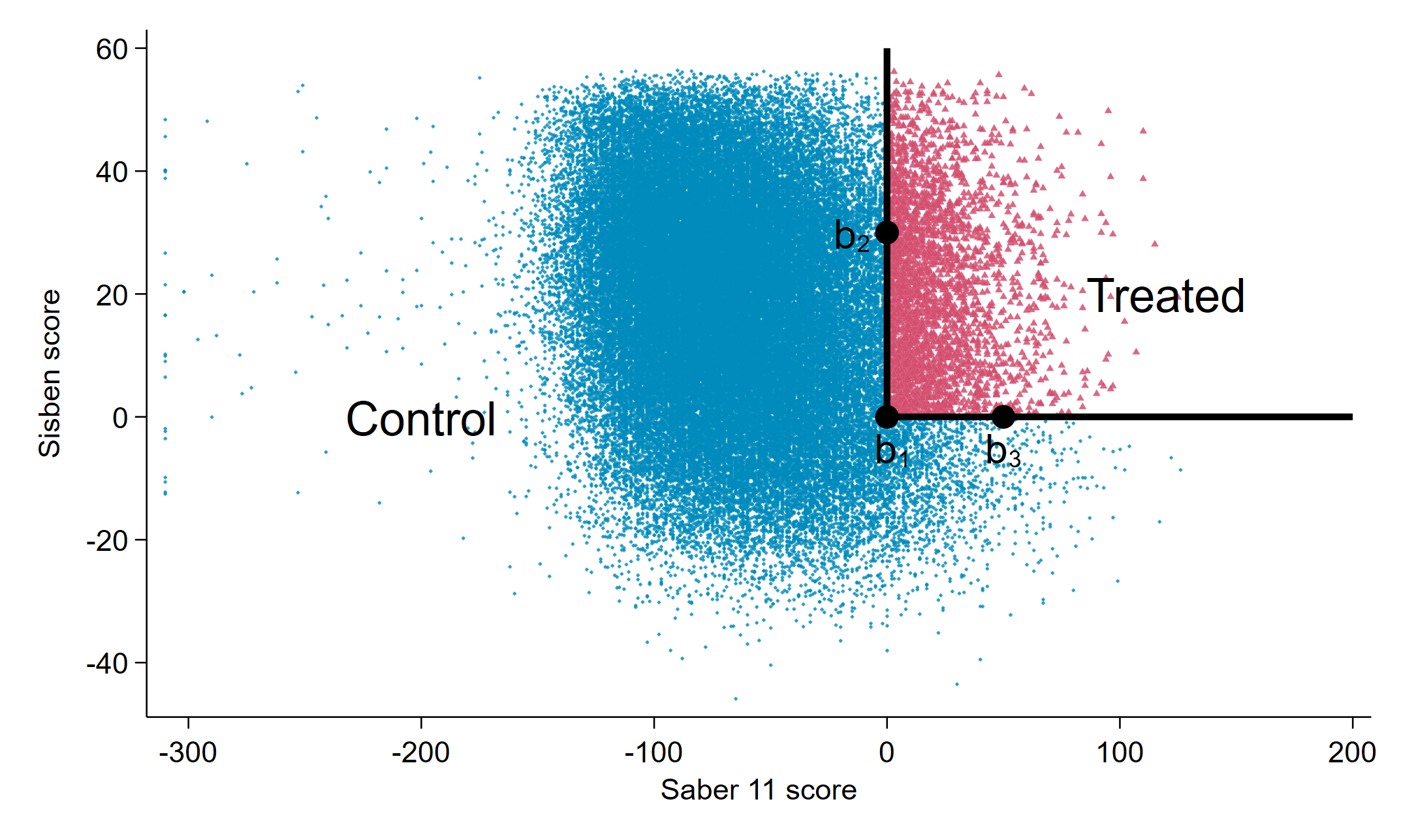}
	\caption{Treated and Control Regions---SPP data}\label{fig:RDMultiScore-SPPdata}
\end{figure}

We first illustrate a non-geographic Multi-Score RD design using both dimensions in the \cite{LondonoVelezRodriguezSanchez_2020_AEJ} study. Recall that eligibility for the SPP program was determined by both academic merit and economic need, and the score is a vector of two dimensions, where the first dimension is the SABER11 score ($X_{1i}=\mathtt{SABER11}_i$), and the second dimension is the SISBEN wealth score ($X_{2i}=\mathtt{SISBEN}_i$), so that the bivariate score is $\mathbf{X}_i=(\mathtt{SABER11}_i,\mathtt{SISBEN}_i)$. We normalized both scores so that all students face a cutoff of zero in each dimension. Figure \ref{fig:RDMultiScore-SPPdata} shows a scatterplot of one score against the other, using the full sample of students in the $2014$ cohort (a total of $574,269$ observations) and using different symbols for eligible (``treated'') and ineligible (``control'') students. The two-dimensional nature of the RD score creates a boundary in the $(\texttt{SABER11},\texttt{SISBEN})$-plane along which assignment to treatment changes discontinuously. In this case, the boundary set is $\mathcal{B}=\{(\texttt{SABER11},\texttt{SISBEN}): (\texttt{SABER11},\texttt{SISBEN}) \in \{\texttt{SABER11} \geq 0 \text{ and } \texttt{SISBEN} = 0\} \cup \{\texttt{SABER11}=0 \text{ and }\texttt{SISBEN}\geq 0\}\}$.

We begin by analyzing location-specific RD treatment effects along the boundary, taking care to choose points where the density of observations is sufficiently high. We consider three points in $\mathcal{B}$: $\mathbf{b}_1 = (0,0)$, $\mathbf{b}_2 = (30,0)$, and $\mathbf{b}_3 = (0,50)$. Students with scores near $\mathbf{b}_1$ are barely above/below the cutoff in both dimensions, while students near $\mathbf{b}_2$ are well above the SABER11 cutoff but just above/below the SISBEN cutoff, and students near the point $\mathbf{b}_3$ are well above the wealth cutoff but just above/below the merit cutoff. These points are all in $\mathcal{B}$, but the subpopulations near each one of them differ and thus lead to potentially heterogeneous effects.

For continuity-based estimation and inference at the points $\mathbf{b}_1$, $\mathbf{b}_2$ and $\mathbf{b}_3$, we use the \texttt{rdms} command. When we call the command, we include the outcome (\texttt{spadies\_any}), the two normalized scores (\texttt{running\_sisben} and \texttt{running\_saber11}), a variable that indicates which observations are assigned to treatment versus control (\texttt{tr}), and the coordinates of each boundary point (which must be passed via the \texttt{C} and \texttt{C2} options, corresponding to the two dimensions).

\labelsnippet{rdmsA}
\rsnip{Vol-2-R_LRS_rdms_threepoints.txt}{\Rlink{\thesection}{\therdmsA}}
\statasnip{Vol-2-STATA_LRS_rdms_threepoints}{\Slink{\thesection}{\therdmsA}}

The results indicate, once again, that students who are barely eligible for the SPP program enroll in a HEI at much higher rates than students who are barely ineligible, with some heterogeneity across boundary points. While the effects at $\mathbf{b}_1 = (0,0)$ and $\mathbf{b}_2 = (30,0)$ are roughly similar at roughly $32$ percentage points, the effect at $\mathbf{b}_3 = (0,50)$ is substantially smaller (around $23$ percentage points), suggesting that the effects are greater for students who are above the merit cutoff but marginal in terms of economic need than for poorer students who are marginal in terms of merit (students at $\mathbf{b}_3$ are further from and above the wealth cutoff and are thus poorer than students at $\mathbf{b}_1$ and $\mathbf{b}_2$).

We can replicate the results from \texttt{rdmc} by following the steps outlined above: first, calculate the Euclidean distance of every observation to the point, and then use that distance in a one-dimensional RD analysis. We implement this for $\mathbf{b_1}=(30,0)$,

\labelsnippet{rdmsB}
\rsnip{Vol-2-R_LRS_rdms_vs_rdrobust.txt}{\Rlink{\thesection}{\therdmsB}}
\statasnip{Vol-2-STATA_LRS_rdms_vs_rdrobust}{\Slink{\thesection}{\therdmsB}}
which leads to the same results given by \texttt{rdms} for this point.

In addition to location-specific RD effects, we can investigate the normalizing-and-pooling RD effect as we did for the Multi-Cutoff RD case. We first calculate, for each student, the distance between their two-dimensional score value $\mathbf{X}_i=(X_{1i}, X_{2i})$ and the point on the boundary closest to $\mathbf{X}_i$. In this case, because the boundary is composed of two straight lines, we can calculate the perpendicular distance between $\mathbf{X}_i$ and the closest of the two straight lines that comprise the boundary. There are two straight lines, one $X_1\equiv\mathtt{SABER11}=0$ and the other $X_2\equiv\mathtt{SISBEN}=0$, and the perpendicular distances from the point $(x_1,x_2)$ to each of them are $|x_1|=|\texttt{SABER11}|$ and $|x_2|=|\texttt{SISBEN}|$, respectively.

All treated observations have $X_{1i}\geq 0$ and $X_{2i}\geq 0$ (upper right corner of Figure \ref{fig:RDMultiScore-SPPdata}). For these observations, the closest distance to the boundary is therefore $\min\{|x_1|,|x_2|\}$. For the control observations, the closest distance to the boundary depends on where they are located: for students with $\mathtt{SABER11}_i\leq 0$ and $\mathtt{SISBEN}_i\geq 0$, the closest distance to the boundary is the distance to the line $\mathtt{SABER11}=0$; for students with $\mathtt{SABER11}_i\geq 0$ and $\mathtt{SISBEN}_i\leq 0$, the closest distance to the boundary is the distance to the line $\mathtt{SISBEN}=0$; and for students with $\mathtt{SABER11}_i\leq 0$ and $\mathtt{SISBEN}_i\leq 0$, the closest distance to the boundary is the distance to the point $(0,0)$. We summarize these distances in Table \ref{tab:RDMS-distances}. 

\begin{table}[ht]
    \centering
    \begin{tabular}{c@{\hspace{0.3in}}l}
        \toprule
        \multicolumn{1}{c}{Subgroup of observations} & 	\multicolumn{1}{c}{$d(\mathbf{X}_i)=\min_{\mathbf{b}\in\mathcal{B}}d(\mathbf{X}_i,\mathbf{b}) $}\\
        \midrule
        $\mathtt{SABER11}_i\geq 0$ and $\mathtt{SISBEN}_i\geq 0$ & $\min\{|\mathtt{SABER11}_i|,|\mathtt{SISBEN}_i|\}$\\
        $\mathtt{SABER11}_i\leq 0$ and $\mathtt{SISBEN}_i\geq 0$ & $|\mathtt{SABER11}_i|$ \\
        $\mathtt{SABER11}_i\geq 0$ and $\mathtt{SISBEN}_i\leq 0$ & $|\mathtt{SISBEN}_i|$  \\
        $\mathtt{SABER11}_i\leq 0$ and $\mathtt{SISBEN}_i\leq 0$ & $\sqrt{\mathtt{SABER11}_i^2+\mathtt{SISBEN}_i^2}$\\
        \bottomrule
    \end{tabular}
	\caption{Shortest distance to the SPP assignment boundary\label{tab:RDMS-distances}}
\end{table}

For the analysis, we use the information in Table \ref{tab:RDMS-distances} to calculate the shortest distance to the boundary for every observation, $d(\mathbf{X}_i)=\min_{\mathbf{b}\in\mathcal{B}}d(\mathbf{X}_i,\mathbf{b})$.

\labelsnippet{rdmsD}
\rsnip{Vol-2-R_LRS_rdms_perpendicular_dist_step1.txt}{\Rlink{\thesection}{\therdmsD}}
\statasnip{Vol-2-STATA_LRS_rdms_perpendicular_dist_step1}{\Slink{\thesection}{\therdmsD}}

We then modify this distance by multiplying it by -1 for all control values--- that is, we calculate the scalar normalized score $\mathfrak{D}_{i}=d(\mathbf{X}_i)(2a(\mathbf{X}_i)-1)$ with associated cutoff $c=0$. 

\labelsnippet{rdmsE}
\rsnip{Vol-2-R_LRS_rdms_perpendicular_dist_step2.txt}{\Rlink{\thesection}{\therdmsE}}
\statasnip{Vol-2-STATA_LRS_rdms_perpendicular_dist_step2}{\Slink{\thesection}{\therdmsE}}

Finally, we run a one-dimensional RD analysis using $\mathfrak{D}_{i}$ as the score (and zero as the cutoff).
\labelsnippet{rdmsF}
\rsnip{Vol-2-R_LRS_rdms_perpendicular_dist_rdrobust.txt}{\Rlink{\thesection}{\therdmsF}}
\statasnip{Vol-2-STATA_LRS_rdms_perpendicular_dist_rdrobust}{\Slink{\thesection}{\therdmsF}}

The same result can be obtained with \texttt{rdms}, passing $\mathfrak{D}_{i}$ as an argument with the option \texttt{xnorm}.
\labelsnippet{rdmsG}
\rsnip{Vol-2-R_LRS_rdms_perpendicular_dist.txt}{\Rlink{\thesection}{\therdmsG}}
\statasnip{Vol-2-STATA_LRS_rdms_perpendicular_dist}{\Slink{\thesection}{\therdmsG}}

\subsubsection{The Effect of Political Television Advertising on Voter Turnout}

We now analyze a Geographic RD design, originally studied by \cite{Keele-Titiunik_2015_PA}. This application studies the effect of political television advertising on voter turnout in New Jersey, United States, using the large differences in the volume of advertising that occur in different designated market areas (DMAs)---also known as ``media markets''. DMAs are areas created by Nielsen Media Research to measure television ratings. Political campaigns that seek to advertise on television often buy television ads by DMAs and do so strategically based on whether the DMA is located in an area where the election is expected to be close. For example, in recent U.S. presidential elections, the state of Pennsylvania has been considered a ``battleground state'' because statewide elections between Democratic and Republican candidates are typically close. Thus, during election season, both parties buy a large volume of television advertising in DMAs that are located in the state of Pennsylvania. In contrast, because the state of New York is not considered competitive, the level of advertising seen in DMAs located in New York is very low. 
	
\cite{Keele-Titiunik_2015_PA} study the West Windsor-Plainsboro (WWP) school district in New Jersey, which is split between the Philadelphia DMA and the New York City DMA (it is common for DMAs to include counties in more than one state). The authors report that during the $2008$ presidential campaign, New Jersey residents in the Philadelphia DMA saw an average of $177$ presidential campaign ads in the two months before the election, while New Jersey residents in the New York DMA saw no ads in the same period. The geographic RD design compares two adjacent areas: the part of the WWP school district contained within the Philadelphia DMA, and the part of this district contained within the New York DMA. From the New Jersey voter registration file, the authors collected the list of citizens in the WWP school district who were registered to vote by $2008$, including an indicator of whether each person voted in the $2008$ presidential election. This file also contained the residential address of each registered citizen, which allowed the authors to geolocate each person. After geolocation, each person in the registration file was associated with two geographic coordinates (latitude and longitude) which together indicate the person's residential address. Descriptive statistics for the main variables are presented in Table \ref{tab:GeoRDdescriptive}. 
	
In this geographic RD design, the unit of analysis is the individual who appears in the 2008 registration file and lives in the WWP school district, the outcome of interest is an indicator equal to one if the individual voted in the 2008 general election, the score is the latitude-longitude vector that stores the geographic coordinates of the individual's residence, $\textbf{X}_i=(\mathtt{latitude}_i,\mathtt{longitude}_i)$, the treatment of interest is political television advertisements, and the treatment assignment rule is $a(\textbf{X}_i)=\I((\mathtt{latitude}_i,\mathtt{longitude}_i) \in \mathcal{A}_\mathtt{PAdma})$, where the set $\mathcal{A}_\mathtt{PAdma}$ collects all the geographic coordinates corresponding to locations inside the Philadelphia DMA. The data was pre-processed using GIS software to include the following information:
\begin{itemize}
    \item Geographic coordinates for each observation (corresponding to each person's residence).
    \item Geographic coordinates of a collection of points along the boundary separating treated and control areas.
    \item Distance between each observation's coordinates and the closest point on the boundary.
\end{itemize}

\begin{table}[ht]
	\centering
	\resizebox{0.9\textwidth}{!}{\begin{tabular}{lcccccc}
			\toprule
			\multicolumn{1}{c}{Variable} & Mean & Median & Std. Dev. & Min. & Max. & N\\
			\midrule
			Voted in 2008 & $0.70$ & $1.00$ & $0.46$ & $0.00$ & $1.00$ & $24,461$ \\
Treated dummy & $0.61$ & $1.00$ & $0.49$ & $0.00$ & $1.00$ & $24,461$ \\
Latitude (in degrees) & $40.31$ & $40.31$ & $0.03$ & $40.25$ & $40.36$ & $24,461$ \\
Longitude (in degrees) & $-74.60$ & $-74.61$ & $0.03$ & $-74.67$ & $-74.55$ & $24,461$ \\
Age & $48.27$ & $49.00$ & $16.47$ & $18.00$ & $101.00$ & $24,461$ \\
Black dummy & $0.05$ & $0.00$ & $0.21$ & $0.00$ & $1.00$ & $24,461$ \\
Hispanic dummy & $0.03$ & $0.00$ & $0.18$ & $0.00$ & $1.00$ & $24,461$ \\
Democratic dummy & $0.33$ & $0.00$ & $0.47$ & $0.00$ & $1.00$ & $24,461$ \\
Female dummy & $0.48$ & $0.00$ & $0.50$ & $0.00$ & $1.00$ & $24,461$ \\
Chordal distance to cutoff 1 (in km) & $4.02$ & $4.15$ & $1.57$ & $0.34$ & $9.04$ & $24,461$ \\
Chordal distance to cutoff 2 (in km) & $3.67$ & $3.75$ & $1.41$ & $0.31$ & $8.87$ & $24,461$ \\
Chordal distance to cutoff 3 (in km) & $3.64$ & $3.57$ & $1.38$ & $0.22$ & $8.70$ & $24,461$ \\
Perpendicular distance to the border (in km) & $2.67$ & $2.67$ & $1.34$ & $0.01$ & $8.33$ & $24,461$ \\ \bottomrule

	\end{tabular}}
	\caption{Descriptive Statistics---DMA data}\label{tab:GeoRDdescriptive}
\end{table}

\begin{figure}[ht]
    \centering
    \includegraphics[scale=0.60]{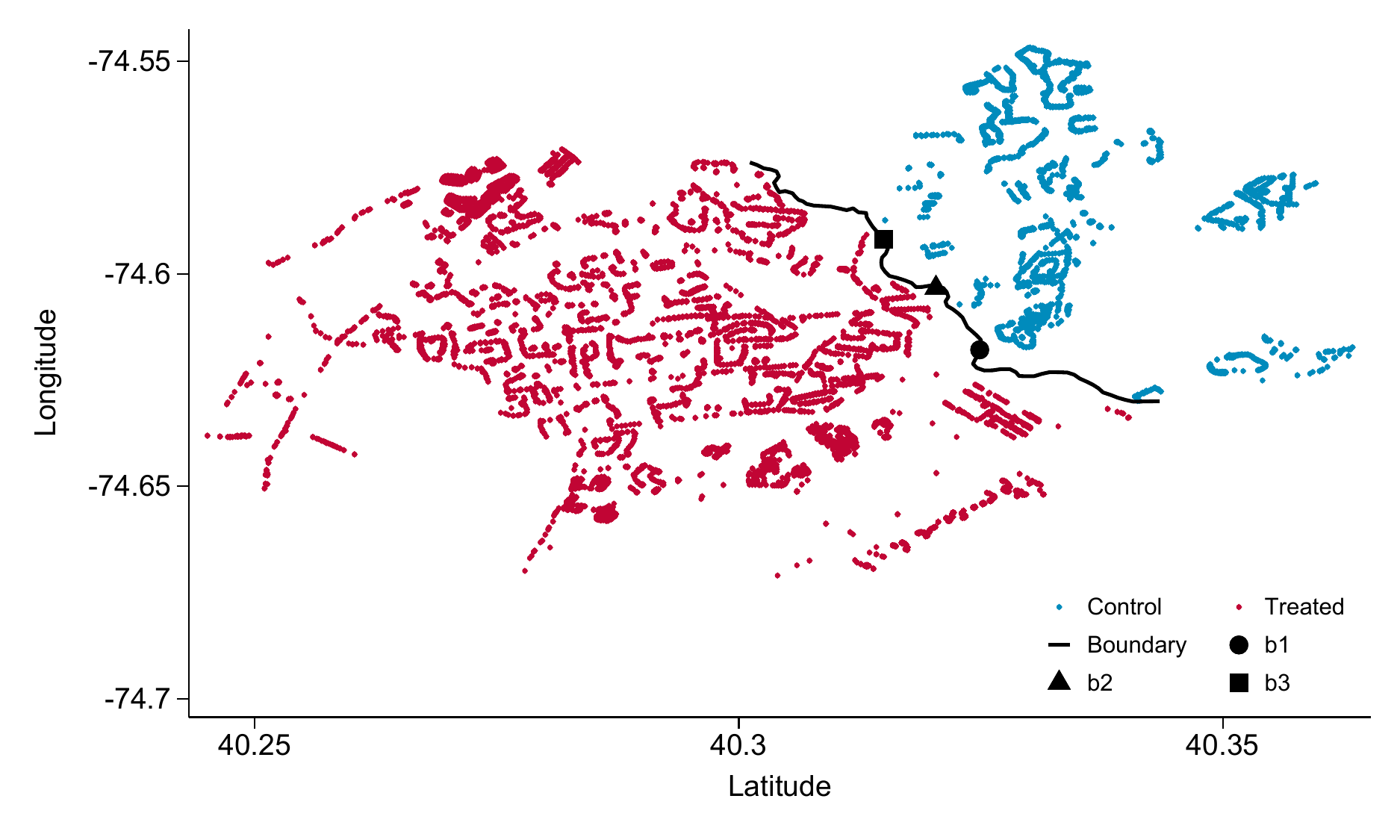}
    \caption{Treated and Control Geographic Areas---Media Market data}\label{fig:RDGeo}
\end{figure}

Figure \ref{fig:RDGeo} shows the raw scatter plot of longitude against latitude for all observations in the replication dataset; the plot also shows the boundary that separates the treated and control areas. Figure \ref{fig:RDGeo} depicts a real data version of Figure \ref{fig:RDMultiScore-geo}, where the assignment boundary is irregular. In contrast to non-geographic applications of the Multi-Score RD Design such as the SPP example illustrated in Figure \ref{fig:RDMultiScore-SPPdata}, the boundary that separates treated and control areas in a Geograhic RD design does not typically have a closed form expression. The boundaries between geographic units (counties, school districts, DMAs) are typically decided by governmental or other administrative units; their precise location is given via a collection of files, sometimes referred to collectively as shape files, that store the location (and also attributes such as elevation) of geographic features (points, lines, and polygons). These files are analyzed using geographic information systems (GIS) software. Thus, instead of deriving the set $\mathcal{B}$ from the treatment assignment rule, as we did in the SPP example, obtaining the boundary in a Geographic RD design requires external information. In this case, the authors obtained the shape files with the polygons representing the New York and Philadelaphia DMAs and the West Windsor-Plainsboro school district, and using GIS software they obtained a set of 80 latitude-longitude points that are on the border between the New York and Philadelaphia DMAs and inside the West Windsor-Plainsboro district. Adding these 80 points as a line in the scatter plot in Figure \ref{fig:RDGeo} produces the boundary.
 
The authors selected three latitude-longitude points on the boundary where RD treatment effects were estimated: $\mathbf{b}_1=(40.32489,-74.61789)$, $\mathbf{b}_2=(40.32037,-74.60335)$, and $\mathbf{b}_3=(40.31497,-74.59191)$. These points are represented by the circle, square and triangle in Figure \ref{fig:RDGeo}. The points $\mathbf{b}_1$ and $\mathbf{b}_3$ were chosen to split the boundary into three equal segments roughly 2.3 kilometers long; the point $\mathbf{b}_2$ is the midpoint between $\mathbf{b}_1$ and $\mathbf{b}_3$.

\begin{figure}[ht]
	\begin{subfigure}{0.48\textwidth}
		\centering
		\includegraphics[scale=0.45]{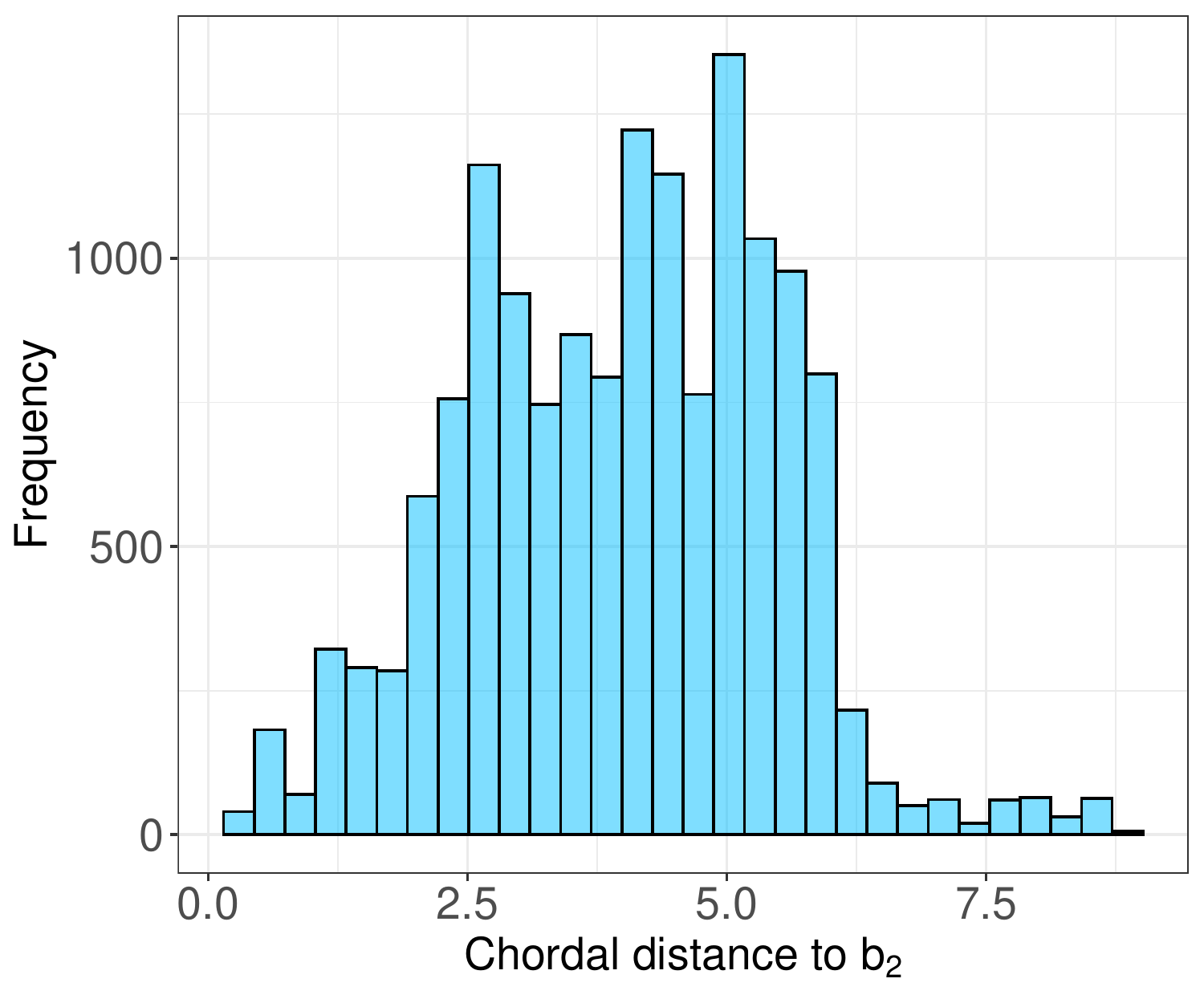} 
		\subcaption{Treated observations} 
	\end{subfigure}
	\begin{subfigure}{0.48\textwidth}
		\centering
		\includegraphics[scale=0.45]{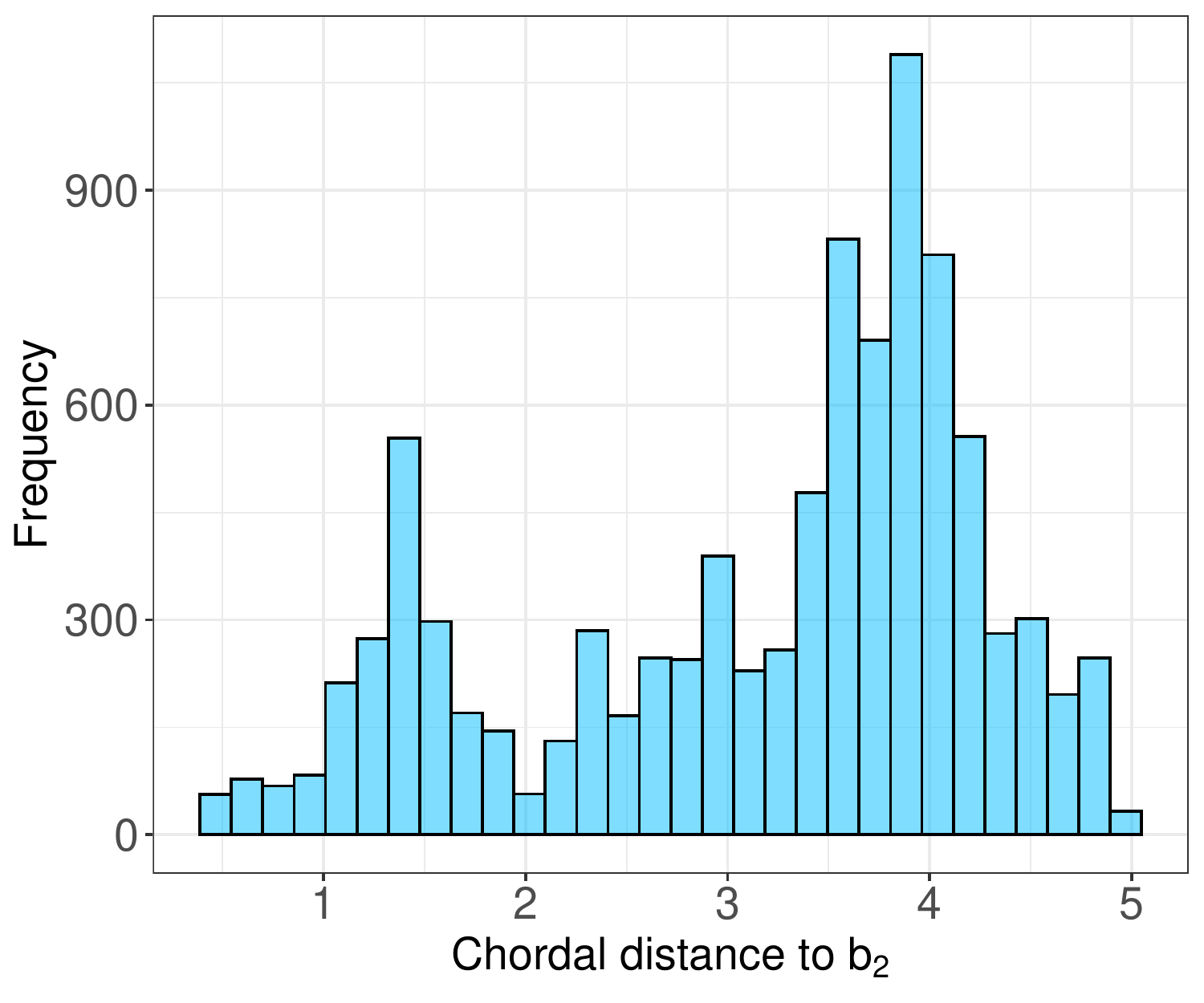} 
		\subcaption{Control observations}
	\end{subfigure}
	\caption{Histogram of Chordal Distance to Boundary Point $\mathbf{b}_2$---DMA data}	\label{fig:ChorDistHist}
\end{figure}

We illustrate how to estimate point-specific effects by analyzing the middle point, $\mathbf{b}_2$. Using the latitude and longitude coordinates of every observation as inputs, we start by calculating the chordal distance from each observation to $\mathbf{b}_2$. Figure \ref{fig:ChorDistHist} shows the histogram of the chordal distance to boundary point $\mathbf{b}_2$ measured in kilometers, separately for treated and control observations. The density of observations near the boundary is low, which is typical in geographic RD applications where the boundary splits less populated areas. We also see that the distances do not get all the way to zero. For example, the minimum distance in the treatment group is $0.30689$ km, and the minimum distance in the control group is $0.4239642$ km. It is important for researchers to check the density of the distance measure; very few observations with distances near zero may indicate that the areas surrounding the boundary are not sufficiently populated, which will result in excessive extrapolation when estimating RD effects. 

The outcome of interest is an indicator equal to one if the person voted in the $2008$ general election. Recall that the treatment assignment indicator is equal to one if the person's residence is located in the Philadelphia DMA, where political TV ads were plentiful, and zero if the person's residence is in the New York DMA, where the volume of ads was very low. Under appropriate assumptions, the RD effect thus captures the effect of a high volume of political TV ads on voter turnout, for voters in the WWP school district who reside right at the boundary between the Philadelphia and New York DMAs.

We estimate the RD effect at $\mathbf{b}_2$ with local polynomials implemented with \texttt{rdrobust}, using the chordal distance from each person's residence to this point as the score. (We omit the code that calculates the chordal distance to conserve space, but provide it in the accompanying replication materials.)

\labelsnippet{geocutoffB}
\rsnip{Vol-2-R_kt_rdrobust_cutoff2.txt}{\Rlink{\thesection}{\thegeocutoffB}}
\statasnip{Vol-2-STATA_kt_rdrobust_cutoff2}{\Slink{\thesection}{\thegeocutoffB}}

The local linear point estimate is very small, $-0.002$, with a robust p-value of $0.838$ and a robust confidence interval roughly centered around zero. Thus, we see no effect of residing in the Philadelphia DMA on voter turnout in $2008$. The effects at $\mathbf{b}_1$ and $\mathbf{b}_3$ can be estimated analogously. 

The default implementation with \texttt{rdrobust} used above chooses the bandwidth optimally; when the analysis is performed for each boundary point, this strategy may result in some observations being included in the analysis of more than one point. If researchers want to avoid reusing observations between boundary points, they can choose the bandwidth manually. For example, in this application, the distance between $\mathbf{b}_2$ and each of the boundary points $\mathbf{b}_1$ and $\mathbf{b}_3$ is roughly $1.15$ km; thus, to ensure that each observation is used in exactly one boundary point, researchers could set the bandwidth manually to $0.575$ km (assuming that there are enough observations). We omit this analysis because of space considerations.

We can also use the \texttt{rdms} command for analysis, but this will only be useful if the researcher is interested in using Euclidean distance, as \texttt{rdms} only uses this metric to calculate the distances between each observation's location and each boundary point.
\labelsnippet{rdmsAA}
\rsnip{Vol-2-R_kt_rdms_basic.txt}{\Rlink{\thesection}{\therdmsAA}}
\statasnip{Vol-2-STATA_kt_rdms_basic}{\Slink{\thesection}{\therdmsAA}}

The middle row corresponds to the effect for the boundary point $\mathbf{b}_2$, which is $0.034$. This effect is different from the point estimate of $-0.002$ estimated above for the same point. The discrepancy occurs because \texttt{rdms} uses the Euclidean metric to calculate the distances between the raw latitude and longitude inputs passed to the function and the boundary point. In contrast, our result above obtained with \texttt{rdrobust} used the chordal distance as the score, which we calculated manually. Despite the difference in point estimate, the conclusions are the same, as both robust p-values are similar (between $0.8$ and $0.9$), and both robust confidence intervals are roughly symmetrical around zero.

Finally, we use \texttt{rdrobust} to calculate the pooled RD effect for all observations together, using the perpendicular distance to the boundary as the score (calculated by Keele and Titiunik using GIS software). 
\labelsnippet{rdrobustABC}
\rsnip{Vol-2-R_kt_rdrobust_perpdist.txt}{\Rlink{\thesection}{\therdrobustABC}}
\statasnip{Vol-2-STATA_kt_rdrobust_perpdist}{\Slink{\thesection}{\therdrobustABC}}

\subsection{Further Reading}

\citet{Cattaneo-Keele-Titiunik-VazquezBare_2016_JOP} introduce an RD framework based on potential outcomes and continuity conditions to analyze the Multi-Cutoff RD design, studying the interpretation of the normalizing-and-pooling estimator and introducing the distinction between cumulative and non-cumulative cutoffs. They also provide analogous results for Fuzzy and Kink RD designs, and discuss the connections between Multi-Score and Multi-Cutoff RD designs in the supplemental appendix. \citet{Cattaneo-Keele-Titiunik-VazquezBare_2021_JASA} propose to use a Multi-Cutoff RD framework for extrapolation of treatment effects; they present multi-cutoff extrapolation for both the continuity-based and the local randomization approaches (the latter is covered in supplemental appendix). \citet{PapayWillettMurnane2011-JoE}, \citet{ReardonRobinson2012-JREE}, and \citet{Wong-Steiner-Cook_2012-JEBS} discuss generic Multi-Score RD settings, and \citet{Keele-Titiunik_2015_PA} discuss a geographic continuity-based RD design. \cite{Banerjee2005-Biometrics} discusses different metrics appropriate for measuring distance between geographic locations on Earth. \cite{Keele-Titiunik_2018_CESIFO} discuss the application of a Geographic RD design to the study of all-mail voting in Colorado, considering the possibility of spillovers between treated and control areas. \citet{Cattaneo-Titiunik-Yu_2024_wp} develop uniform inference methods for boundary discontinuity designs and study the properties of two approaches---one using the raw two-dimensional score and the other using the collapsed one-dimensional score. Further references are provided in \citet{Cattaneo-Titiunik_2022_ARE}.

\clearpage
\section{Final Remarks}

This monograph continues the practical discussion of RD analysis that we started in \textit{Foundations}. In that first volume, we focused on the canonical RD setup where the running variable has a single dimension and is continuously distributed, there is only one cutoff, compliance with the treatment assignment is perfect, and all effects are defined at the cutoff and estimated via local polynomials based on extrapolation and continuity assumptions. This volume explored the implications of relaxing these assumptions.

Section \ref{sec:localrand} presented the local randomization framework as an alternative way of analyzing and interpreting RD designs. Instead of focusing on the cutoff at which the assignment switches from control to treatment, this approach defines a window around the cutoff and deploys assumptions akin to those in a randomized controlled experiment to define and analyze treatment effects. Because these assumptions are stronger than the standard RD continuity assumptions, we presented this alternative approach as a complement to the continuity-based methods in \textit{Foundations}. Still, local randomization methods are an important part of the RD toolkit because it is common to justify the RD assumptions by invoking a similarity between the RD treatment assignment and the way the treatment is assigned in a randomized experiment. This similarity was invoked by \citet{Thistlethwaite-Campbell_1960_JEP} to justify the RD design in their foundational paper. We hope that our discussion of the advantages and limitations of the local randomization approach is useful in clarifying the analogy between RD designs and randomized experiments that is so often invoked in practice.

Section \ref{sec:FuzzyRD} focused on Fuzzy RD designs and discussed best practices for analysis when compliance with treatment is imperfect. This is relevant to many real-world applications of the RD design. For example, in social programs and other policies that are assigned via RD rules, individuals are encouraged to receive a treatment when their score is above a cutoff rather than coerced to take a treatment. We introduced and discussed several treatment effects in the context of Fuzzy RD designs, and explained how continuity-based and local randomization methods can be effectively deployed in such a context. We also highlighted the role of validation and falsification methods.

Section \ref{sec:discrete} discussed the features and limitations of continuity-based and local randomization methods when the running variable is discrete and thus different observations have the same value of the score. Continuity-based methods are not applicable to the analysis of RD designs with discrete running variables without further assumptions allowing extrapolation, while local randomization methods are often more appropriate. Our discussion illustrated how local randomization concepts and methods can be effectively deployed to estimate useful treatment effects in this case, focusing on both standard parameters as well as new parameters arising from the discreteness of the score.

Finally, Section \ref{sec:multiRD} studied Multidimensional RD designs, covering Multi-Cutoff, Multi-Score, and Geographic RD designs. These are RD designs where there is more than one cutoff or the score has more than one dimension, or both. Although most of the concepts and assumptions remain unchanged, the introduction of multiple dimensions leads to many different parameters of potential interest. Our discussion outlined how to define, analyze, and interpret such parameters. We also offered several empirical illustrations highlighting some of the particularities of each multidimensional RD design.

It is our hope that the combination of \textit{Foundations} and \textit{Extensions} provides a useful practical guide for empirical researchers, and contributes to the transparency and replicability of RD analyses across all disciplines.

\clearpage
\bibliographystyle{econometrica}
\bibliography{CIT_2024_CUP--bib.bib}

@article{Abadie-Cattaneo_2018_ARE,
	title   = {Econometric Methods for Program Evaluation},
	author  = {Abadie, Alberto and Matias D. Cattaneo},
	journal = {Annual Review of Economics}, volume={10}, number={}, pages={465--503}, year={2018}
}

@article{Andrews-Stock-Sun_2019_ARE,
  title={Weak instruments in instrumental variables regression: Theory and practice},
  author={Andrews, Isaiah and Stock, James H and Sun, Liyang},
  journal={Annual Review of Economics},
  volume={11},
  pages={727--753},
  year={2019},
  publisher={Annual Reviews}
}

@article{Arai-Hsu-Kitagawa-Mourifie-Wan_2022_QE,
	title={Testing Identifying Assumptions in Fuzzy Regression Discontinuity Designs},
	author={Arai, Yoichi and Hsu, Yuchin and Kitagawa, Toru and Mourifi{\'e}, Ismael and Wan, Yuanyuan},
	journal = {Quantitative Economics}, volume={13}, number={1}, pages={1--28}, year={2022}
}

@article{Banerjee2005-Biometrics,
	title={On geodetic distance computations in spatial modeling},
	author={Banerjee, Sudipto},
	journal={Biometrics},
	volume={61},
	number={2},
	pages={617--625},
	year={2005}
}

@Article{Barreca-Lindo-Waddell_2016_EI,
  Title                    = {Heaping-Induced Bias in Regression-Discontinuity Designs},
  Author                   = {Barreca, Alan I. and Jason M. Lindo and Glen R. Waddell},
  Journal                  = {Economic Inquiry},
  Year                     = {2016},
  Number                   = {1},
  Pages                    = {268--293},
  Volume                   = {54}
}

@article{Calonico-Cattaneo-Farrell_2018_JASA,
	author  = {Calonico, Sebastian and Matias D. Cattaneo and Max H. Farrell},
	title   = {On the Effect of Bias Estimation on Coverage Accuracy in Nonparametric Inference},
	journal = {Journal of the American Statistical Association}, volume={113}, number={522}, pages={767--779}, year={2018}
}

@article{Calonico-Cattaneo-Farrell_2020_ECTJ,
	author  = {Calonico, Sebastian and Matias D. Cattaneo and Max H. Farrell},
	title   = {Optimal Bandwidth Choice for Robust Bias Corrected Inference in Regression Discontinuity Designs},
	journal = {Econometrics Journal}, volume={23}, number={2}, pages={192--210}, year={2020}
}

@article{Calonico-Cattaneo-Farrell_2022_Bernoulli,
	author  = {Calonico, Sebastian and Matias D. Cattaneo and Max H. Farrell},
	title   = {Coverage Error Optimal Confidence Intervals for Local Polynomial Regression},
	journal = {Bernoulli}, volume={28}, number={4}, pages={2998--3022}, year={2022}
}

@article{Calonico-Cattaneo-Farrell-Titiunik_2017_Stata,
  author  = {Calonico, Sebastian and Matias D. Cattaneo and Max H. Farrell and Rocio Titiunik},
  title   = {\texttt{rdrobust}: Software for Regression Discontinuity Designs},
  journal = {Stata Journal}, volume={17}, number={2}, pages={372--404}, year={2017}
}

@article{Calonico-Cattaneo-Farrell-Titiunik_2019_RESTAT,
  author  = {Calonico, Sebastian and Matias D. Cattaneo and Max H. Farrell and Rocio Titiunik},
  title   = {Regression Discontinuity Designs using Covariates},
  journal = {Review of Economics and Statistics}, volume={101}, number={3}, pages={442--451}, year={2019}
}

@article{Calonico-Cattaneo-Titiunik_2014_ECMA,
  author  = {Calonico, Sebastian and Matias D. Cattaneo and Rocio Titiunik},
  title   = {Robust Nonparametric Confidence Intervals for Regression-Discontinuity Designs},
  journal = {Econometrica}, volume={82}, number={6}, pages={2295--2326}, year={2014}
}

@article{Calonico-Cattaneo-Titiunik_2014_Stata,
  author  = {Calonico, Sebastian and Matias D. Cattaneo and Rocio Titiunik},
  title   = {Robust Data-Driven Inference in the Regression-Discontinuity Design},
  journal = {Stata Journal}, volume={14}, number={4}, pages={909--946}, year={2014}
}

@article{Calonico-Cattaneo-Titiunik_2015_R,
  author  = {Calonico, Sebastian and Matias D. Cattaneo and Rocio Titiunik},
  title   = {\texttt{rdrobust}: An R Package for Robust Nonparametric Inference in Regression-Discontinuity Designs},
  journal = {R Journal}, volume={7}, number={1}, pages={38--51}, year={2015}
}

@article{Calonico-Cattaneo-Titiunik_2015_JASA,
  author  = {Calonico, Sebastian and Matias D. Cattaneo and Rocio Titiunik},
  title   = {Optimal Data-Driven Regression Discontinuity Plots},
  journal = {Journal of the American Statistical Association}, volume={110}, number={512}, pages={1753--1769}, year={2015}
}

@Article{Cattaneo-Frandsen-Titiunik_2015_JCI,
  Title                    = {Randomization Inference in the Regression Discontinuity Design: An Application to Party Advantages in the U.S. Senate},
  Author                   = {Cattaneo, Matias D. and Brigham Frandsen and Rocio Titiunik},
  Journal                  = {Journal of Causal Inference},
  Year                     = {2015},
  Number                   = {1},
  Pages                    = {1--24},
  Volume                   = {3}
}

@Book{Cattaneo-Idrobo-Titiunik_2020_Vol1,
  Title                    = {A Practical Introduction to Regression Discontinuity Designs: Foundations},
  Author                   = {Cattaneo, Matias D. and Nicol\'{a}s Idrobo and Roc\'{i}o Titiunik},
  Publisher                = {Cambridge Elements: Quantitative and Computational Methods for Social Science, Cambridge University Press},
city = {Cambridge, UK},
  Year                     = {2020}
}

@Article{Cattaneo-Jansson-Ma_2018_Stata,
  Title                    = {Manipulation Testing based on Density Discontinuity},
  Author                   = {Cattaneo, Matias D. and Michael Jansson and Xinwei Ma},
  Journal                  = {Stata Journal},
  Year                     = {2018},
  Number                   = {1},
  Pages                    = {234--261},
  Volume                   = {18}
}

@article{Cattaneo-Jansson-Ma_2020_JASA,
	author  = {Matias D. Cattaneo and Michael Jansson and Xinwei Ma},
	title   = {Simple Local Polynomial Density Estimators},
	journal = {Journal of the American Statistical Association}, volume={115}, number={531}, pages={1449--1455}, year={2020}
}

@incollection{Cattaneo-Keele-Titiunik_2023_HandbookCh,
	author    = {Cattaneo, Matias D. and Luke Keele and Rocio Titiunik},
	title     = {Covariate Adjustment in Regression Discontinuity Designs},
	booktitle = {Handbook of Matching and Weighting in Causal Inference}, editor = {J. R. Zubizarreta, E. A. Stuart, D. S. Small and P. R. Rosenbaum},
	chapter   = {8}, pages = {153--168}, publisher={Chapman \& Hall}, address={Boca Raton, FL}, year={2023}
}

@article{Cattaneo-Keele-Titiunik_2023_SIM,
	author  = {Cattaneo, Matias D. and Luke Keele and Rocio Titiunik},
	title   = {A Guide to Regression Discontinuity Designs in Medical Applications},
	journal = {Statistics in Medicine}, volume={42}, number={24}, pages={4484-4513}, year={2023}
}

@article{Cattaneo-Keele-Titiunik-VazquezBare_2016_JOP,
  author  = {Cattaneo, Matias D. and Luke Keele and Rocio Titiunik and Gonzalo Vazquez-Bare},
  title   = {Interpreting Regression Discontinuity Designs with Multiple Cutoffs},
  journal = {Journal of Politics}, volume={78}, number={4}, pages={1229--1248}, year={2016}
}

@article{Cattaneo-Keele-Titiunik-VazquezBare_2021_JASA,
  author  = {Cattaneo, Matias D. and Luke Keele and Rocio Titiunik and Gonzalo Vazquez-Bare},
  title   = {Extrapolating Treatment Effects in Multi-Cutoff Regression Discontinuity Designs},
  journal = {Journal of the American Statistical Association}, volume={116}, number={536}, pages={1941--1952}, year={2021}
}

@article{Cattaneo-Titiunik_2022_ARE,
  author  = {Cattaneo, Matias D. and Rocio Titiunik},
  title   = {Regression Discontinuity Designs},
  journal = {Annual Review of Economics}, volume={14}, number={}, pages={821--851}, year={2022}
}

@article{Cattaneo-Titiunik-VazquezBare_2016_Stata,
  author  = {Cattaneo, Matias D. and Rocio Titiunik and Gonzalo Vazquez-Bare},
  title   = {Inference in Regression Discontinuity Designs under Local Randomization},
  journal = {Stata Journal}, volume={16}, number={2}, pages={331--367}, year={2016}
}

@article{Cattaneo-Titiunik-VazquezBare_2017_JPAM,
  author  = {Cattaneo, Matias D. and Rocio Titiunik and Gonzalo Vazquez-Bare},
  title   = {Comparing Inference Approaches for RD Designs: A Reexamination of the Effect of Head Start on Child Mortality},
  journal = {Journal of Policy Analysis and Management}, volume={36}, number={3}, pages={643--681}, year={2017}
}

@article{Cattaneo-Titiunik-VazquezBare_2019_Stata,
	author  = {Cattaneo, Matias D. and Rocio Titiunik and Gonzalo Vazquez-Bare},
	title   = {Power Calculations for Regression Discontinuity Designs},
	journal = {Stata Journal}, volume={19}, number={1}, pages={210--245}, year={2019}
}

@incollection{Cattaneo-Titiunik-VazquezBare_2020_BookCh,
	author    = {Cattaneo, Matias D. and Rocio Titiunik and Gonzalo Vazquez-Bare},
	title     = {The Regression Discontinuity Design},
	booktitle = {Handbook of Research Methods in Political Science and International Relations}, editor = {L. Curini and R. J. Franzese},
	chapter   = {44}, pages = {835--857}, publisher={Sage Publications}, address={London, UK}, year={2020}
}

@article{Cattaneo-Titiunik-VazquezBare_2020_Stata,
	author  = {Cattaneo, Matias D. and Rocio Titiunik and Gonzalo Vazquez-Bare},
	title   = {Analysis of Regression Discontinuity Designs with Multiple Cutoffs or Multiple Scores},
	journal = {Stata Journal}, volume={20}, number={4}, pages={866--891}, year={2020}
}

@article{Cattaneo-Titiunik-Yu_2024_wp,
  author  = {Cattaneo, Matias D. and Rocio Titiunik and Rae Yu},
  title   = {Estimation and Inference in Boundary Discontinuity Designs},
  journal = {working paper}, volume={}, number={}, pages={}, year={2024}
}

@Article{Dong_2015_JAE,
  Title                    = {Regression Discontinuity Applications with Rounding Errors in the Running Variable},
  Author                   = {Dong, Yingying},
  Journal                  = {Journal of Applied Econometrics},
  Year                     = {2015},
  Number                   = {3},
  Pages                    = {422--446},
  Volume                   = {30}
}

@article{Dong_2018_OxfordBull,
  title={Alternative Assumptions to Identify LATE in FFuzzy Regression Discontinuity Designs},
  author={Dong, Yingying},
  journal={Oxford Bulletin of Economics and Statistics},
  volume={80},
  number={5},
  pages={1020--1027},
  year={2018}
}

@Article{Ernst_2004_SS,
  Title                    = {Permutation Methods: A Basis for Exact Inference},
  Author                   = {Ernst, Michael D},
  Journal                  = {Statistical Science},
  Year                     = {2004},
  Number                   = {4},
  Pages                    = {676--685},
  Volume                   = {19}
}

@Article{Feir-Lemieux-Marmer_2016_JBES,
  Title                    = {Weak identification in fuzzy regression discontinuity designs},
  Author                   = {Feir, Donna and Lemieux, Thomas and Marmer, Vadim},
  Journal                  = {Journal of Business \& Economic Statistics},
  Year                     = {2016},
  Number                   = {2},
  Pages                    = {185--196},
  Volume                   = {34}
}

@Article{Hahn-Todd-vanderKlaauw_2001_ECMA,
  Title                    = {Identification and Estimation of Treatment Effects with a Regression-Discontinuity Design},
  Author                   = {Hahn, Jinyong and Petra Todd and Wilbert van der Klaauw},
  Journal                  = {Econometrica},
  Year                     = {2001},
  Number                   = {1},
  Pages                    = {201--209},
  Volume                   = {69}
}

@article{Hyytinen-etal-_2018_QE,
  title={When does regression discontinuity design work? Evidence from random election outcomes},
  author={Hyytinen, Ari and Meril{\"a}inen, Jaakko and Saarimaa, Tuukka and Toivanen, Otto and Tukiainen, Janne},
  journal={Quantitative Economics},
  volume={9},
  number={2},
  pages={1019--1051},
  year={2018},
  publisher={Wiley Online Library}
}

@Book{Imbens-Rubin_2015_Book,
  Title                    = {Causal Inference in Statistics, Social, and Biomedical Sciences},
  Author                   = {Imbens,~{Guido W.} and Donald B. Rubin},
  Publisher                = {Cambridge University Press},
  Year                     = {2015}
}

@Article{Keele-Titiunik_2015_PA,
  Title                    = {Geographic Boundaries as Regression Discontinuities},
  Author                   = {Keele, Luke~J. and Roc\'{i}o Titiunik},
  Journal                  = {Political Analysis},
  Year                     = {2015},
  Number                   = {1},
  Pages                    = {127--155},
  Volume                   = {23}
}

@article{Keele-Titiunik_2018_CESIFO,
	title={Geographic natural experiments with interference: The effect of all-mail voting on turnout in Colorado},
	author={Keele, Luke and Titiunik, Roc{\'\i}o},
	journal={CESifo Economic Studies},
	volume={64},
	number={2},
	pages={127--149},
	year={2018},
	publisher={Oxford University Press}
}

@Article{Lee_2008_JoE,
  Title                    = {Randomized Experiments from Non-random Selection in U.S. House Elections},
  Author                   = {Lee, David S.},
  Journal                  = {Journal of Econometrics},
  Year                     = {2008},
  Number                   = {2},
  Pages                    = {675--697},
  Volume                   = {142}
}

@Article{Lee-Card_2008_JoE,
  Title                    = {Regression discontinuity inference with specification error},
  Author                   = {Lee, David S and Card, David},
  Journal                  = {Journal of Econometrics},
  Year                     = {2008},
  Number                   = {2},
  Pages                    = {655--674},
  Volume                   = {142}
}

@Article{Lindo-Sanders-Oreopoulos_2010_AEJ,
  Title                    = {Ability, Gender, and Performance Standards: Evidence from Academic Probation},
  Author                   = {Lindo, Jason M. and Nicholas J. Sanders and Philip Oreopoulos},
  Journal                  = {American Economic Journal: Applied Economics},
  Year                     = {2010},
  Number                   = {2},
  Pages                    = {95--117},
  Volume                   = {2}
}

@article{LondonoVelezRodriguezSanchez_2020_AEJ,
	title={Upstream and downstream impacts of college merit-based financial aid for low-income students: Ser Pilo Paga in Colombia},
	author={Londo{\~n}o-V{\'e}lez, Juliana and Rodr{\'\i}guez, Catherine and S{\'a}nchez, Fabio},
	journal={American Economic Journal: Economic Policy},
	volume={12},
	number={2},
	pages={193--227},
	year={2020}
}

@Article{McCrary_2008_JoE,
  Title                    = {Manipulation of the Running Variable in the Regression Discontinuity Design: A Density Test},
  Author                   = {McCrary, Justin},
  Journal                  = {Journal of Econometrics},
  Year                     = {2008},
  Number                   = {2},
  Pages                    = {698--714},
  Volume                   = {142}
}

@Article{PapayWillettMurnane2011-JoE,
  Title                    = {Extending the regression-discontinuity approach to multiple assignment variables},
  Author                   = {Papay, John P and Willett, John B and Murnane, Richard J},
  Journal                  = {Journal of Econometrics},
  Year                     = {2011},
  Number                   = {2},
  Pages                    = {203--207},
  Volume                   = {161}
}

@Article{ReardonRobinson2012-JREE,
  Title                    = {Regression discontinuity designs with multiple rating-score variables},
  Author                   = {Reardon, Sean F and Robinson, Joseph P},
  Journal                  = {Journal of Research on Educational Effectiveness},
  Year                     = {2012},
  Number                   = {1},
  Pages                    = {83--104},
  Volume                   = {5},
  Publisher                = {Taylor \& Francis}
}

@Book{Rosenbaum_2010_Book,
  Title                    = {Design of Observational Studies},
  Author                   = {Rosenbaum, Paul R.},
  Publisher                = {Springer},
  Year                     = {2010},
  Address                  = {New York}
}

@Article{Sekhon-Titiunik_2016_ObsStud,
  Title                    = {Understanding Regression Discontinuity Designs as Observational Studies},
  Author                   = {Sekhon, Jasjeet S. and Titiunik, Roc\'{i}o},
  Journal                  = {Observational Studies},
  Year                     = {2016},
  Pages                    = {174--182},
  Volume                   = {2}
}

@InCollection{Sekhon-Titiunik_2017_AIE,
  Title                    = {On Interpreting the Regression Discontinuity Design as a Local Experiment},
  Author                   = {Sekhon, Jasjeet S. and Titiunik, Roc\'{i}o},
  Booktitle                = {Regression Discontinuity Designs: Theory and Applications (Advances in Econometrics, volume 38)},
  Publisher                = {Emerald Group Publishing; distributed by Turpin Distribution, Ashland, OH},
  Year                     = {2017},
city = {Bingley, UK},
  Editor                   = {Cattaneo, Matias D. and Juan Carlos Escanciano},
  Pages                    = {1--28}
}

@Article{Thistlethwaite-Campbell_1960_JEP,
  Title                    = {Regression-Discontinuity Analysis: An Alternative to the Ex-Post Facto Experiment},
  Author                   = {Thistlethwaite, Donald L. and Campbell, Donald T.},
  Journal                  = {Journal of Educational Psychology},
  Year                     = {1960},
  Number                   = {6},
  Pages                    = {309--317},
  Volume                   = {51}
}

@incollection{Titiunik_2021_HandbookCh,
	author = {Titiunik, Rocio},
	booktitle = {Advances in Experimental Political Science},
	editor = {J. N. Druckman and D. P. Gree},
	publisher = {Cambridge University Press},
	title = {Natural Experiments}, chapter=6, pages={103-129},
	year = {2021}
}

@Article{Wong-Steiner-Cook_2012-JEBS,
  Title                    = {Analyzing Regression-Discontinuity Designs With Multiple Assignment Variables A Comparative Study of Four Estimation Methods},
  Author                   = {Wong, Vivian C. and Steiner, Peter M. and Cook, Thomas D.},
  Journal                  = {Journal of Educational and Behavioral Statistics},
  Year                     = {2013},
  Number                   = {2},
  Pages                    = {107--141},
  Volume                   = {38}
}

\newpage
\appendix

\end{document}